\documentclass[
 reprint,=
 amsmath,amssymb,
 aps,unsortedaddress]{revtex4-1}
 
\usepackage[pdftex]{graphicx,color}
\usepackage{dcolumn}
\usepackage{bm}

\usepackage{amsfonts}
\usepackage{amssymb}
\usepackage{amsmath}
\usepackage{amsthm}

\usepackage{rotate}

\let\oldmarginpar\marginpar
\renewcommand\marginpar[1]{\-\oldmarginpar[\raggedleft\tiny #1]%
{\raggedright\tiny #1}}

\def\KeyWord#1{$\backslash$\IfColor{$\!\!$\textRed{#1}\textBlack}{#1}$\!\!$}

\newcommand{\be}{\begin{equation} }
\newcommand{\ee}{\end{equation} }
\newcommand{\ba}{\begin{eqnarray} }
\newcommand{\ea}{\end{eqnarray} }
\newcommand{\n}{\nonumber \\ }

\newcommand{\mac}{\mathcal}

\newcommand{\bit}{\begin{itemize}}
\newcommand{\eit}{\end{itemize}}

\graphicspath{{FinalFigures/}}

\begin{document}

\title{Fractonic Chern-Simons and BF theories}
\author{Yizhi You}
\affiliation{Princeton Center for Theoretical Science, Princeton University, 
NJ, 08544, USA}

\author{Trithep Devakul}
\affiliation{Department of Physics, Princeton University, 
NJ, 08544, USA}

\author{S.~L. Sondhi}
\affiliation{Department of Physics, Princeton University, 
NJ 08544, USA}

\author{F.~J. Burnell}
\affiliation{Department of Physics, University of Minnesota Twin Cities, 
MN, 55455, USA}

\date{\today}
\begin{abstract}
 Fracton order is an intriguing new type of order which shares many common features with topological order, such as topology-dependent ground state degeneracies, and excitations with mutual statistics.  However, it also has several distinctive geometrical aspects, such as excitations with restricted mobility, which naturally lead to effective descriptions in terms of higher rank gauge fields.
In this paper, we investigate possible effective field theories for 3D fracton order, by presenting a general philosophy whereby topological-like actions for such higher-rank gauge fields can be constructed.  Our approach draws inspiration from Chern-Simons and BF theories in 2+1 dimensions, and imposes constraints binding higher-rank gauge charge to higher-rank gauge flux.  
We show that the resulting fractonic Chern-Simons and BF theories reproduce many of the interesting features of their familiar 2D cousins. 
We analyze one example of the resulting fractonic Chern-Simons theory in detail, and show that upon quantization it realizes a gapped fracton order with  quasiparticle excitations that are mobile only along a sub-set of 1-dimensional lines, and display a form of fractional self-statistics. The ground state degeneracy of this theory is both topology- and geometry- dependent, scaling exponentially  with the linear system size when the model is placed on a 3-dimensional torus.  By studying the resulting quantum theory on the lattice, we show that it describes a $\mathbb{Z}_s$ generalization of the Chamon code.

\end{abstract}

\maketitle

\section{Introduction}

Topological quantum field theories (TQFTs) have been a powerful tool in developing our understanding of the possible strongly interacting, gapped phases of matter.  In particular, they exhibit behaviors not perturbatively accessible from either weakly interacting or semi-classical limits, in which particles interact statistically and systems exhibit a ground-state degeneracy that depends on the topology of the underlying spatial manifold.  This behavior, known as topological order, has drawn a tremendous amount of interest, and our understanding of where it may manifest itself in nature\cite{wen1990topological,wen2003quantum,willett1987observation,dijkgraaf1990topological,wen1992classification}, its interplay with symmetry\cite{bernevig2006quantum,Fu2007-xo,chen2011two,chen2012symmetry}, and possible applications to quantum computing\cite{kitaev2003fault,Haah2011-ny} have developed rapidly in recent years.  

Recently, a new class of phases, known as fractonic phases, have been discovered in the context of exactly solvable lattice models \cite{Haah2011-ny,Halasz2017-ov,Vijay2016-dr,Vijay2015-jj,Chamon2005-fc,hsieh2017fractons,Slagle2017-ne,Halasz2017-ov,Hsieh2017-sc,shirley2018fractional,yoshida2013exotic,Yoshida2013-of}.  Fractonic phases\cite{Vijay2016-dr,Slagle2017-ne,Ma2017-qq,Halasz2017-ov,Hsieh2017-sc,Vijay2017-ey,Slagle2017-gk,Ma2017-cb,Chamon2005-fc,yoshida2013exotic,Haah2011-ny,Slagle2017-la,shirley2017fracton,pretko2017fracton,ma2018fracton,prem2017emergent,pretko2017subdimensional,Ma2017-qq,bulmash2018higgs,Prem2017-ql,bulmash2018generalized,you2018subsystem,devakul2018fractal,you2018symmetric,shirley2018foliated,shirley2017fracton,shirley2018fractional,song2018twisted,bulmash2018braiding,Slagle2017-la,shirley2017fracton,prem2018pinch,cho2015condensation,pretko2017fracton,Pretko2017-ej,slagle2018symmetric,gromov2017fractional,2018arXiv180711479P,pai2018fractonic,bulmash2018generalized,pretko2017subdimensional,ma2018higher,pretko2017generalized,pretko2017finite,Ma2017-qq,Ma2017-cb,yan2019rank,you2018majorana} exhibit behaviors in some respects similar topological order, such as robust ground state degeneracies, and statistical interactions between point-like quasiparticles.  However, they are also qualitatively different from topologically ordered phases in several respects: the ground state degeneracy is not topological, but rather sensitive to geometric aspects such as system sizes and aspect ratios, and excitations are generally {\it subdimensional}, meaning that they are either immobile, or that their motion is restricted to lines or planes in a 3 dimensional system.  

Given the power of topological quantum field theory to study topological orders in 2 dimensions, it is natural to ask whether there is a class of quantum field theories which capture fractonic behavior. 
Clearly, such field theories must be both similar to, and qualitatively different from, TQFTs.  Specifically, a TQFT describes an infrared limit---the topological scaling limit---in which the details of the underlying lattice (or regularization) are unimportant, and universal topological physics emerges. Interestingly the lattice also recedes in another scaling limit, this time near critical points, whose properties are famously captured by universal critical field theories. Fracton models, however, do not have a continuum limit in this strong sense: their ground state degeneracies depend explicitly on the lattice size, and the sub-dimensional mobility of their excitations means that rotational symmetry can never emerge at long wavelengths. Finding a quantum field theory appropriate to describing fracton phases thus represents an interesting theoretical challenge.

A number of possible approaches to this challenge have been discussed in the literature thus far \cite{Slagle2017-ne,Ma2017-qq,Slagle2017-gk,pretko2017fracton,ma2018fracton,prem2017emergent,pretko2017subdimensional,bulmash2018higgs,Prem2017-ql,bulmash2018generalized,you2018symmetric,gromov2018towards,gromov2017fractional}.  
For certain models, a connection to a continuous field theory can be made via a Higgs transition\cite{pretko2017fracton,ma2018fracton,bulmash2018higgs,Slagle2017-gk}, though a continuum version of the action describing the infrared fixed point of these theories is not known in general\cite{pretko2017fracton,Pretko2017-ej,slagle2018symmetric,gromov2017fractional,2018arXiv180711479P,pretko2017subdimensional,pretko2017generalized}.
Moreover, this formulation only describes a subset of the known Fracton models; other models, such as the Chamon code\cite{Chamon2005-fc,hsieh2017fractons}, do not admit a Higgs-type description.

A second approach is to work directly from a known fracton lattice model, and derive the constraints that it must impose on a resulting gauge theory.   Once these constraints are known, often they can be imposed via a gauge-invariant continuum action, leading to actions reminiscent of $2+1$D BF theory.  Drawing inspiration from similar results in 2-dimensional  topologically ordered systems \cite{hansson2004superconductors}, Ref.~\cite{Slagle2017-gk} used this approach to derive a BF-like continuum field theory for the X-$cube$ model\cite{Vijay2016-dr}.  A more general framework, relating constraints in excitations' mobility to generalized Gauss' laws, was developed by Ref. \cite{bulmash2018generalized}.  Finally, Gromov \cite{gromov2018towards} has recently proposed a framework within which ``matter'' fields which exhibit a set of multipole conservation laws stemming from polynomial shift symmetries can be gauged to obtain fracton models.

In this work, we adopt a different approach, by ``generalizing"  2+1 D TQFT's to the context of higher-rank gauge theories with a single time component $A_0$ (so-called ``scalar" charge theories).  That is, rather than considering usual vector gauge fields, we seek possible TQFT-like actions for tensor gauge fields, whose spatial gauge transformations may involve products of 2 derivatives \footnote{More generally, the term `higher rank gauge theory' refers to any symmetric gauge structure whose gauge transformation contains higher order differential forms.}.
This is a natural choice if one desires to replicate some features of fractonic phases of matter\cite{pretko2018fracton,pretko2017fracton,pretko2017finite,pretko2017generalized,Pretko2017-ej,pretko2017finite,prem2017emergent,bulmash2018generalized,slagle2018foliated}.

Our main focus is on what we will call fractonic Chern-Simons theory.
Specifically, we will take as our starting point an action inspired by $2+1$D Chern-Simons theory, which  imposes a constraint binding charge to the flux of a higher-rank gauge field.   We also comment briefly on the possibility of similar theories inspired by $2+1$D BF theory (or mutual Chern-Simons theory), which are particularly interesting in the context of general higher-rank gauge theories, whose gauge transformations can contain mixed first- and second-order polynomials in derivatives\cite{bulmash2018generalized,gromov2018towards}. In both cases, we restrict our attention to abelian ($U(1)$) theories, which are tecnhically simpler to deal with than their non-abelian counterparts.

The behavior of the resulting theories depends sensitively on the number of gauge fields present, since in a scalar charge theory our construction gives only a single Chern-Simons constraint.  We will primarily discuss a gapped field theory that emerges naturally when we require our rank-2 gauge fields to transform in representations of $C_3$ rotations about a fixed $(1,1,1)$ axis.  Since the appropriate representations are 2-dimensional this leads to a theory with 2 spatial gauge fields, whose single propagating degree of freedom can be eliminated by our Chern-Simons constraint, leading to a fully gapped theory.  

We discuss in detail both a continuum classical version of this model, and a lattice-regularized quantum version.  At the classical level, we find a theory whose gauge transformations imply that charged excitations (lineons) are mobile along only discrete sets of lines, and identify non-local (Wilson-line like) gauge invariant observables exhibiting a strong sensitivity to both the topology and the geometry of the spatial manifold. In particular, we show that though imposing the Chern-Simons constraint does reduce the number of independent Wilson operators, this number grows with the linear system size.  

Strikingly, upon quantizing our theory, we find that it has all of the expected hallmarks of Type I fracton order.  Specifically, it has a ground state degeneracy that is sensitive to both the topology (periodic boundary conditions are required) and the geometry (aspect ratios and system sizes) of the system.  Furthermore, its lineon excitations have non-trivial statistical interactions of the type exhibited in certain fractonic lattice models \cite{bulmash2018generalized,you2018subsystem,devakul2018fractal,you2018symmetric,pai2019fracton}, in which pairs of particles propagating along different lines in the same plane may have mutual statistics.   In fact, we show that by first quantizing this theory on a lattice, and then applying the Chern-Simons constraint, we are naturally lead to a lattice Hamiltonian that can be viewed as a $\mathbb{Z}_s$ generalization of the Chamon code\cite{Chamon2005-fc}.

We also discuss an analogue of Maxwell-Chern Simons theory model with three spatial gauge fields, corresponding to the off-diagonal elements of a symmetric rank-2 tensor.  In this case the single Chern-Simons constraint is insufficient to fully gap the theory.  One interesting feature of this model is that in the absence of the Chern-Simons term it has been shown to be necessarily confined \cite{xu2008resonating}, whereas with our higher-rank  Chern-Simons term confiement is suppressed and we find a deconfined U(1) phase with dipolar excitations mobile in 2-dimensional planes.

Our approach highlights that, although our fractonic Chern-Simons theories are clearly not TQFTs, it is possible to construct field theories for higher-rank gauge fields that share several important features of the {\it chiral} 2+1D Chern-Simons theories.  First, our fractonic Chern-Simons term creates self- statistical interactions between charged excitations.  
Second, our fractonic Chern-Simons action is gauge invariant only up to a boundary term,  implying that their boundaries host gapless surface states that cannot be realized in 2 dimensions with subsystem symmetry.  These are closely related to the surface states of subsystem-symmetry protected models described in Ref. \cite{you2018symmetric}.

The presence of such anomalous surfaces is  surprising in light of the correspondence between our field theories and exactly solvable lattice models, which is not expected for systems with topologically protected gapless boundary modes.  This is one of several hints that the regularization may play a more fundamental role in quantizing our higher-rank Chern-Simons theories than it does for TQFTs or critical theories.  Indeed, it is not clear whether it is possible to construct a well-defined continuum version of our compact $U(1)$ theory that correctly captures the low-energy behavior of the lattice model.

The paper is organized as follows. In Sec.~\ref{Sec:GeneralCS}, we introduce a general formulation for Chern-Simons -like actions appropriate to models with 3-component gauge fields $(A_0, A_1, A_2)$ and a single scalar charge.  This formulation applies both to vector gauge theories, whose gauge transformations are linear in derivatives, and tensor gauge theories whose gauge transformations are quadratic in derivatives.  
In Sec. \ref{dipole}, we discuss a particular realization of such a rank-2 theory, with $2$ spatial gauge fields transforming under $C_3$ rotations about the $(1,1,1)$ direction.  We discuss the possible gauge invariant operators in this case, and show that the associated quadratic (in derivatives) gauge transformations lead to matter fields that are restricted to move on lines, and gauge invariant ``cage-net" operators similar to those previously discussed in the context of lattice fracton models \cite{huang2018cage,song2018twisted,you2018symmetric}. 
In Sec.~\ref{Sec:ClassicalCS}, we scrutinize the classical Chern-Simons theory of this rank-2 theory.  In particular, we show that the Chern-Simons constraint fixes all gauge invariant operators except non-contractible loop operators, and discuss the number of independent loop operators of this type for the 3-torus.  We also show that the Chern-Simons action is gauge invariant only up to a boundary term, and discuss the nature of the resulting boundary theory. 

Sec.~\ref{dis} describes a lattice regularization of our rank-2 gauge theory, which we use to discuss two distinct routes to quantization.  In Sec. \ref{Sec:Quantum}, we discuss quantizing the constrained lattice model, derive the resulting ground state degeneracy on the $L \times L \times L$ torus, and describe the self- and mutual- statistics that follow from our Chern-Simons action.  In Sec.~\ref{Sec:LattQuantum}, we first quantize the lattice gauge fields, and then impose the Chern-Simons constraint.  We see that this leads to a lattice Hamiltonian that is a $\mathbb{Z}_s$ generalization of the Chamon code\cite{Chamon2005-fc}.  
Finally, in Sec.~\ref{GaplessCS},  we discuss adding a  Chern-Simons term to the Maxwell action of a symmetric tensor gauge theory with $4$ components, $(A_0, A_{xy}, A_{xz}, A_{yz}).$  We argue that though the resulting theory is gapless, it is nonetheless interesting as the Chern-Simons term appears to overcome the theory's expected confinement \cite{xu2008resonating} in a manner very similar to the case of compact $U(1)$ Maxwell-Chern-Simons theory in $2+1$ dimensions \cite{fradkin1991chern}.

\section{General higher rank Chern-Simons gauge theories} \label{Sec:GeneralCS}

Our starting point is a theory with 2 spatial gauge fields $A_1$ and $A_2$, which will allow us to obtain a fully gapped Chern-Simons theory with a single constraint.  Consider gauge transformations of the form
\be 
A_1 \rightarrow A_1 + D_1 \alpha  \ , \ \ A_2 \rightarrow A_2 + D_2 \alpha \ 
\ee
where $ D_1$ and $D_2$ are differential operators, whose form we will leave unspecified for now.   
Since we only have 2 gauge fields, the magnetic field defined has a single component
\be \label{Eq:Bfield}
B = D_2 A_1 - D_1 A_2 \ .
\ee
Note that the magnetic field (\ref{Eq:Bfield}) is always gauge invariant; however it is not necessarily the most relevant gauge invariant magnetic field that we can write down.  If $D_1$ and $D_2$ share a common factor $\partial_\ell$, the operator $\partial_{\ell}^{-1} B$ is also gauge invariant. Throughout the paper, we will focus on the cases where $D_1,D_2$ do not have common factor and the lowest order gauge invariant term is the magnetic flux.

The gauge-invariant electric fields have the form
\be \label{Eq:Efield}
E_i = \partial_t A_i - D_i A_0
\ee
where we have introduced the usual time component of the gauge field, which transforms as
\be
A_0 \rightarrow A_0 + \partial_t \alpha
\ee
under gauge transformations.  

The generalized Chern-Simons action we consider is
\be \label{LCS0}
\mac{L}_{CS} = \frac{s}{4 \pi} \left( A_1 E_2 - A_2 E_1 - (-1)^\eta A_0 B \right)
\ee
where $\eta = 1$ if $D_i$ contain only even numbers of derivatives, and $\eta = 2$ if they contain only odd numbers of derivatives.   
Under gauge transformations, we have
\ba  \label{Eq:CSGtrans}
\delta \mac{L}_{CS} &=& 
\frac{s}{4 \pi} ( D_1 \alpha E_2 - D_2 \alpha E_1 - (-1)^\eta \partial_t \alpha B )\n
&=&\frac{s}{4 \pi} (  D_1 \alpha \partial_t A_2 + (-1)^\eta \partial_t \alpha D_1 A_2 \n
&& -( D_2 \alpha \partial_t A_1 + (-1)^\eta \partial_t \alpha D_2 A_1) \n
&& +  D_2 \alpha D_1 A_0 -  D_1 \alpha D_2 A_0 )
\ea
In the absence of boundaries, we may freely integrate by parts, to obtain:
\be
\delta \mac{L}_{CS; \text{Bulk}} = 
0
\ee
The boundary terms in general do not vanish, implying the existence of gapless boundary modes, whose precise nature depends on the choice of $D_i$.  We will return to this point later when we discuss specific examples.  

Irrespective of the choice of $D_i$, the Chern-Simons action (\ref{LCS0}) has several commonalities with the standard vector Chern-Simons theory in $2+1$ dimensions.
First, in the absence of sources the constraint simply sets $B=0$.  Since there is only one component of the magnetic field, this one constraint is sufficient to eliminate the possibility of any propagating gauge degrees of freedom, leading to a gapped theory whose physics is entirely determined by operators describing pure gauge degrees of freedom\footnote{This only applies to the case where $D_1,D_2$ do not share any common factor. Otherwise, even the magnetic flux fluctuation is fixed, there might exist some local operator with lower order exhibiting a dispersive gapless mode.}.In ordinary Chern-Simons theory these are the holonomies, or gauge-invariant Wilson lines along non-contractible curves.  We will discus the analogue of Wilson line operators  for specific examples of $D_i$ in detail presently; these have the general form $e^{i\int_s A_i }$ with the submanifold $s$ chosen to ensure the operator is gauge invariant.

Second,  irrespective of the choice of $D_i$, the gauge fields $A_1$ and $A_2$ are canonically conjugate.  If both gauge fields are compact, this implies that a generalized Wilson operator of the form $e^{i\int_s A_i }$ must be discrete as well as compact.  Thus each of the generalized Wilson operators can take on only a finite, discrete set of values, which fully specify the states allowed in the absence of sources.  
On closed manifolds this can gives either a finite or a countable ground state degeneracy.

Finally, in the presence of matter fields, the Chern-Simons action (\ref{LCS0}) has the effect of binding charge to flux.  To see this, we add matter fields to our Chern-Simons action in the standard way, by adding a term
\be
\mac{L}_{\text{Matter}} = A_0 \rho - A_i J^i
\ee
where the currents obey the conservation law:
\be
D_i J^i = \partial_t \rho
\ee
Depending on the specific form of the differential operator $D_i$, the theory might contain additional subsystem charge conservation law and charge multipole conservation\cite{gromov2018towards}.
In the presence of sources the Chern-Simons constraint is
\be
B = D_2 A_1 - D_1 A_2 = \frac{2 \pi}{s} \rho
\ee
which binds the generalized magnetic flux to charge.  One might anticipate that a generalized Aharonov-Bohm effect may endow these charge-flux bound states with fractional statistics.  Indeed, as gauge invariant operators involving $A_1$ do not commute with gauge-invariant operators involving $A_2$, we will usually find at least some excitations with nontrivial mutual statistics.

However, as we will see the choice of $D_i$ does have profound implications for the final theory, and is key to determining the nature and mobility of the sources, as well as the ground state degeneracy.  This is because it is the form of $D_i$, and not the action, that determines the gauge-invariant operators and conservation laws, which play an essential role in both of these physical properties.  We therefore now discuss a few examples in detail.

\subsection{ Example 1: $D_1$ and $D_2$ are linear in derivatives: Stacking of 2D Chern-Simons theory}

As a warm-up, we consider the case where $D_1$ and $D_2$ are linear in derivatives. 
  In this case, we can always write $D_1 = \partial_{l_1}, D_2 = \partial_{l_2}$, with $l_1, l_2$ being two non-parallel directions.  We will see that in 2 spatial dimensions this always yields the conventional 2D Chern-Simons theory, while in 3 spatial dimensions it behaves like a stack of decoupled Chern-Simons theories.
  
To understand this theory, let us first understand its symmetries.  
First, theories of this type will be rotationally invariant in the plane perpendicular to $l_1, l_2$.  
This is because $A_{i}$  transform like vectors under rotations in the $l_1,l_2$ plane.   The gauge-invariant magnetic field $B$ is thus a scalar under in-plane rotations, as is the combination $A_1 E_2 - A_2 E_1$.  Thus our Chern-Simons action is fully rotationally invariant within the $l_1 , l_2$ planes.  (Indeed, it is easy to check that in this case $\mac{L}_{CS}$ has full Lorentz invariance).  

Second, the gauge transformations dictate that this theory has a conserved charge in each 2D plane.  To see this we couple our gauge fields to matter currents in the usual way:
\be
\mac{L}_{\text{Matter}} = A_0 \rho - A_1 J_1 - A_2 J_2
\ee
Gauge invariance requires that the current is conserved, i.e. 
\be
\partial_t \rho =  \partial_{l_1} J_1 + \partial_{l_2} J_2
\ee
If we integrate the right-hand side over any plane spanned by $(l_1, l_2)$ (in periodic boundary conditions), we obtain zero, implying charge conservation in each plane.  

Next, let us examine the gauge invariant operators.  First, consider open line segments of the form $\int A_1 d l_1$, $\int A_2 d l_2$, where the lines run along the $\hat{l}_1$ and $\hat{l}_2$ directions, respectively.  Under gauge transformations we have
\be
\int_x^y A_1 d l_1 \rightarrow   \int_x^y A_1 d l_1 + \alpha |_x^y  \ , \ \  \int_x^y A_2 d l_2 \rightarrow  \int_x^y A_2 d l_2 + \alpha |_x^y \ .
\ee
Thus with periodic boundary conditions, closed lines of either type are gauge invariant.  Further, we can see that a corner between a line along $\hat{l}_2$ and a line along $\hat{l}_1$ is gauge invariant.  Thus in addition to $E_i$ and $B$, there are also gauge invariant contractible closed loops.   
Indeed defining $\tilde{\vec{l}}_1, \tilde{\vec{l}}_2$ such that
\be
 \tilde{\vec{l}}_i \cdot  \vec{l}_j = \delta_{ij}
\ee
we see that integrals of the form
\be
\oint( A_1 d \tilde{l}_1 + A_2 d \tilde{l}_2 )
\ee
 are gauge invariant
for any closed curve in the $(l_1, l_2)$ plane.

Next, we examine how the Chern-Simons constraint $B=0$ restricts our possible choices of gauge-invariant operators.  
First, note that for a contractible closed curve bounding a region $\mac{R}$, we have
\be
\oint( A_1 d \tilde{l}_1 + A_2 d \tilde{l}_2 ) \propto \int_{\mac{R}} B \ \ .
\ee
(This can be shown by expressing $\tilde{l}_i, l_i$ in terms of a set of orthonormal basis vectors, and applying Stoke's theorem.  Note that the coefficient of proportionality is not $1$ unless $l_i$ and $l_j$ are orthogonal.)  We conclude that the constraint $B=0$ ensures that all contractible Wilson line operators are trivial.

Next, suppose we have periodic boundary conditions along the $l_1, l_2$ directions, such that there are also non-contractible gauge invariant line operators.  Since the two line operators concern lines in different directions, we will apply the Chern-Simons constraint to each set of lines individually. With this logic the Chern-Simons constraint then gives 
\be 
D_2 \int_1 A_1 =0
\ee
where $\int_1$ runs along a non-contractible curve in the $\hat{l}_1$ direction.   In 2D this tells us that once we have fixed one $A_1$ line we have fixed them all; a similar argument applies for $A_2$.  This is simply the familiar result that if there is no magnetic flux through the surface of the torus, the only degrees of freedom are the fluxes through its two non-contractible curves.  In 3D we are free to choose one line in each $(l_1, l_2)$ plane, exactly as for a stack of decoupled Chern-Simons theories. 

It is easy to see that quantizing such a theory also gives a result that is identical to a stack of decoupled Chern-Simons theories.  Hence we conclude that choosing $D_1, D_2$ to be linear in derivatives is essentially the same as choosing ordinary Chern-Simons theory in 2D, or a stack of ordinary Chern-Simons theories in 3D.

\subsection{BF generalizations}

Before moving on to our second example, which will lead to a fractonic Chern-Simons theory whose behavior we will analyze in detail, it is worth pointing out that a similar generalization of mutual Chern-Simons, or BF, theories can be carried out.  This generalization allows us to consider a wider variety of higher-rank gauge theories since, unlike the Chern-Simons construction described above, it can be applied to gauge fields whose gauge transformations are described by {\it arbitrary} polynomials in momenta.  Gauge transformations of this type are necessary to capture current conservation laws arising from general subsystem symmetries \cite{gromov2018towards}, including Type II fracton orders \cite{bulmash2018generalized}.

We begin with two gauge fields $(A_0, A_1, A_2)$ and $(B_0, B_1, B_2)$, which transform under gauge transformations according to
\ba
A_0 \rightarrow &A_0 + \partial_t \alpha  \ , \ \ &B_0 \rightarrow B_0 - \partial_t \alpha \\
A_1 \rightarrow &A_1+ D_1 \alpha \ , \ &B_1 \rightarrow B_1+ \tilde{D}_1 \alpha  \n
A_2 \rightarrow& A_2+ D_2\alpha \ , \ \ &B_2 \rightarrow B_2+ \tilde{D}_2 \alpha  \nonumber 
\ea
where
\be
D_i = D_i^{(e)} + D_i^{(o)} \ , \ \ \tilde{D}_i = D_i^{(e)} - D_i^{(o)}
 \ee
 with $D_i^{(e)}, D_i^{(o)}$ are the differential polynomials containing even and odd numbers of derivatives, respectively\footnote{For general differential polynomial $D_i^{(e)}, D_i^{(o)}$, the coefficient in each differential term is dimensionful so such operator is only well-defined on the lattice}.
The higher-rank BF action has the form
\ba
\mac{L}_{BF} &=& A_0 (\tilde{D}_1 B_2 - \tilde{D}_2 B_1 ) + A_1 ( \tilde{D}_2 B_0 + \partial_t B_1) \n && + A_2 (- \partial_t B_1 - \tilde{D}_1 B_0 )
\ea
It is easy to check that this action is gauge invariant up to a boundary term.  In the presence of sources, it imposes the constraints
\ba
 (\tilde{D}_1 B_2 - \tilde{D}_2 B_1 ) = \rho_A  \n
  (D_1 A_2 - D_2 A_1 ) = \rho_B 
  \ea
where $\rho_{A}, \rho_B$ are the charges coupled to the $A$ and $B$ gauge fields, respectively.  Provided $D_1,D_2$ does not share any common factor, these constraints are sufficient to eliminate any propagating modes, leading to a gapped theory describing a stable infrared fixed point.  Further,
It is clear that $A_1$ and $B_2$ (and $A_2$ and $B_1$) are canonically conjugate, such that the sources of $A$ and $B$ will acquire mutual statistics upon quantizing the theory. 
In this way, a wide variety of higher-rank gapped fractonic actions can be constructed.  We defer a discussion of the many interesting examples to future work, except to note that the BF-like field theory of the X-cube model proposed by Ref.~\cite{slagle2017fracton} involves a construction of this type, albeit modified to work with gauge fields with 3 spatial components, whose gauge transformations share common factors.  

\section{Example II: Fractonic Chern-Simons theory with dipole excitations}
\label{dipole}

Let us now consider an example that will lead to a fractonic Chern-Simons theory.  
To obtain this, we will take $D_i$ to be quadratic in derivatives.
In this case (unlike for linear $D_i$), we have several choices, distinguished by their transformation under spatial symmetries. 

For our example, consider a system with a cubic geometry, with the cubic axes $x,y,$ and $z$.  We will not require our action to have full cubic symmetry-- indeed we will see later that full cubic symmetry is compatible with our Chern-Simons action only for a specific choice of the coupling constant.   Instead, we require  invariance under $C_3$ rotations about the (111) direction. 
Thus $D_i$ (and consequently $A_i$) must transform in an irreducible representation $\Gamma$ such that $\mathbf{1} \in \Gamma \otimes \Gamma$, to ensure that the quantities $B$ and $A_1 E_2 - A_2 E_1$ transform as scalars under $C_3$ rotations.  We will also require that the $D_i$ transform non-trivially under $C_3$ rotations; this ensures that both $A_1$ and $A_2$ are required to construct a symmetric action.  

\begin{figure}[h]
  \centering
      \includegraphics[width=0.15\textwidth]{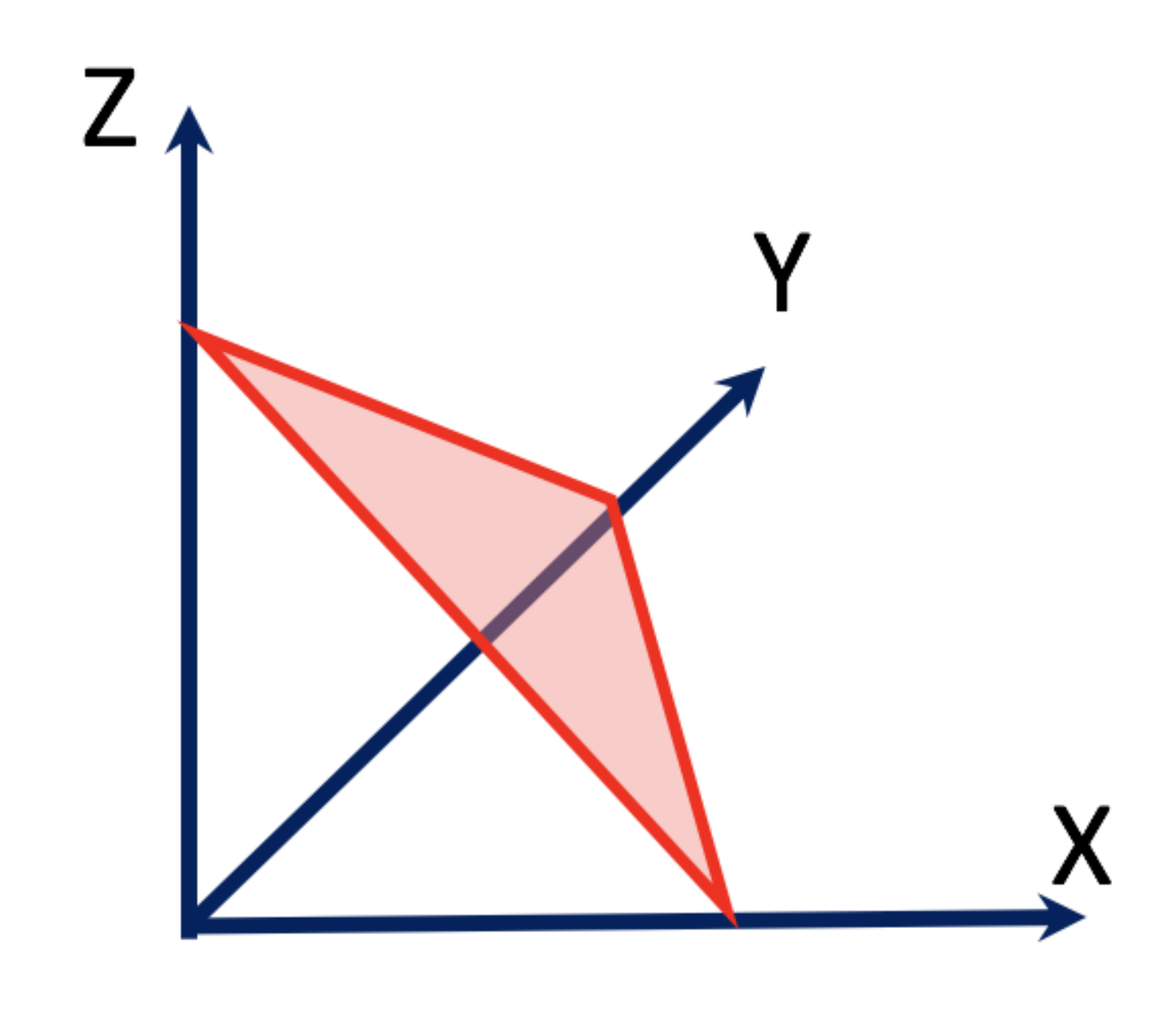}
  \caption{Illustration of the $x,y,z$ and $u,v,w$ coordinates. The red plane is the 111 plane and the three boundary lines of the red triangle denote the $u,v,w$ direction. The $C_3$ rotation on the 111 plane rotates the red triangle by $2\pi/3$ and permutes the axis along $x,y,z$.} 
  \label{Fig:axis}
\end{figure}

The full set of irreducible representations of $C_3$ are given in Appendix \ref{irrepApp}.  
If $D_i$ are quadratic in derivatives, we have two choices for the irreducible representation $\Gamma$, which we denote $\Gamma^a$ and $\Gamma^b$:
\begin{align} \label{Eq:Dirreps}
& D_1^a=\frac{1}{\sqrt{3}} \partial_{\ell} \partial_u, 
~D_2^a=\frac{1}{\sqrt{3}} \partial_{\ell} \partial_v
\nonumber\\
& D_1^b=\frac{\sqrt{2}}{\sqrt{3}} \partial_{u} \partial_{u_{\perp}}, ~ D_2^b=\frac{1}{2 \sqrt{2}} (\partial_u^2 -\partial_{u_{\perp}}^2)-\frac{1}{ \sqrt{6}} \partial_u \partial_{u_{\perp}}
\end{align}
where relative to the cubic axes $x,y,z$, we have defined (see Fig. \ref{Fig:axis})
\begin{align} \label{Eq:CoordTrans}
\hat{\ell} = \frac{1}{\sqrt{3} }  \left( \hat{x} + \hat{y} + \hat{z} \right ) \n 
\hat{u} =  \frac{1}{\sqrt{2}} \left(\hat{y} - \hat{z} \right ) \n
v = 
 \frac{1}{\sqrt{2} } (z-x)  \n
 w =    ( -u - v)  =
\frac{1}{\sqrt{2} } (x-y) 
\end{align}
in terms of which $u_{\perp}$, the direction orthogonal to $u$, is 
\be
\hat{u}_{\perp} =\frac{1}{\sqrt{3}} (-2 v - u ) =\frac{1}{\sqrt{6}} \left(2 \hat{x} - \hat{y} -\hat{z} \right ) \ \ .
\ee
Under arbitrary rotations about the $\ell$ axis, $\Gamma^a$ transforms like a vector, with an angular momentum of $1$ along the $\ell$ axis, while $\Gamma^b$ transforms like a rank 2 tensor, with angular momentum $2$.  For $C_3$ rotations, however, where the angular momentum $3$ representation transforms like a scalar, these are effectively two vector representations.

Thus we have $\mathbf{1} \in \Gamma^\alpha \otimes \Gamma^\beta$ for all combinations of $\alpha,\beta$. Consequently we may choose $D_i$ to be an arbitrary linear combination of $D_i^a$ and $D_i^b$.  We will see later that, provided both irreducible representations appear with non-zero coefficients, the resulting theories are closely analogous.  Thus we will take
\begin{align} \label{Gtrans}
&D_1= D_1^a+  D_1^b = \partial_x \partial_u = \frac{1}{\sqrt{2}} \partial_x ( \partial_y - \partial_z ) \nonumber \\
&D_2=  D_2^a+  D_2^b =  \partial_y \partial_v = \frac{1}{\sqrt{2}} \partial_y ( \partial_z - \partial_x ) \nonumber \\
&-D_1 - D_2 = \partial_z \partial_w =\frac{1}{\sqrt{2}} \partial_z ( \partial_x - \partial_y ) 
\end{align}

Under gauge transformations, we have
\ba \label{Eq:AuvwGauge}
A_{1} \rightarrow& A_1 +  
\partial_x \partial_u \alpha 
\\
A_{2} \rightarrow &
A_{2} +\partial_y \partial_{v} \alpha
\n
-A_1 - A_2  \rightarrow &   
-A_1 - A_2  + \partial_z \partial_{w}\alpha
\ea
The $C_3$ rotations about the $(1,1,1)$ direction permute $(u,v,w)$ and $(x,y,z)$ directions.  Since $\alpha$ is a scalar, this implies that $C_3$ rotations permute $A_1, A_{2},$ and $- A_{1}-A_2$ (and similarly for $E_1, E_2$):
\begin{align}
&C_3: x \rightarrow y, \ \ y \rightarrow z, \ \ z \rightarrow x \n
&~~~~~A_1\rightarrow A_2, A_2\rightarrow -(A_1+A_2), -(A_1+A_2) \rightarrow A_1\nonumber\\
&~~~~~D_1\rightarrow D_2, D_2\rightarrow -(D_1+D_2), -(D_1+D_2) \rightarrow D_1
\end{align}

The gauge-invariant magnetic field given in Eq. (\ref{Eq:Bfield}) is 
\be \label{Beq}
B =  \left( \partial_x \partial_u A_2 - \partial_y  \partial_v A_1 \right )
\ee
It is easy to check that $B$ (as well as the combination $A_1 E_2 - A_2 E_1$) transforms as a scalar under $C_3$ rotations, as it must if our theory is to be symmetric with $A_0$ rotationally invariant.

\subsection{Conservation laws} \label{Sec:conservation}

Before studying Chern-Simons theory per se, let us understand what properties of the theory are dictated solely by the struture of its gauge transformation laws.  The canonical coupling of gauge fields to sources
\be
\mac{L}_{\text{matter}} = A_0 \rho - A_{1} J^{1} - A_2 J^2
\ee
implies the current conservation relation 
\be
\partial_t \rho - D_1 J_{1}- D_2 J_2 = 0
\ee
Note that in order for $\mac{L}_{\text{matter}}$ to be invariant under $C_3$ rotations, we must have
\be
C_3 : \ J_1 \rightarrow J_2 - J_1 \ , \ \ J_2 \rightarrow - J_1
\label{c3s}
\ee
Evidently, 
\be
\partial_t \int d^3{\bf r} \rho = \int d^3{\bf r} (D_1 J_{1}+ D_2 J_2) = 0
\ee
so that charge is conserved in the system as a whole.  Note that here we assume periodic boundary conditions and single-valuedness of all currents, such that the integral of any derivative over all space is zero.

In addition, however, from Eq. (\ref{Gtrans}) we see that both operators $D_i$ contain only terms with at least one derivative in each $u-v$ plane.  Thus, we also have
\be
\partial_t \int du d v  \rho =  0
\ee
and charge is conserved in each $u-v$ plane.  In addition,  all terms in $D_1$ and $D_2$ contain either $\partial_x$ or $\partial_y$ (or both), such that the charge is conserved in each $x-y$ plane, and similarly for the $x-z$ and $y-z$ planes.  

Finally, since charge conservation in an individual plane automatically implies that the dipole moment orthogonal to that plane is conserved, our theory has conserved dipole moments along the $x,y,z$, and $\ell$ (or $x+ y + z$) directions.  The net result of these four  dipole conservation laws is that dipole mobility in our system is severely restricted.  Consider a dipole oriented along the $z$ direction, which has a conserved dipole moment orthogonal to both the $(x,y)$ and the $(u,v)$ planes.  As a consequence it can move only along the  line $x+y =0$, which lies in the intersection of these two planes. Similar considerations apply to other dipole orientations: all dipole excitations in this theory are restricted to move along lines.   

\subsection{Gauge invariant cage-net operator} \label{Sec:CageNet}

Having understood the theory's conservation laws, we now  
consider the nature of the gauge invariant line operators.  We will see that these reflect the one-dimensional motion of dipolar excitations, as well as introducing a cage-net structure similar to that noted in other fracton models \cite{huang2018cage}.

To identify what types of operators we should study, observe that one obvious difference relative to the usual vector gauge theory is the dimension of the gauge field $A$: if $\alpha$ is dimensionless (as is natural, since it appears as a $U(1)$ phase rotation of the matter fields), then $A$ has mass dimension $2$.  Thus dimensionless gauge invariant operators in these theories require integrating along surfaces.  

To determine which surfaces to examine, we will begin with dimensionful gauge-invariant {\it line} operators, whose end-points transform as $\partial_{i} \alpha$ for some direction $i$.  This implies that the end-points of these open lines do not harbor charges, but instead are associated with the derivative of the charge along the $\hat{i}$ direction.   To make physical sense of this, we define dimensionless  gauge-invariant ribbon operators by integrating over a surface of width $a_i$ transverse to the line's orientation; the end-points of such ribbons harbor a dipole oriented along the $\hat{i}$ direction.  We will see presently that the associated dipole moments are conserved, suggesting that such a fixed length scale is not unnatural.

From the gauge transformations, we see that the theory admits 6 types of gauge-invariant line operators. Three of these are in-plane line operators:
\ba
&\Gamma_u = \int d u A_1 \ , \ \ \Gamma_{v} = \int d v A_{2} \ , \ \ \  \Gamma_{w} = \int d w (- A_1 - A_2) \n
\ea
while three are line operators extending along the cubic axes $\hat{x},\hat{y}, \hat{z}$.  
\be
\Gamma_x = \int d x A_1 \ , \ \ \Gamma_{y} = \int d y A_2 \ , \ \ \  \Gamma_{z} = \int d z (-A_1 -A_{2} )
\ee
The associated dimensionless  gauge invariant ribbon operators have the form
\be \label{Eq:Wdef}
W_{u} = e^{i \int_{x}^{x+a} \Gamma_u dx }  \ , \ \ W_{x} = e^{i  \int_{u}^{u+\sqrt{2} a} \Gamma_x du }
\ee
and similarly for other pairs of directions.  Here $a$ is the fundamental dipole scale of our problem, as discussed above. Note that we choose the length of the fundamental dipoles in the $u,v,w$ directions to be $\sqrt{2}$ that of the fundamental dipoles along the $x,y,z$ directions.  With this choice, a dipole along $u$ can be viewed as a combination of a dipole along $y$ and a dipole along $-z$.

Naively, the Wilson lines $\{ \Gamma_i \}$ appear similar to line operators we would expect from a stack of 2D vector gauge theories discussed in the previous example.  However, in the present theory the lines can run {\it only} along one of the 6 directions specified above.  (See Appendix \ref{LineApp}).  Further, a corner between a $\Gamma_u$ line and a $\Gamma_v$ line must have a charge, by gauge invariance.  To see this, consider the operator:
\be
T = \int_{\vec{r}_0}^{ \vec{r}_1 } A_1 d u + \int_{\vec{r}_1 }^{ \vec{r}_2} A_{2} d v + \int_{\vec{r}_2 }^{ \vec{r}_0} ( - A_1 - A_2 ) d w
\ee 
where the points $\vec{r}_0, \vec{r}_1, \vec{r}_2$ form a triangle in one of the $(u,v)$ planes, with edges along the $u, v,$ and $w$ directions respectively.  For a usual rank 1 gauge theory, $T$ would be a gauge invariant line operator.  In our case, however, it is not: under gauge transformation, we find that 
\be
T \rightarrow 
T + ( \partial_x -\partial_z)  \alpha(  \vec{r}_1 ) +  ( \partial_z -\partial_y)  \alpha(  \vec{r}_2 ) +  ( \partial_y -\partial_x)  \alpha(  \vec{r}_0 )
\ee
 This shows that in order for $T$ to be gauge invariant, we must attach an infinitesimal dipole to each of the triangle's corners.  Thus though we do have Wilson line like operators in-plane, these lines cannot bend without creating new dipolar excitations.

 \begin{figure}[h]
  \centering
      \includegraphics[width=0.4\textwidth]{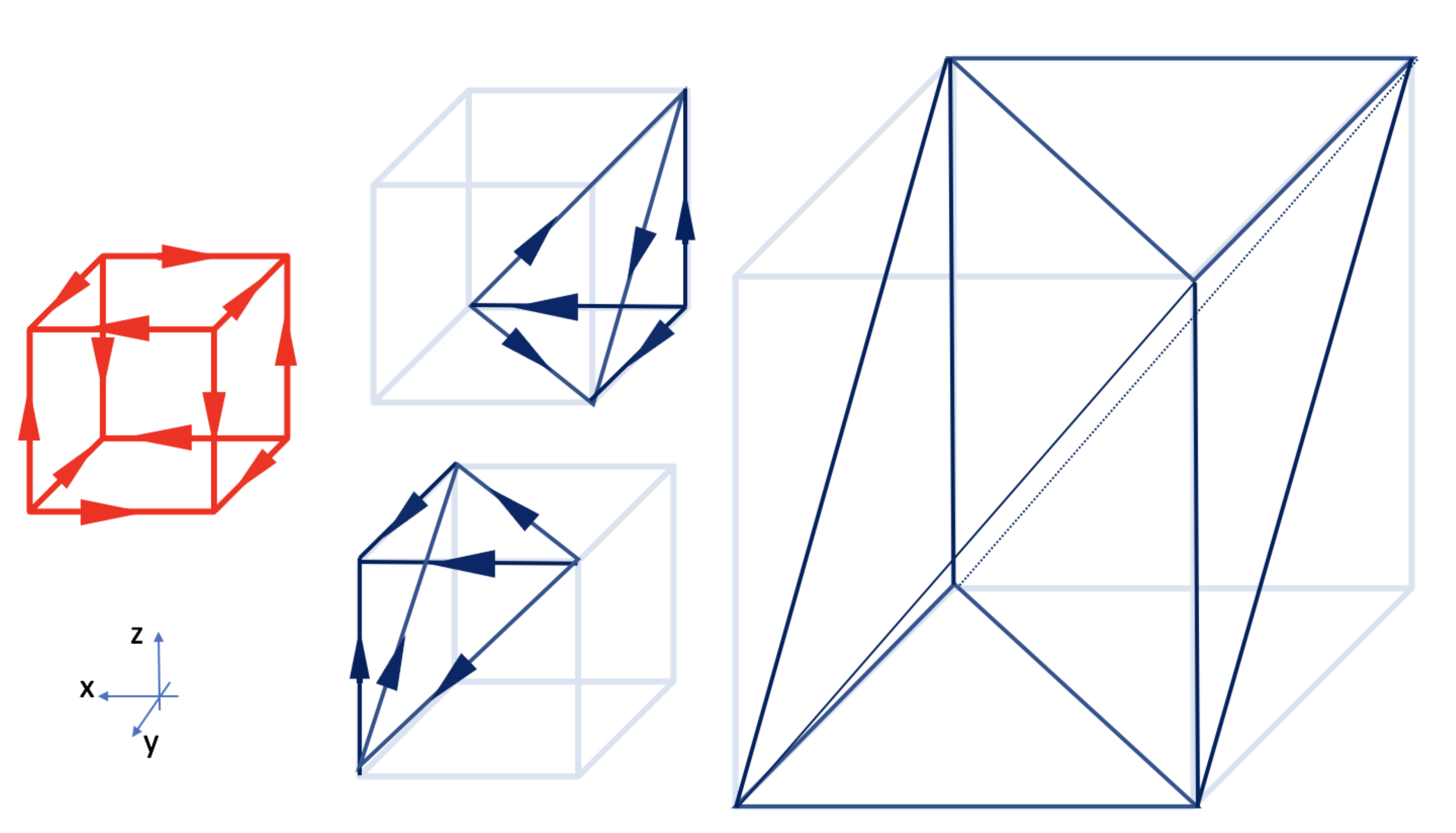}
  \caption{Gauge invariant cage nets of the symmetric rank 2 theory.  With differential operators given in Eq. (\ref{Gtrans}), the cages can have edges along the $x,y,z$ or $u,v,w$ directions.} 
  \label{cage}
\end{figure}

 Similarly, lines running along the cubic axes cannot turn, either into other cubic directions or into the $(u,v)$ planes.  However, because 
 \be
 ( \partial_u + \partial_{v} + \partial_{w}) \alpha = 0 
 \ee
three lines running along orthogonal cubic axes can meet at a point without creating extra charges, as shown in Fig.~\ref{cage}.  Similarly there is no charge at a trivalent vertex  between $\sqrt{2} \Gamma_y, \Gamma_u$, and $\Gamma_{w}$ (and its appropriately rotated analogues), since 
\be \label{Eq:Corner2}
\sqrt{2} \partial_{v} \alpha -( \partial_x \alpha - \partial_y \alpha )= 0 \ \ .
\ee
Note that Eq. (\ref{Eq:Corner2})  requires the lines to be correctly oriented at the vertex; the correct orientations are shown in Fig.~\ref{cage}.  

\begin{figure}[h]
  \centering
      \includegraphics[width=0.4\textwidth]{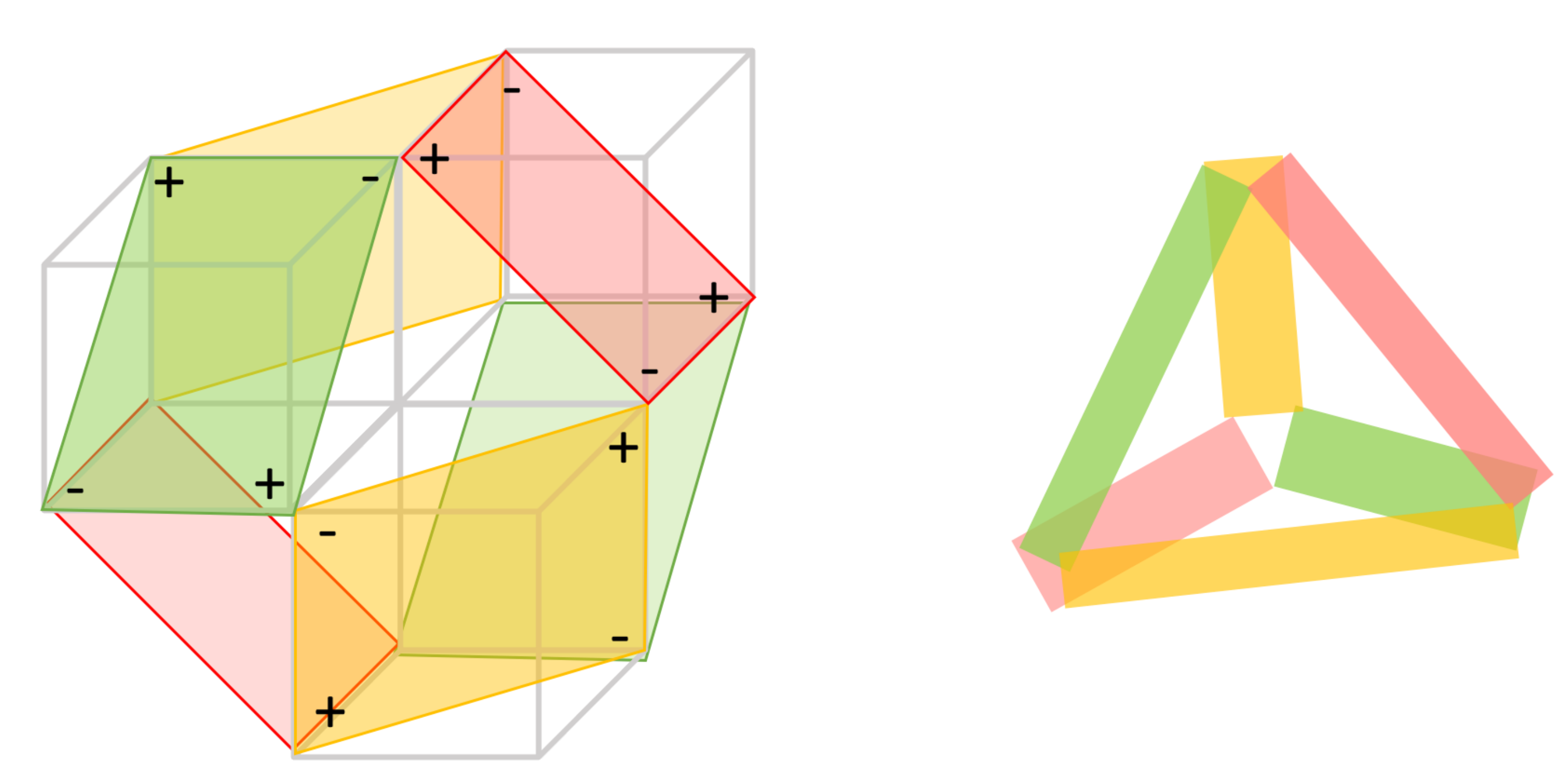}
  \caption{Left: The red/yellow/green sheets represent thick ribbon operators.  An isolated ribbon must harbor a charge at each corner; however, with appropriately oriented gauge fields these corner charges cancel in the configuration shown. 
  Right: By lengthening and thinning the ribbons, we obtain a Wilson ribbon  cage-net.  Note that the Wilson ribbon cage nets are gauge invariant only if we choose the width of ribbons along the $u,v,w$ directions to be longer by a factor of $\sqrt{2}$ than those along the $x,y,z$ directions.} 
  \label{diamond}
\end{figure}

From the above arguments, it is easy to see that similar gauge-invariant cage-net structures can be formed of our dimensionless Wilson ribbons (see Fig. \ref{diamond}), provided that we choose the dipole scales as specified in Eq. (\ref{Eq:Wdef}).  The relative factor of $\sqrt{2}$ in the ribbons' widths ensures that the three dipoles can annihilate at the corresponding corners, so that the operators are gauge invariant.       

In summary, the structure of the gauge transformations (\ref{Gtrans}) leads to a qualitatively different Fractonic Chern-Simons theory, in which the analogue of the Wilson line is a gauge invariant dimensionless ribbon operator.  These ribbons are
not free to turn, but can meet at certain trivalent corners.  This results in gauge-invariant ``cage net" operators, which can be tetrahedral, prismatic, or cubic as shown in Fig. \ref{cage}.

In addition to cage-nets we can also form gauge-invariant closed membranes, by widening our ribbons to some multiple of the fundamental dipole scale, and then forming a closed surface along which all ribbons share an edge.  In the extreme limit these are 2-dimensional membrane operators that can extend in the $x-u, y-v$ or $z-w$ planes, and the associated cage nets are   stretched into the diamond configuration shown in Fig.~\ref{diamond}.   We will not discuss these membrane operators in our analysis, however, since they can always be viewed as arrays of the fundamental ribbon operators $W_\alpha$ described above.

\begin{figure}[h]
  \centering
      \includegraphics[width=0.4\textwidth]{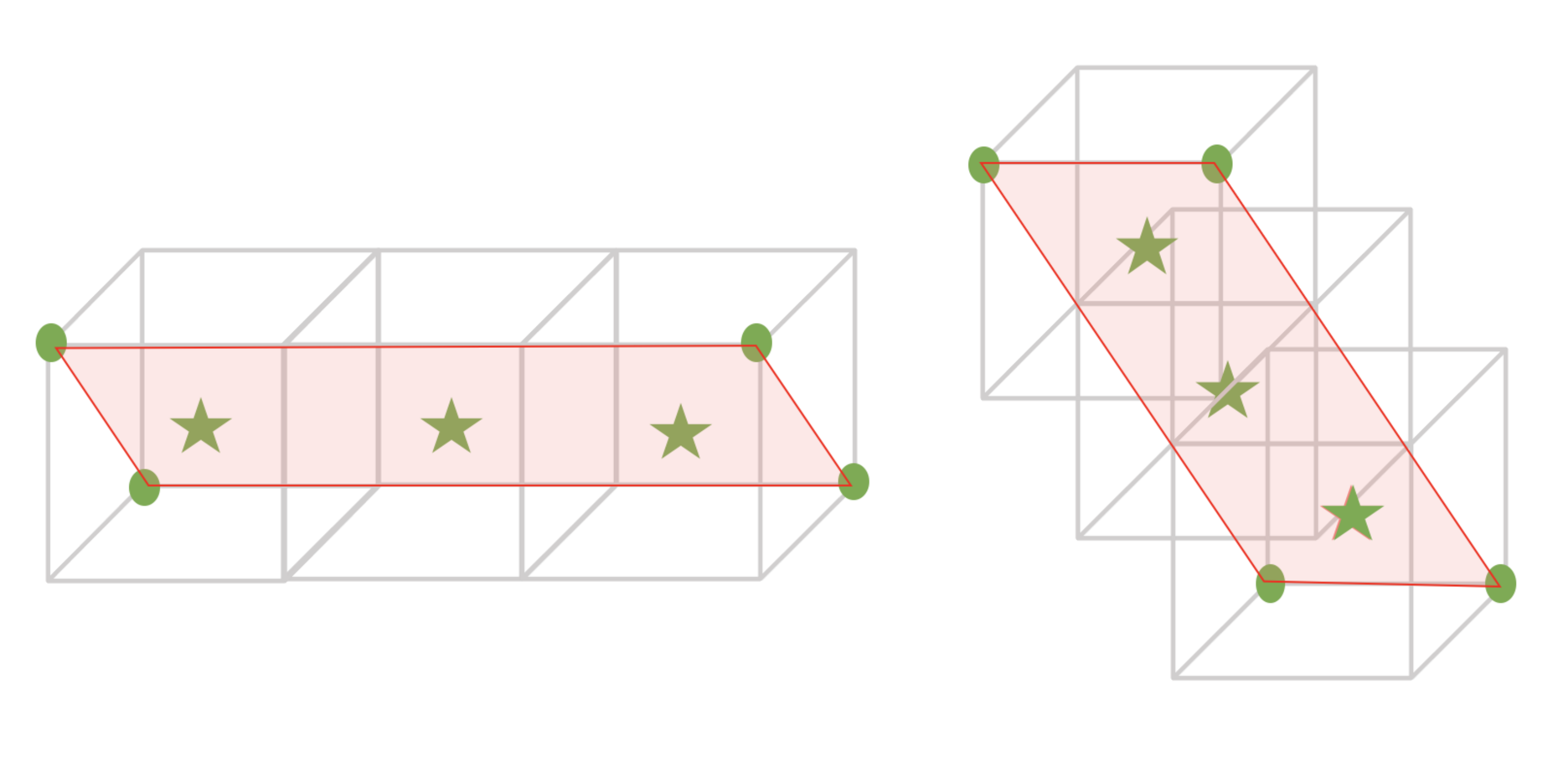}
  \caption{Lineon excitation configurations: Each end of an open Wilson ribbon extended along the $x$ direction must terminate on a dipole (or anti-dipole) with dipole moment oriented along $u$. Likewise, a Wilson ribbon extended along the $u$ direction hosts a dipole oriented along $x$ at each end-point.  Since Wilson ribbons cannot turn, these dipoles are mobile only along the direction of the ribbon, and are hence 1-dimensional lineon excitations. Here star represent gauge fields $A_1, A_2$ while dot represent charge $\rho$.  
 } 
  \label{lattice}
\end{figure}

Finally, if we allow sources in our theory, open ribbon operators can also be gauge invariant, provided that we attach an appropriately oriented dipole at each endpoint (see Fig. \ref{lattice}).  Each of the 6 possible line directions is thus associated with a dipolar source which, due to dipole conservation, is mobile only along a specific linear direction; these thus behave like the ``lineons" typical of type I fracton theories.  To strengthen the connection to fracton orders, we can also consider open membrane operators whose width is some multiple of the fundamental dipole scale. 
The charges appearing at the corners of each membrane are then immobile fractons.  

\subsection{Related models with the same symmetry }\label{c3section}

It is clear from the above discussion that (irrespective of our choice of action) a theory with the operators $D_1, D_2$ described above cannot be topological.  In particular the cage net structure requires us to choose a fundamental dipole scale, and the theory is manifestly not scale invariant.  The existence of this scale follows from the fact that the theory conserves dipole moment perpendicular to each individual $u-v, x-y, y-z$ and $x-z$ plane.  Further, the structure of the cage nets is invariant only under a discrete set of $C_3$ rotations, which are naturally viewed as a subset of the rotational symmetries of the cubic lattice.

We may nonetheless ask whether related theories exist, which share the same $C_3$ rotation symmetry, and conserve dipole moment along four families of planes.

To see that they do, we consider a more general form of the operators 
$D_1, D_2$ :
\begin{align} \label{Eq:Dgens}
&D_1=\alpha D_1^a+ \beta D_1^b, ~D_2=\alpha D_2^a+ \beta D_2^b 
\end{align}
where $D_i^{a,b}$ are given in Eq. (\ref{Eq:Dirreps}), and  $\alpha, \beta$ are parameters which only takes discrete values if we put the theory on the cubic lattice. Regardless of the choice of $\alpha, \beta$, the theory is $C_3$ rotation invariant; essentially this is because both of the 2-dimensional irreducible representations are vector-like from the point of view of $C_3$ rotation symmetry.

\begin{figure}[h]
  \centering
      \includegraphics[width=0.4\textwidth]{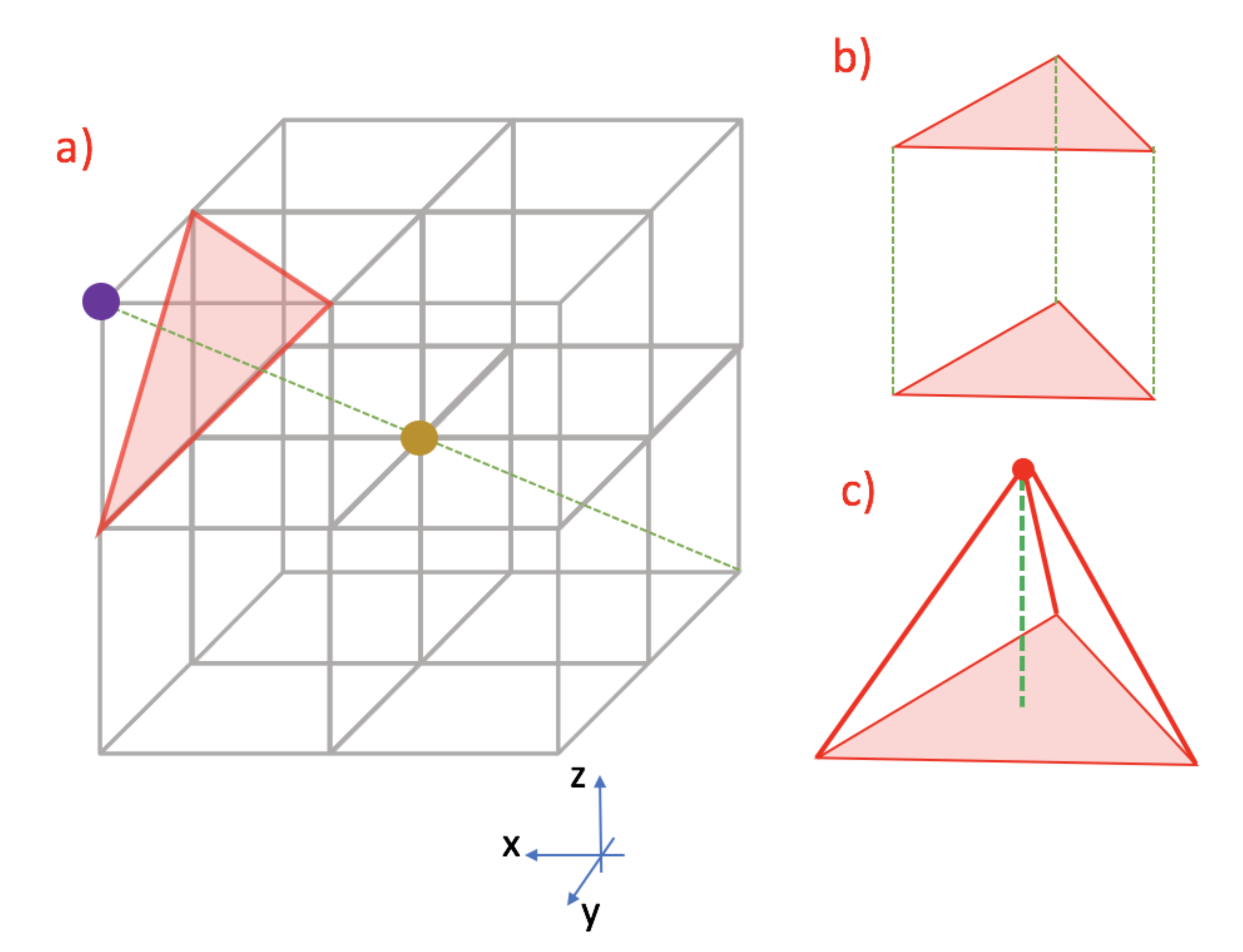}
  \caption{ By tuning $\alpha,\beta$ in Eq. (\ref{Eq:Dgens}), we are changing the height of the tetrahedron in our cage-net configuration. There are an infinite number of equivalent cage-net configurations related by this deformation.  c) A rotated view of the cage-net tetrahedron. 
  The dimensions of the base of the tetrahedron (indicated with a shaded red triangle) is fixed, while its height along the $(111)$ direction (represented by the green dotted line) varies depending on $\arctan{(\alpha/\beta)}$.
  a) Examples of equivalent cage net tetrahedra.  The purple dot corresponds to the case where $\alpha,\beta=1$. 
  The yellow dot corresponds to $\alpha=2,\beta=1$.
  b) When $\beta=0$, the height of the tetrahedron goes to infinity and the cage-net configurations becomes a prism.}
  \label{gen}
\end{figure}

For general nonzero $\alpha, \beta$,  the differential operators have the explicit form, 
\begin{align}
&D_1\sim (\partial_x+(\alpha- \beta) \partial_\ell)(\partial_y-\partial_z) \nonumber\\
& D_2\sim (\partial_y+(\alpha- \beta) \partial_\ell)(\partial_z-\partial_x)
\end{align}
Thus we obtain gauge-invariant ribbon operators extended along $6$ non-parallel directions.  At the end-points of any open ribbon, there is a dipole. With an appropriate choice of the dipole scales in each direction, these 6 lines are allowed to meet at trivalent corners, leading to tetrahedral cage-net configurations as shown in Fig.\ref{gen}.  One face of the tetrahedron is an equilateral triangle in the $u-v$ plane; the remaining three have edges in the $x+(\alpha- \beta) \ell,y+(\alpha- \beta) \ell,z+(\alpha- \beta) \ell$ directions.  

Following the arguments of Sec. \ref{Sec:conservation}, it is not hard to check that for general $\alpha$ and $\beta$, the charge is conserved in each $u-v$ plane, as well as in the other three families of planes spanned by two of the three vectors $x+(\alpha- \beta) \ell, y+(\alpha- \beta) \ell,z+(\alpha- \beta) \ell$. These three planes are related by $C_3$ rotation along 111-axis.  These conservation laws ensure that the dipoles described above can only move in 1 dimension. In particular, when $\alpha=2,\beta=1$, the cage-net configuration becomes a regular tetrahedron and the Chern-Simons coupling is odd under cubic rotation.

We note that the choices  $\alpha=0$ and $\beta=0$ {\it do} lead to qualitatively different theories.  Taking $\alpha =0$ clearly leads to a stack of 2-dimensional theories, since $D_1, D_2$ only involve derivatives in the $u-v$ plane.   Taking $\beta =0$ gives $D_1= \partial_l \partial_v , D_2 =\partial_l \partial_u$. 
As discussed in Appendix \ref{StackedApp}, the line operators in the resulting theory are  similar to those of a stack of ordinary 2D gauge theories, with
both 2d particles mobile in each u-v plane and a 1d particle mobile along the $\ell$-direction.

\section{Classical Chern-Simons theory of $C_3$ -symmetric rank 2 model} \label{Sec:ClassicalCS}

Thus far, we have focused only characteristics of our field theory that result solely from the form of the operators $D_1, D_2$, which dictate the nature of the conservation laws, and the geometry of the gauge-invariant operators.  We now turn to the implications of the Chern-Simons action.  In the following sections we will analyze this from a classical perspective, turning to the quantum theory in Section \ref{Sec:Quantum}.

\subsection{The Chern-Simons constraint}

Classically, the main role of the Chern-Simons Lagrangian (\ref{LCS0}) is to enforce the constraint 
\be
B \equiv   \left( \partial_x \partial_u A_2 - \partial_y  \partial_v A_1 \right ) = \frac{2 \pi}{s} \rho \ \ .
\ee
This has two important effects on our classical theory: first, it fixes the value of all closed cage nets in our 3-dimensional system. Second, it imposes conditions on parallel Wilson lines, drastically reducing the number of such operators that are independent.

First, 
we observe that the constraint $B = \frac{2 \pi}{s} \rho$ fixes the value of all cage net operators.  This can be checked directly by 
integrating the magnetic field over the volume enclosed by the cage net, which gives exactly the series of line operators associated with the cage net itself.  
A similar result holds for cage nets bounded by dimensionless ribbon operators.  Hence in the Chern-Simons theory, our gauge-invariant cage net operators are constrained to take the value $0$. 

Second, let us determine the effect of the Chern-Simons constraint on non-contractible Wilson operators.  Consider operators of the type $W_x$, in the absence of sources.  A priori there are $L_y L_z/a^2$ non-overlapping operators of this type (where $a$ is the dipole scale), and similarly for other directions.  However,  integrating the constraint over $x$ gives
\begin{align}
0 = - \oint B dx = \partial_z \partial_y \oint d x A_1=\partial_z \partial_y \Gamma_x \n
0 =  \oint B dy = \partial_z \partial_x \oint d y A_2=\partial_z \partial_x \Gamma_y \n
0 = \oint B dz = \partial_x \partial_y \oint d z (A_1+ A_2) = - \partial_x \partial_y \Gamma_z
\end{align}
where we have assumed periodic boundary conditions, and that $A$ is single-valued.  Thus we may fix the value of $\Gamma_x$ (and hence $W_x$) along the boundaries of the $yz$ plane, by specifying one function of $y$ and one function of $z$ -- but having done this, the value of $\Gamma_x$ elsewhere in the plane is fixed.  For an $L_x \times L_y \times L_z$ system this gives $(L_y + L_z)/a -1$ independent non-overlapping ribbon operators $W_x$, and similarly for $W_y$. 
Further, once $\Gamma_x$ and $\Gamma_y$ are fixed everywhere, the condition that the cubic cage nets (see Fig.~\ref{cage}) must all be trivial then fixes $\Gamma_z$ along any line, such that the ribbon operators $W_z$ are fixed.

\begin{figure}[h]
  \centering
      \includegraphics[width=0.45\textwidth]{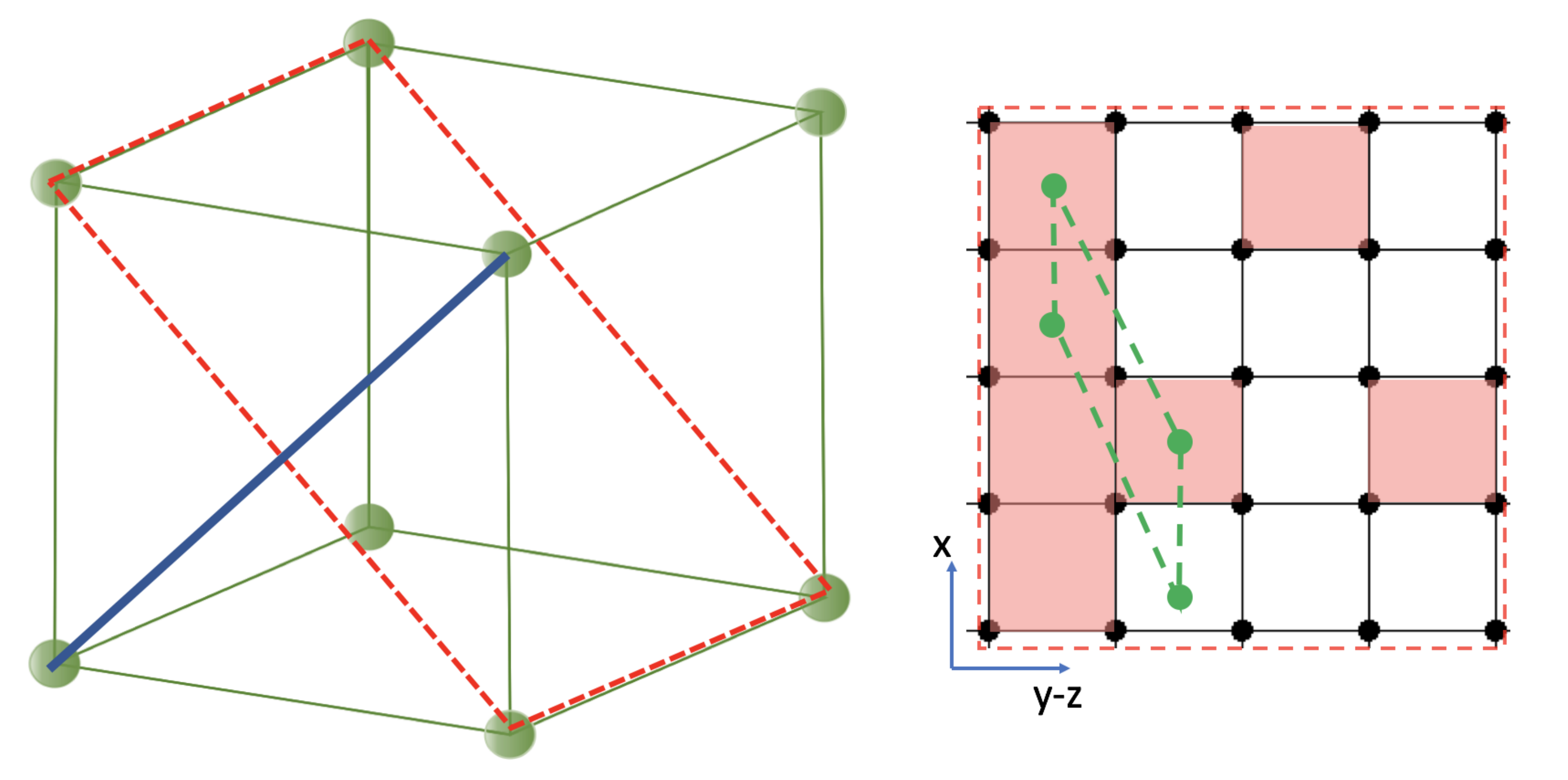}
  \caption{Counting of independent non-overlapping Wilson ribbons on an $L \times L \times L$ system.  Right panel: Each square has side length $a$, representing the width of the fundamental Wilson ribbon; pink squares indicate the number of independent ribbon operators $W_u$ on an $(x, y+z)$ plane.  The trajectories of the diagonal Wilson ribbons on a unit cell of the cubic lattice are illustrated in the left panel.  }  
  \label{Fig:wil}
\end{figure}

Similarly, we have:
\ba
0 = \oint B d u 
&=& \partial_v  \partial_y \oint d u  A_1 = \partial_v \partial_y \Gamma_u
\ea
 This again allows two types of solutions: either $\Gamma_u$ is constant in the $\hat{y}$ direction, or in the $\hat{v}$ direction (meaning that it satisfies $\partial_x \Gamma_u = \partial_z \Gamma_u$).  Since 
 \ba
 \partial_y = \frac{1}{2}( \sqrt{2} \partial_u + \partial_y  + \partial_z )\n
 \partial_v = - \frac{1}{2} ( \partial_u + \sqrt{3} \partial_{u_{\perp}} )\ ,
 \ea
 and $\partial_u \Gamma_u =0$,
 this effectively tells us that we may choose $\Gamma_u$ (and hence $W_u$) to be a function of $y+z$ {\it or} a function of $x$, but not both (Fig.~\ref{Fig:wil} illustrates the relevant geometry).  For $L_i \equiv L$ this naively gives another $2 L/a -1$ independent, non-overlapping ribbon operators. However, not all of these can be independent of the $\Gamma_x$, since 
\be
\oint \Gamma_x d u = \oint \Gamma_u d x \ .
\ee
Thus classically, on an $N a  \times N a \times N a $ system with periodic boundary conditions along the $x,y,$ and $z$ directions, we anticipate $4N-3$ independent line integrals for $A_1$, and the same number for $A_2$.  Since the Chern-Simons constraint in the absence of sources requires that the cage-net configuration in Fig.~\ref{cage} are trivial, the remaining line operators $\Gamma_z, \Gamma_w$ (containing integrals of $(A_1+A_2)$) 
are also fixed.  Thus, there are $8N-6$ independent global flux operators.

For general values of $L_x, L_y, L_z$  the counting of the number of independent line operators is more involved.  For example, 
if the spatial lengths $ L_y/a,L_z/a$ are coprime, a closed $\Gamma_u$ ribbon crosses every point in the $y-z$ plane; clearly there is a maximum of one such operator for each $y-z$ plane, or a total of $L_x$ such operators, of which $L_x -1$ are independent of the $\Gamma_x$.  This gives a total of $(L_y + L_z)/a + L_x/a - 2 $ independent line operators involving $A_1$.  More generally, the number of lines along the $u$ direction in each $y-z$ plane is given by gcd$(L_y/a, L_z/a)$, and we obtain $(L_y + L_z)/a + L_x/a +\text{gcd}(L_y/a, L_z/a) - 3 $ independent line operators involving $A_1$.  Similar considerations apply for line operators involving $A_2$.  

Similarly, the number of independent non-overlapping Wilson ribbons is sensitive to the boundary conditions, since the non-contractible lines can go only along specific directions.  Thus twisting the boundary conditions can lead to dramatically different line operator counting.  The dependence of the number of independent operators on both the aspect ratio and twist of the boundary conditions reflects the fact that our higher-rank Chern-Simons theory is not a topological field theory, but rather is sensitive to both the geometry and the topology of the system.

\subsection{Gauge invariance and gapless boundary modes}

As we saw above, quite generally the Lagrangian (\ref{LCS0}) is not gauge invariant in the presence of boundaries.  For theories where $D_i$ are linear in derivatives, this leads to the chiral boundary modes familiar from usual Chern-Simons theories (or stacks thereof).   
We will now show that gauge invariance requires similar gapless boundary modes in the case at hand.

To be concrete, consider a lattice with a single spatial boundary at $x=0$, and with all fields vanishing as $t \rightarrow \pm \infty$.  (With these boundary conditions we may freely integrate by parts in time without incurring additional boundary terms).  From Eq. (\ref{Eq:CSGtrans}) we deduce that under gauge transformations, the action transforms as:
\begin{align} \label{Eq:deltaS}
\delta S
=& -\int_{ x=0}  \frac{s}{2 \pi}  (  \partial_u \partial_t \alpha  A_2-  \partial_t \partial_{y} \alpha A_1- \partial_u \partial_z \alpha \partial_y A_0) 
\end{align}

To cancel the resulting gauge anomaly, 
we must add a boundary scalar field $\phi$ to our theory, which transforms as $\phi \rightarrow \phi + \alpha$ under gauge transformations.  
Upon adding the term:
\be
\mac{L}_{\text{Bdy}, \phi} = -\frac{s}{2 \pi} \left [  A_{2} \partial_t \partial_u \phi -A_{1} \partial_t \partial_{y } \phi - \partial_u \partial_z \phi \partial_y A_0 \right ] 
\ee
the total action is explicitly gauge invariant.  

We now enforce the constraint $B=0$ by writing
\be
A_1 = \partial_x \partial_u \alpha \ , \ \ A_2 = \partial_y \partial_v \alpha
\ee
Substituting these into the formula (\ref{Eq:deltaS}), and choosing $A_0 =0$,  we obtain the contribution of the gauge fields to boundary action for the scalar field $\alpha$:
\be \label{Eq:Lbdy}
 S_{\text{Bdy}, \alpha}
= -\int_{ x=0}  \frac{s}{2 \pi}  ( 
- \partial_t \partial_u \alpha \partial_y \partial_z \alpha   ) \\
\ee
Note that here we have assumed that we can integrate by parts freely in $y$ and $z$, in order to set
\be
\int_{ x=0} \partial_u \partial_t \alpha \partial_x \partial_{y} \alpha - \partial_{y} \partial_t \alpha \partial_x \partial_{u} \alpha = 0 \ .
\ee
Adding the two contributions to the boundary effective action together, and integrating over the gauge parameter $\alpha$, we obtain the effective action for our boundary scalar field
\begin{align} \label{Eq:Lbdy}
\mac{L}_{\text{Bdy}} =& -\frac{s}{2 \pi}  (\partial_y \partial_z \phi \partial_t \partial_u \phi)
\end{align}

To understand what Eq.~(\ref{Eq:Lbdy}) means for the boundary, let us define
\be
\chi_i = \partial_i \phi
\ee
which is exactly the dipole charge along the $i^{th}$ direction.  
In terms of the $\chi_i$ fields, the boundary action can be expressed:
\begin{align} \label{Eq:Lbdy2}
\mac{L}_{\text{Bdy}} =& -\frac{s}{2 \pi}  (\partial_y \chi_z \partial_t \chi_z-\partial_z \chi_y \partial_t \chi_y)
\end{align} 
This describes two chiral dipole currents -- a $y$-oriented dipole propagating along the $+\hat{z}$ direction, and an $z$-oriented dipole propagating along the $-\hat{y}$ direction -- at the boundary.   However, since the two currents come from the same underlying scalar field $\phi$, the boundary modes are not truly 1-dimensional in their propagation, and this description must be used with some care.

The discussion above shows that in the presence of a boundary our higher-rank Chern-Simons theory is incomplete, and extra fields must be added at the boundary to ensure gauge invariance.  By definition, the action associated with these fields is also not gauge invariant without the bulk, such that no two-dimensional theory that is invariant under the relevant rank-2 gauge symmetry can exist without the bulk.  This is reminiscent of the situation in 2+1 dimensional Chern-Simons theories, where gauge invariance requires chiral boundary modes that are necessarily gapless.  There are two important differences, however.  First, in the case at hand, rank-2 gauge symmetry in a 2-dimensional system requires charge conservation along individual lines, rather than in the system as a whole.  Thus  for a boundary at $x=0$, our result implies that no 2-dimensional theory in which the total charge is preserved along each $y$ and $z$ line can be described by an action of the form (\ref{Eq:Lbdy}).  This suggests that if we take this conservation law to be sacred (meaning that we require subsystem symmetry to be preserved at the boundary), then our rank 2 Chern-Simons theory necessarily has gapless surface states.  Indeed, a theory of this form was previously used to describe gapless boundary modes protected by subsystem symmetry \cite{2018arXiv180302369Y,you2018symmetric,2018arXiv180506899S}.  

Second, for two-dimensional quantum Hall systems, the boundary modes in the absence of the bulk not only violate charge conservation, they also violate energy conservation.  This raises the question of whether there may also be a rank 2 analogue of the thermal Hall effect, associated with our surface dipolar flow.  As we will see, quantizing our theory on the lattice suggests that this is not the case, though we leave a more thorough discussion of this issue for future work.

\section{Discretizing to the lattice}\label{dis}
Before we quantize our theory, we first explicitly write down a discretization of our theory to the simple cubic lattice.  This regularization leads to a quantum theory with a fracton-like ground state degeneracy, and is closely related to a known fracton lattice model, the Chamon code \cite{Chamon2005-fc}.  We will leave the interesting question of whether other regularizations lead to qualitatively different quantum theories for future investigation.

\begin{figure}[h]
  \centering
      \includegraphics[width=0.3\textwidth]{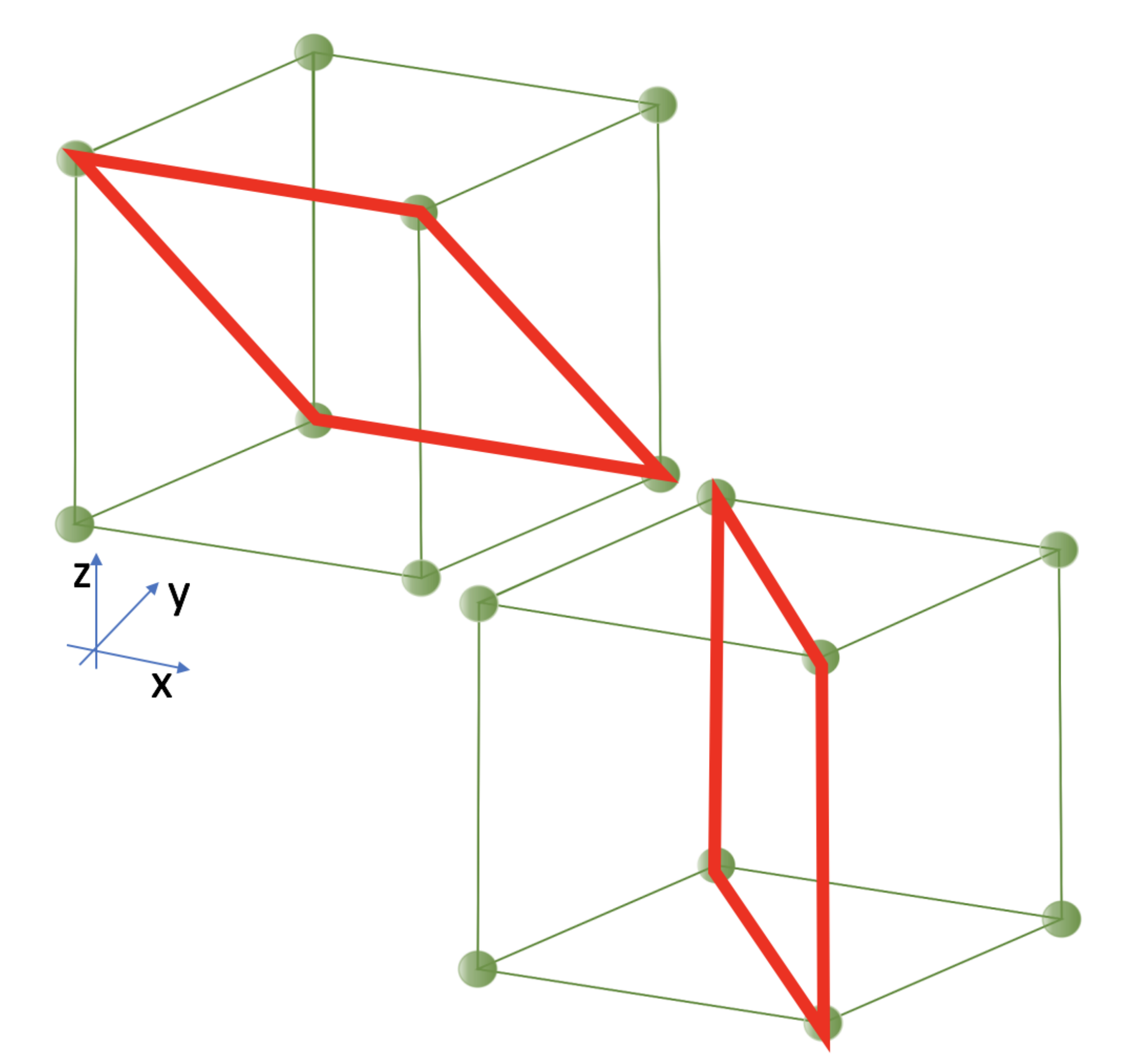}
  \caption{Plaquettes along which ring-exchange interactions occur in our subsystem-symmetric two gauge field model. }
  \label{Fig:RingEx}
\end{figure}

\begin{figure}[h]
  \centering
      \includegraphics[width=0.2\textwidth]{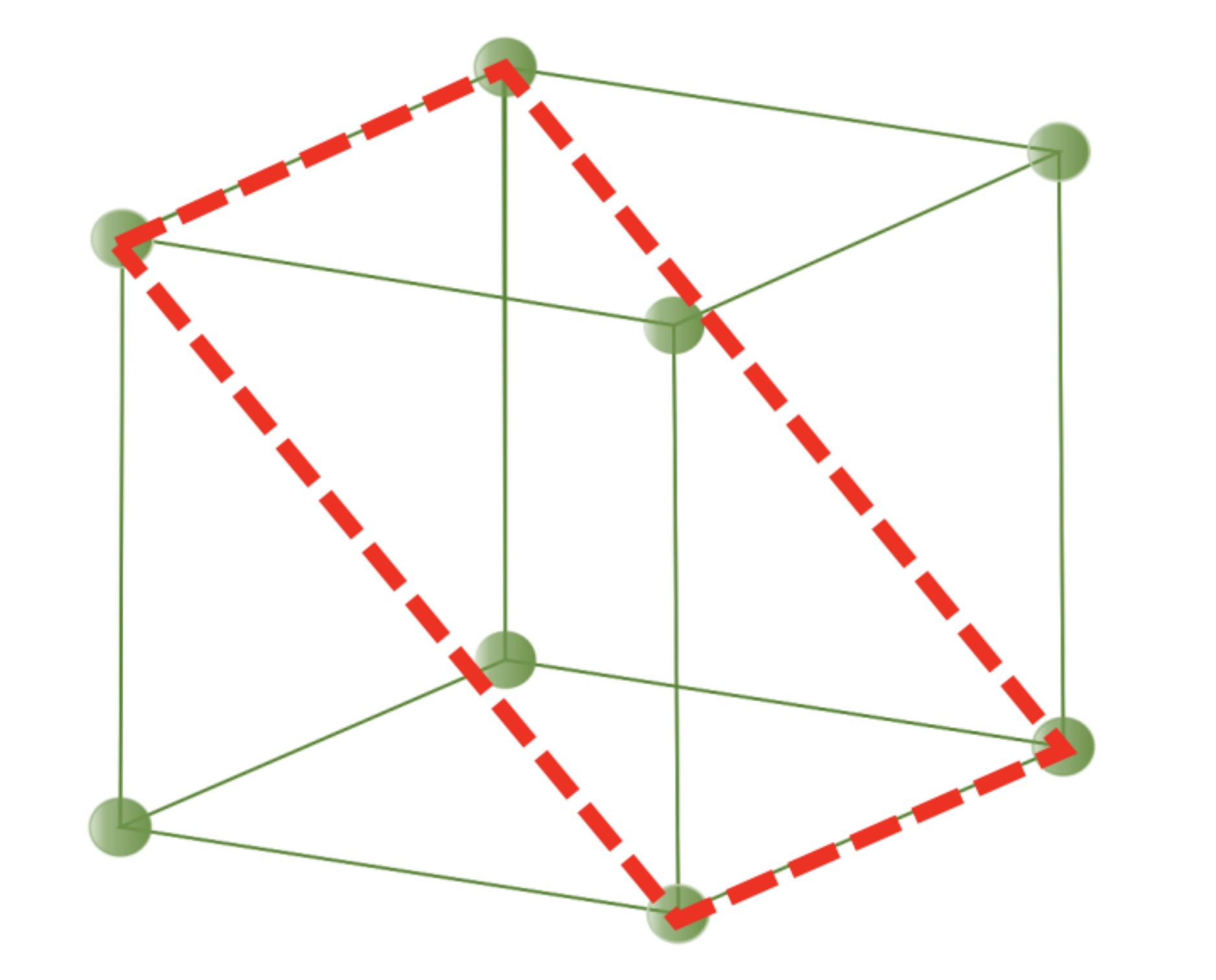}
  \caption{The product of the two ring-exchange processes shown in Fig. \ref{Fig:RingEx} gives a ring exchange process along a plaquette perpendicular to the $(1,1,0)$ direction, times boson number operators at the remaining corners of the cube.} 
  \label{Fig:RingEx2}
\end{figure}

We begin our discussion by showing how the gauge field content and gauge transformations of our model can arise by gauging a model with an appropriate set of planar U(1) subsystem symmetries.
We begin with a model of
charged bosons on the cubic lattice, whose Hamiltonian consists of ring exchange couplings on the three red plaquettes shown in Fig.~\ref{Fig:RingEx} and ~\ref{Fig:RingEx2}, perpendicular to the  $(0,1,1)$, $(1,0,1)$, and $(1,1,0)$ directions.
Specifically, 
\ba
H&=& 
\sum_{\vec{r} } (\phi^\dagger_{\vec{r}+\hat{y}}\phi_{\vec{r} + \hat{x}+\hat{y}} \phi^\dagger_{\vec{r} + \hat{x} + \hat{z}} \phi_{\vec{r} + \hat{z}} \n
&&+  \phi^\dagger_{\vec{r}+\hat{x}}\phi_{\vec{r} + \hat{y}+\hat{x}} \phi^\dagger_{\vec{r} + \hat{y} + \hat{z} } \phi_{\vec{r} + \hat{z} } \n
&& +  \phi^\dagger_{\vec{r}+\hat{x}}\phi_{\vec{r} + \hat{z}+\hat{x}} \phi^\dagger_{\vec{r} + \hat{z} + \hat{y}} \phi_{\vec{r} + \hat{y} })
\ea
 where $\hat{x},\hat{y},\hat{z}$ are the three cubic unit vectors.
 For convenience, we have set the lattice constant to be $1$, and label sites on the cubic lattice via $\vec{r}=(x,y,z) \in\mathbb{Z}^3$.

These ring-exchange interactions preserve the U(1) charge on each $x$-$y$,$y$-$z$,and $x$-$z$ plane, as well as on the family of lattice planes 
 perpendicular to the $(1,1,1)$-direction. Thus, there are four independent subsystem symmetries.

To obtain our desired higher-rank gauge theory, we  place a spatial gauge field $A_1, A_2$ at the center of each of the two types of plaquette in Fig.~\ref{Fig:RingEx}, and a time-like gauge field $A_0$ at each lattice site.  We label these gauge fields via $A_{0,1,2}(\vec{r},t)$, where 
$t$ is a continuous time variable.
We use the vector $\vec{r}$ in both cases, even though 
$A_{1,2}(\vec{r},t)$ are in fact associated with the dual lattice site at $\vec{r}+\frac{\hat{x}+\hat{y}+\hat{z}}{2}$.

We follow the prescription of Ref. \cite{Vijay2015-jj,Vijay2016-dr} to obtain the minimal coupling between these plaquette gauge fields and matter.  On a plaquette perpendicular to $(0,1,1)$, this gives
\be  
\phi^\dagger_{\vec{r}+\hat{y}}\phi_{\vec{r} + \hat{x}+\hat{y}} \phi^\dagger_{\vec{r} + \hat{x} + \hat{z}} \phi_{\vec{r} + \hat{z}} e^{i A_1 (\vec{r})} \ .
\ee
For plaquettes normal to  $(1,0,1)$ the coupling is analogous, with $A_1$ replaced by $A_2$.  However, because the product of the two ring-exchange terms in Fig.~\ref{Fig:RingEx} gives rise to a ring-exchange process of the type shown in Fig.~\ref{Fig:RingEx2} (times some charge-neutral boson number operators), the gauge connection on plaquettes perpendicular to the $(1,1,0)$ direction is just $(A_1+A_2)$.  Thus our model has only two independent gauge fields on the cubic lattice.

Let us define the forward difference operator in the $x$ direction, 
\begin{equation} \label{Eq:ForwardDiff}
    d_x f(\vec{r},t) = f(\vec{r}+\hat{x},t)-f(\vec{r},t)
\end{equation}
and similarly for $d_y$ and $d_z$.
We also define the backwards difference operator,
\begin{equation}
    \hat{d}_x f(\vec{r},t) = f(\vec{r},t)-f(\vec{r}-\hat{x},t)
\end{equation}
and similarly $\hat{d}_y$ and $\hat{d}_z$.
Now, we may define the discretized version of our differential operators,
\begin{eqnarray}
D_1 = d_x(d_y-d_z)\\
D_2 = d_y(d_z-d_x)
\end{eqnarray}
and also $\hat{D}_1$ and $\hat{D}_2$, with backwards difference operators instead.
Under a U(1) gauge transformations that takes $\phi_{\vec{r}} \rightarrow e^{ i \alpha_{\vec{r},t}} \phi_{\vec{r}}$, the gauge fields transform as
\begin{eqnarray}
A_0(\vec{r},t) \rightarrow A_0(\vec{r},t)+\partial_t \alpha(\vec{r},t)\\
A_i(\vec{r},t) \rightarrow A_i(\vec{r},t)+D_i \alpha(\vec{r},t)
\end{eqnarray}
 In the continuum limit, this yields the gauge transformations discussed in Sec.~\ref{dipole} up to an overall re-scaling of the gauge field:
\be \label{Eq:CtmGtrans}
A_1 \rightarrow  A_1+ \sqrt{2}\partial_x \partial_{u}  \alpha, A_2 \rightarrow A_2+ \sqrt{2}\partial_y \partial_{v}   \alpha \ \ .
\ee
We note that the generalized theories described in Sec.~III are also naturally described by a model of the type described here, albeit on a distorted cubic lattice, in which the $x-u$ plane is deformed into the  $(x+(\alpha-\beta) l)-u$ plane, and similarly for the other directions.

\begin{figure}[h]
  \centering
      \includegraphics[width=0.3\textwidth]{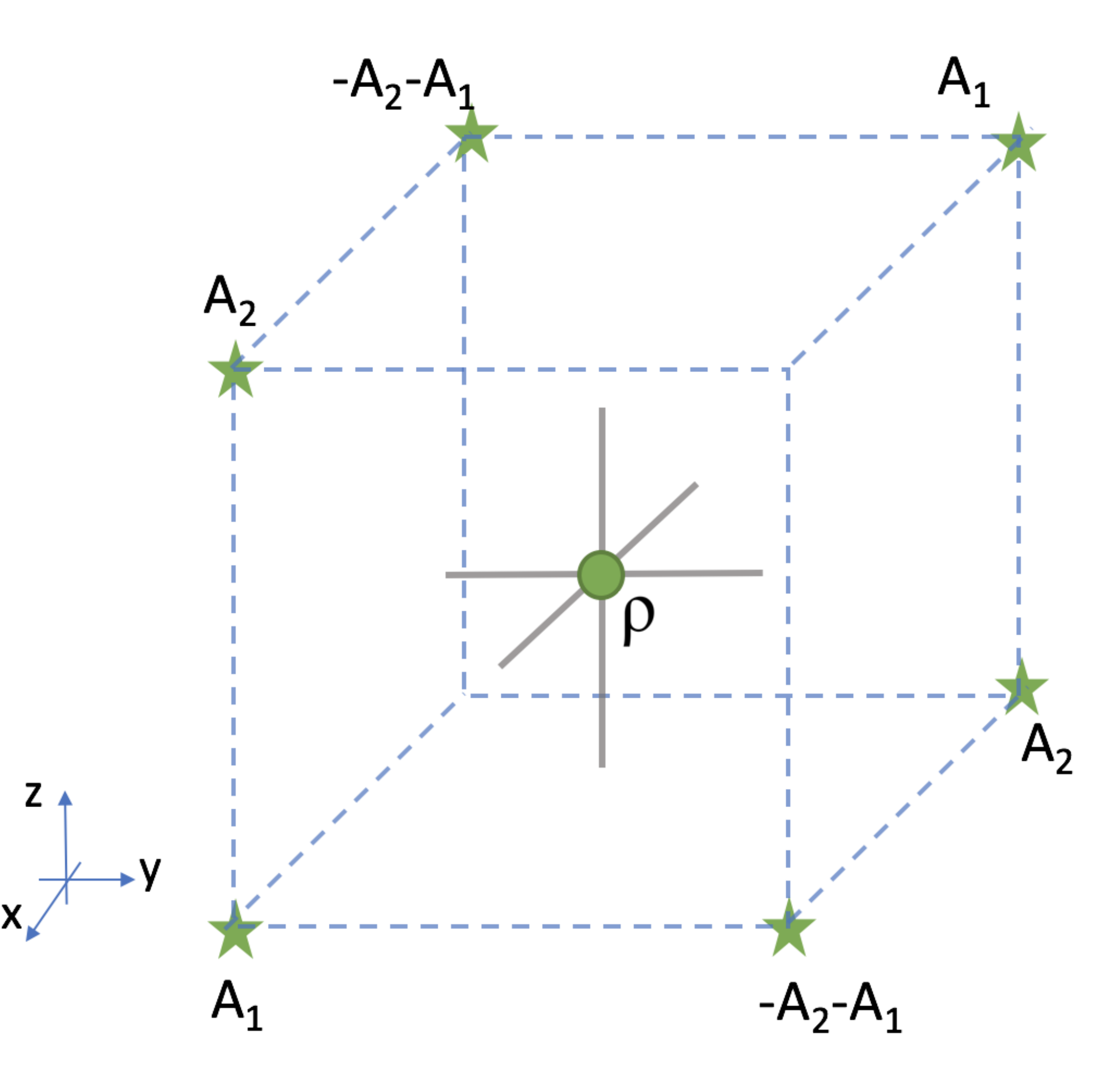}
  \caption{ The Chern-Simons constraint on the cubic lattice. The magnetic field at a given lattice site (green dot at the center of the cube) is given by the linear combination of gauge fields $A_1,A_2$ at 6 of the surrounding dual lattice sites(star). The Chern-Simons constraint sets this magnetic field equal to the charge $(\rho)$ on this lattice site.} 
  \label{Fig:density}
\end{figure}

The gauge-invariant electric fields are defined in the same way as before, using these discretized difference operators,
\begin{equation}
E_i(\vec{r},t) = \partial_t A_i(\vec{r},t) - D_i A_0(\vec{r},t) \ \ .
\end{equation}
Each electric field is associated with a cube in the cubic lattice.
The magnetic field should be defined with backwards difference operators,
\begin{equation} \label{Eq:BLatt}
B(\vec{r},t) = \hat{D}_2 A_1(\vec{r},t) - \hat{D}_1 A_2(\vec{r},t) \ \ ,
\end{equation}
hence each magnetic field is associated with a site, as shown in Fig.~\ref{Fig:density}.
One can verify that the $B$ field is gauge-invariant by noticing that $\hat{D}_i = S_{-x} S_{-y} S_{-z} D_i$, where $S_{-x}$ is a shift operator in the $-x$ direction, $S_{-x}f(\vec{r},t) = f(\vec{r}-\hat{x},t)$.
Then, the variation of $B$ under a gauge transformation is $\delta B = S_{-x}S_{-y} S_{-z} (D_2 D_1 - D_1 D_2) \alpha(\vec{r},t) = 0$

In the absence of matter sources, the lattice fractonic Chern-Simons action then takes the form
\begin{eqnarray} \label{Eq:ScsLat}
\begin{split}
S_{CS} =&  \frac{s}{4 \pi} \int dt \sum_{\vec{r}} \left[ A_1(\vec{r},t) E_2(\vec{r},t) \right.\\
&\left. - A_2(\vec{r},t) E_1(\vec{r},t) 
 + A_0(\vec{r},t)B(\vec{r},t)\right]
\end{split}
\end{eqnarray}
which one can verify is gauge-invariant in the absence of a boundary. 
For this, one must use a summation by parts, which for our difference operators simply amounts to the identity
\begin{equation}
\sum_{\vec{r}} f(\vec{r},t) D_i g(\vec{r},t) = \sum_{\vec{r}} (\hat{D}_i f(\vec{r},t)) g(\vec{r},t)
\end{equation}
up to boundary terms.

Notice that our theory does not run into the subtle issues associated with discretizing and quantizing the regular 2D Chern-Simons theory (see for example Ref~\onlinecite{eliezer199266}).
These subtle issues arise when, for example, canonically conjugate variables do not live on the same location (in 2D CS theory, $A_1$ lives on the $x$ links, while $A_2$ lives on the $y$ links), or when there are multiple natural choices to be made for the charge-vortex binding (a one-to-one correspondence between plaquettes and vertices is required for the discretized 2D CS theory~\cite{PhysRevB.92.115148}).
Our model sidesteps these issues, as the conjugate variables $A_1$ and $A_2$ are both located at the center of cubes, and both the $B$ field and charges are located on the vertices (see Fig.~\ref{Fig:density}). Consequently, the lattice discretization does not attribute any subtlety when quantizing the Fractonic Chern-Simons term.

\subsection{Gauge invariant ribbon operators on the lattice}

It is worth briefly discussing how the gauge invariant line and ribbon operators are manifest on the lattice.  To do this, we re-introduce the lattice constant, and imagine that the lattice gauge field $A^{\text{latt}}_1(\vec{r}, t)$ is related to a continuum gauge field $A^{\text{cont}}_1 (\vec{x},t)$ by integrating over the associated plaquette: 
\be
A^{\text{latt}}_1(\vec{r}, t) = \int_{0}^{a} d x \int_{0}^{\sqrt{2} a} d u A^{\text{cont}}_1 (\vec{r} + x \hat{x} + u \hat{u}, t) 
\ee
Note that in the continuum limit, if we assume that $\tilde{A}_1$ is smooth, this gives
\be
A^{\text{latt}}_1(\vec{r}, t) = \sqrt{2} a^2  A^{\text{cont}}_1(\vec{r}, t)
\ee
explaining the relative factor of $\sqrt{2}$ in Eq.~(\ref{Eq:CtmGtrans}).  The factor of $a^2$ gives the expected relationship between the dimensionless lattice gauge field, and our continuum gauge field with dimensions of $1/$length$^2$.  

Since the lattice gauge field is dimensionless, the dimensionful line integrals do not have a lattice analogue.  However, the dimensionless ribbon operators do, since
\be
\int_{x_0}^{x_0 + l a} d x \int_{u_0}^{u_0+\sqrt{2} a} du A^{\text{cont}}_{1}(\vec{x}, t)  =\sum_{n=0}^{l} A^{\text{latt}}_1(\vec{r}_0 + n a \hat{x} , t)
\ee
and similarly for other directions.  In other words, in our lattice theory a ribbon corresponds to a line of plaquettes, with the dipole scale $a$ set by the lattice constant.  We will henceforth refer to these operators as lattice Wilson ribbons, or simply Wilson ribbons in contexts where the lattice is understood.

Cage nets on the lattice are constructed from the ribbon operators, exactly as described in the continuum case in Sec. \ref{Sec:CageNet}.  It is straightforward to show that these  lattice cage net operators are fixed by the value of the magnetic field they enclose, and hence that constraint $B=0$ completely fixes these.

\section{Quantizing fractonic lattice Chern-Simons theory} \label{Sec:Quantum}

We now discuss quantization of the fractonic Chern-Simons theory, using the lattice regularization introduced in Sec.~\ref{dis}.  
Following Ref.~\cite{witten1991quantization}, we will quantize within the constrained subspace -- meaning that we will first restrict ourselves to configurations where the magnetic field $B({\vec{r},t})$ defined in Eq.~(\ref{Eq:BLatt}) vanishes everywhere.   The remaining gauge-invariant operators are the gauge-invariant ribbon operators, and our focus will be on quantizing these in our lattice theory, bearing in mind that not all of them are independent in the constrained Hilbert space.  

\subsection{Is the Chern-Simons coefficient quantized? }

Before quantizing the theory, it is useful to ask whether, 
if the gauge parameter $\alpha \equiv \alpha + 2 \pi$ is compact, the Chern-Simons coefficient is quantized.  Recall that in ordinary compact U(1) Chern-Simons theory, such quantization is necessary to ensure that large gauge transformations (for example, those that thread a flux of $ 2 \pi$ through one of the non-contractible curves on the torus) do not actually affect the partition function. 

To study this question in more detail, let us consider a gauge transformation of the form
\be
\alpha(x,y,z) = \begin{cases} 2 \pi & x > x_0, z > z_0 \\
 0 & \text{otherwise}
 \end{cases}
 \ee
 Note that this gauge transformation is allowed if $\alpha \equiv \alpha+ 2 \pi$; if not it would be incompatible with our choice of periodic boundary conditions.
This gauge transformation takes  
\be \label{Eq:Holonomy}
A_{1} (x,y,z) \rightarrow A_{1} (x,y,z)- 2 \pi  \delta_{x,x_0} \delta_{z,z_0}
\ee
where $\delta_{x, x_0}$ are Kroenecker $\delta$ functions.  As explained above, the constraint ensures that $A_1$ must be independent of the remaining co-ordinate $y$.  

\begin{figure}[h]
  \centering
      \includegraphics[width=0.45\textwidth]{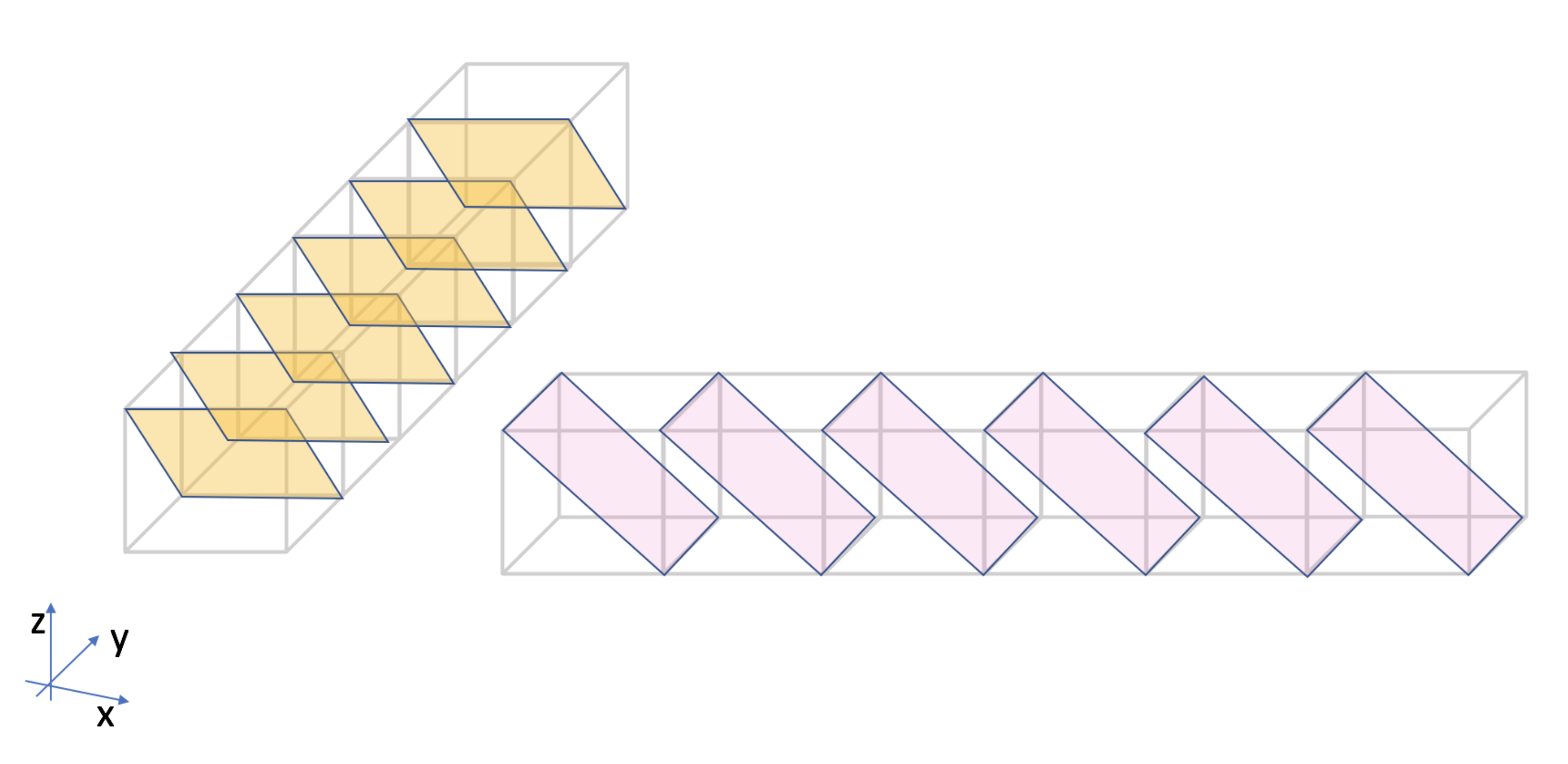}
  \caption{Large gauge transformations on the lattice consist of changing $A_1$ along a line of plaquettes parallel to the $y$ (or $u$) direction, or $A_2$ along a line of plaquettes parallel to the $x$ (or $v$) direction, as shown.}
  \label{lgt}
\end{figure}

In this configuration the gauge field $A_1$ vanishes everywhere except along a line of plaquettes (see Fig.~\ref{lgt}), where it has the value $2 \pi$.  
The Wilson operators are:
\ba
\sum_{n=0}^{N_x} A_{1} (\vec{r} + n a \hat{x}, t)  = -2 \pi \delta_{z_0, r_3} \n
\sum_{n=0}^{N_u} A_{1} (\vec{r} + n a (\hat{y}- \hat{z}), t) =  -2 \pi \delta_{x_0, r_1} 
\ea
indicating that a dipole that encircles the torus along any one of these lines acquires a net phase change of $2\pi$. Similar results hold for ribbons along the $z$ and $w$ directions, which also involve $A_1$.  If we take $\alpha$ to be compact, then this phase of $2 \pi$ should not affect the physics at all, and the configuration in Eq.~(\ref{Eq:Holonomy}) corresponds to a holonomy of our rank-2 gauge field.  

Let us now consider the effect of this gauge transformation on the Chern-Simons action.  To avoid complications due to boundaries of the manifold in time, we consider periodic boundary conditions in time and space.   (Here, as for usual rank 1 Chern-Simons theories, this choice is important since open boundaries require additional fields to preserve gauge invariance).  The net change in our Chern-Simons action is: 
\begin{align}
\mac{S}_{CS} &\rightarrow S_{CS}  - \frac{s}{2 \pi} \sum_{ \vec{r}} \int ( 2 \pi  \delta_{x,x_0} \delta_{z,z_0}  E_2 )  dt \n
&=  - s \sum_{ y} \int  E_2(x_0, y, z_0,t)   dt
\end{align}
Note that to obtain the factor of $2$ here comes from integrating by parts in time, which does not induce boundary terms with our choice of periodic boundary conditions.

To complete our analysis, we must understand the quantization of $\sum_y \int dt E_2$.  
With periodic boundary conditions in time, we may consider only processes in which the initial and final {\it gauge field } configurations are equivalent (up to a gauge transformation).  Thus, let $E_2$ be the electric field generated by
turning on a second holonomy, by taking
\be
A_{2} = \frac{ 2 \pi   t }{ \tau} \delta_{y,y_0} \delta_{z,z_0}
\ee
where $\tau$ is the radius of the circular time dimension.  Then $E_2$ is constant in time, and 
\be
\sum_{ y} \int  E_2(x_0, y, z_0,t)   dt = 2 \pi
\ee
Thus, in order to ensure that the gauge transformation (\ref{Eq:Holonomy}) does not change the partition function, the appropriate quantization for our Chern-Simons coefficient is
\be \label{CsQuant}
s \in \mathbb{Z} \ .
\ee

It is worth noting that the above argument must be modified slightly in the continuum theory, where the gauge field has dimensions $1/$length$^2$, and the Chern-Simons coupling $s$ thus has dimensions of length.   
In this case any the quantization of the Chern-Simons coupling must depend on some fundamental length scale in the problem, suggesting that the quantized theory requires a fixed ultraviolet cut-off.  For this reason, it is natural to quantize the lattice theory, rather than its continuum cousin.

\subsection{Canonical commutation relations}

Having established that for compact $U(1)$ gauge transformations the Chern-Simons couping coefficient is quantized, we are ready to quantize our higher-rank lattice Chern-Simons theory.  
From the lattice action  (\ref{Eq:ScsLat}), the canonical commutation relations of the gauge fields $A_1,A_2$ are
\be \label{Acoms}
[ A_1 (\vec{r}, t) , A_2 (\vec{r}', t) ] = i\frac{ 2 \pi}{ s }  \delta_{\vec{r}, \vec{r}'}
\ee
Formally, we wish to work within the constrained subspace where $B=0$, and quantize the remaining gauge invariant ribbon operators.  As discussed in Sec. \ref{Sec:conservation}, it is sufficient to consider only ribbon operators involving $A_1$ and $A_2$, as the values of the remaining ribbon operators involving the linear combination $-A_1-A_2$ are not independent.  
Two such ribbon operators necessarily intersect on a single cube, and hence the sums involved share only a single site.  
Thus the commutators between intersecting ribbon operators along the major cubic axes are:
\ba \label{WComs}
W_{x} W_{z} = W_{z} W_{x}  e^{-i   \frac{2  \pi }{s} } \n
W_{x} W_{y } = W_{y} W_{x}  e^{i  \frac{2  \pi  }{s} } \n
W_{z} W_{y } = W_{y} W_{z}  e^{-i  \frac{2  \pi  }{s} }
\ea
  Evidently, 
\be
[W_{u}, W_{x}  ] = [ W_{v}, W_{y}] = [ W_{w} , W_{z} ] = 0  \ \ . 
\ee
while
\be \label{Eq:OffDComs}
W_{u} W_{y}   = \left( e^{i  \frac{2  \pi  }{s} } \right)^{b} W_{y} W_{u} 
\ee
where $b $ counts the number of times that a line along the $u$ direction intersects a line along the $y$ direction.  For example, if $L_y \leq L_z$ then $b=1$; if $L_y = m L_z$ with $m\in \mathbb{Z}$, $b=m$.  Similar commutators apply for the remaining directions.  

Eq. (\ref{WComs}), together with the fact that the $W_i$ are compact operators, implies that in the quantized theory they are discrete, with a finite set of eigenvalues:
\be
W_{i,j} = e^{ 2 \pi i l_{ij}/n}
\ee
where $n = s/a$, 
and $ l_{ij} \in \mathbb{Z}$.

 \subsection{Ground state degeneracies}
 
Since this theory is fully gapped, one telling quantity is the number of ground states.  For topological quantum field theories this number can depend only on the topology of the underlying spatial manifold.  In the present case, the ground state degeneracy is sensitive not only to the topology of the underlying manifold, but also to geometrical factors including the system size and twist angle of the boundary conditions.  Here we will examine this dependence.

The ground states are fully characterized by the eigenvalues of the gauge-invariant line operators in the absence of matter fields.  For the case at hand, these are given by the 6 ribbon operators:
\be
W_{x} \ , \ \  W_{u}  \ , \ \ W_{y}  \ , \ \  W_{v} \ , \ \ W_{z}  \ , \ \  W_{w}  \ \ .
\ee
We begin by considering periodic boundary conditions along the $x$, $y$, and $z$ directions.  In this case, as discussed above, there are $L_z/a + L_y/a -1$ operators of the type $W_{x}$, and $ L_x/a + \text{gcd}(L_y/a, L_z/a) -2$ additional independent operators of the type $W_{u}$.  These can all be simultaneously diagonalized.

Let us first diagonalize all line operators $W_{x}$ running parallel to the $x$-axis.    Since every line along $y$ ($z$) intersects at least one straight line along $x$, we cannot simultaneously diagonalize $W_{x}$ and $W_{ y}$ ($W_{z}$).   Because the lines are straight, however, we may simultaneously diagonalize all $L_z/a + L_y/a -1$ operators $W_{x}$, and all independent operators of the form $W_{ y} (x, z_0) ( W_{ y}(x_0, z_0) )^{-1}$.   
Since $\partial_x \partial_z W_{ y} = 0$, there are $L_x/a -1$ independent operators of this type.  (The logic here is that they must be independent of $z$, since the derivative in $x$ cannot vanish.) 
This set fails to commute with any combinations of line operators parallel to the $z$ axis. 
Thus there are a total of $L_x +L_y+ L_z -2$ simultaneously diagonalizeable line operators along the cubic directions.  

Next, we consider whether any of the operators along the $u$ direction can be diagonalized simultaneously with this entire set.  The operator $W_{u}$ commutes with $W_{x}$, but in general not with $W_y$ ribbons with which it intersects. However, one can choose a set of linear combinations $W_{u} (x_0, y_0+z_0) (W_{u}(x_0, y_i+z_i))^{-1}$ which commute with all $W_{ y} (x, z_0) ( W_{ y}(x_0, z_0))^{-1} $ operators and hence these $W_u$ lines {\it can} be simultaneously diagonalized with the full set described above, leading to an additional $L-1$ line operators on an $L \times L \times L$ system.

Finally, we must determine how many eigenvalues each line operator may take.  From taking linear combinations of  Eq. (\ref{WComs}), we see that lines that intersect once change each others' values by $2 \pi /s$, leading to $s$ possible eigenvalues for each ribbon operator in our set.  If we choose $L_x = L_y = L_z \equiv L$ (in which case the diagonal ribbons do not contribute to the ground state degeneracy), we obtain a total degeneracy of $( s)^{ 4L /a -3}$ states.

\subsection{ Statistical interactions}

From the commutation relations between the operators $W_{i}$, it is straightforward to infer the quasiparticle statistics.  In general, statistical interactions between particles with one-dimensional motion (lineons) can be non-trivial only if both particles move in the same plane, such that their world-lines intersect.  In our model not all intersecting line operators fail to commute, meaning that some pairs of lineon excitations have trivial mutual statistics.  An example are the excitations that travel along the $x$ direction, and those travelling along the $y-z$ direction, both of which are associated with integrals of the gauge field $A_1$.

\begin{figure}[h]
  \centering
      \includegraphics[width=0.4\textwidth]{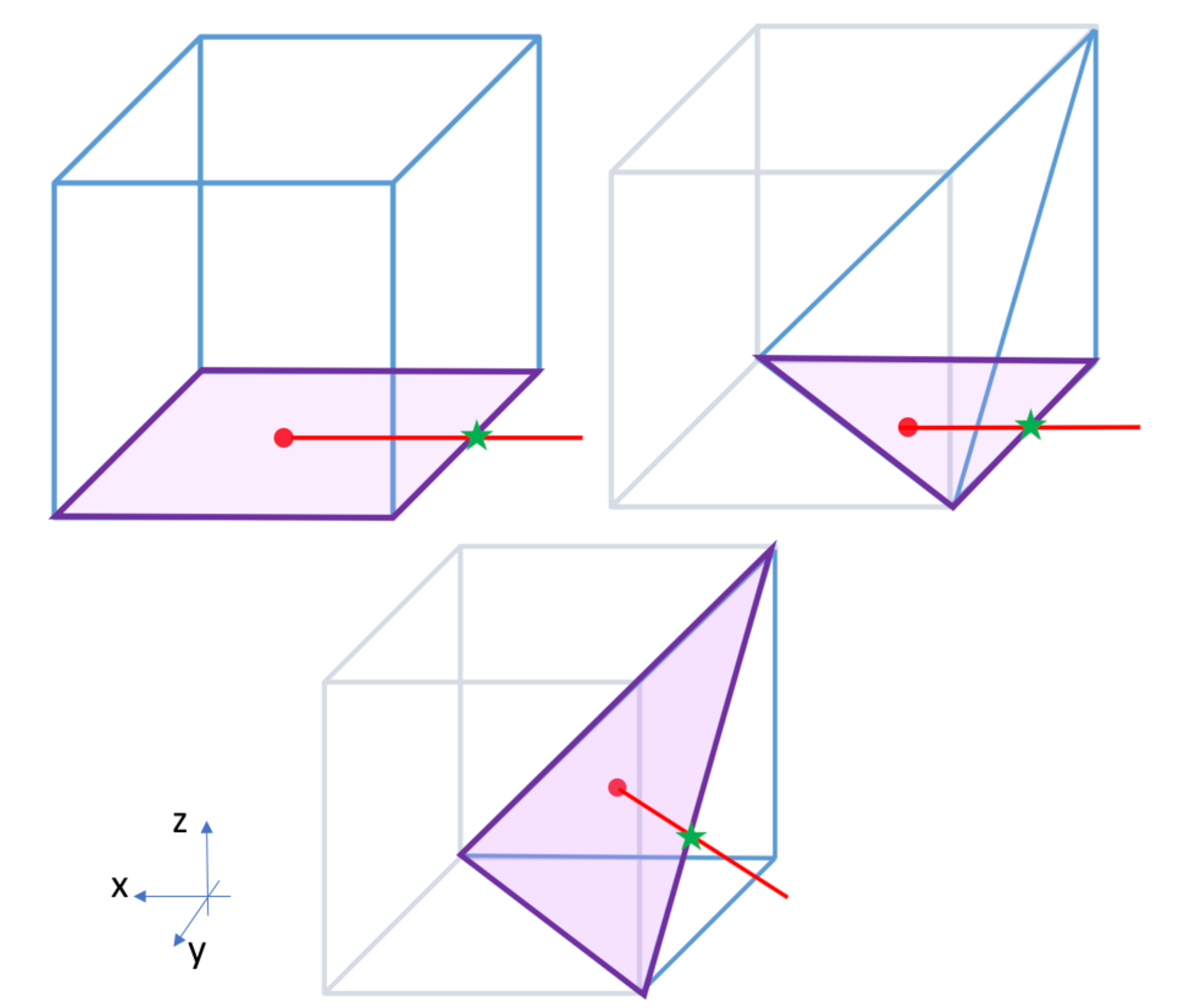}
  \caption{Schematic of the statistical processes of dipoles in the quantized lattice theory.  Red lines represent Wilson ribbons connecting the dipole (represented by a red dot) to a distant anti-dipole; blue lines represent the ribbon operators of a cage net.  Green dots indicate where the dipole and cage-net ribbons cross.} 
  \label{braid}
\end{figure}

World-lines that intersect and fail to commute, such as $W_{x}$ and $W_{y}$, imply that the dipoles have ``lineon mutual statistics".  Following Ref.\cite{huang2018cage,chen2019,pai2019fracton}, we define these statistics by comparing two processes.  In process (a) We first place a dipole at the origin, and then create a dipole-anti-dipole pair and move the anti-dipole around in a plane surrounding the origin.   As discussed above, any turns in the anti-dipole's trajectory create other (anti)-dipoles; hence  to return the system to its ground state these other dipoles must also be brought together and annihilated; the entire process is represented by a cage net, as shown in Fig.~\ref{braid}.  In process (b) we first create, move, and re-annihilate the other dipoles to create the cage-net, and then (after all of these excitations have vanished) we bring our dipole to the origin.  Evidently, the restricted mobility of our dipoles constrains both the planes in which they can encircle each other, and the shapes of the corresponding cage nets.

The braiding phase is determined by the phase difference between processes (a) and (b), which results from the commutator between two intersecting ribbon operators, as shown in Fig.~\ref{braid}.  For example, if the ribbon ending on the dipole runs along the $\hat{x}$ direction, and the cage net has a surface in the $x-y$ plane, we obtain
\begin{align}
&M_{ab} =\frac{ W_x W_y }{W_y W_x} = e^{i  2 \pi/s}
\end{align}

Similarly, as described in Ref. \cite{you2018symmetric,huang2018cage}, the available cage-net moves can be used to define a type of self-statistics for the lineons.   Note that though some aspects of our theory, such as the ground state degeneracy, are explicitly cut-off dependent, these statistical interactions are scale invariant, depending only on the pattern of crossings between the cage frame and the Wilson ribbon associated with our dipole.  

\section{Quantizing and constraining: fractonic Chern-Simons theory and the Chamon code} \label{Sec:LattQuantum}

Since much of our current understanding of fracton order is based on studying commuting projector lattice Hamiltonians, we now turn to the question of what lattice model of this type could potentially be described by our Chern-Simons theory.  To do this, we will consider first quantizing the lattice theory in Sec.~\ref{dis}, and then imposing the constraints.  We will see how at the second step a mass gap for matter fields results in a Hamiltonian that can be viewed as a $\mathbb{Z}_s$ generalization of the Chamon code \cite{Chamon2005-fc}.

We begin with the commutation relations
\be 
[ A_1 (\vec{r}, t) , A_2 (\vec{r}', t) ] = i\frac{ 2 \pi}{ s }  \delta_{\vec{r}, \vec{r}'}
\ee
Recall that here $\vec{r}, \vec{r}'$ refer to sites at sites on the dual cubic lattice, and that gauge fields on different dual lattice sites commute.  If $A_1$ is compact, then this commutation relation implies that $A_2$ is quantized in units of $ 2 \pi /s$; similarly if $A_2$ is compact, then $A_1$ is quantized.  Thus if the gauge fields are compact, our quantized theory is described by an $s$-state spin on each dual lattice site, with
\be \label{Eq:AtoSpin}
e^{ i A_1} = U \ , \ \ \ e^{ i A_2} = V
\ee
where $U, V$ are $s$-state clock matrices, given by
\ba 
U_{mn } = \delta_{mn } e^{ 2 \pi i n/s } \n
V_{mn} = \delta_{m, (n+1 \mod s)} 
\ea
such that $UV = e^{2\pi i / s} VU$.

Next, we must impose the constraint $B = \rho$ at the lattice level.  Recall that the lattice magnetic field at site $\vec{r}$ on the direct lattice  is given by $B(\vec{r}, t) = \hat{D}_2 A_1 (\vec{r}, t) - \hat{D}_1 A_2 (\vec{r},t)$, where the combination of gauge fields is shown in Fig.~\ref{Fig:density}.  We have
\ba
e^{ i B(\vec{r},t) } &=& e^{ -i (A_1(\vec{r}-\hat{x}-\hat{y},t)+ A_2 (\vec{r}-\hat{x}-\hat{y},t))} \n 
&& \times e^{ i A_1 (\vec{r} - \hat{x},t) } e^{ i A_1(\vec{r}- \hat{y} - \hat{z},t)} \n 
&& \times e^{i A_2 (\vec{r}- \hat{y},t)} e^{ i  A_2 (\vec{r}- \hat{x} - \hat{z},t)}   \n 
&& \times  e^{-i(A_1 (\vec{r}-\hat{z},t) )+ A_2(\vec{r}-\hat{z},t))  }
\ea
where we have used the fact that gauge fields on different sites commute.  

\begin{figure}[h]
  \centering
      \includegraphics[width=0.4\textwidth]{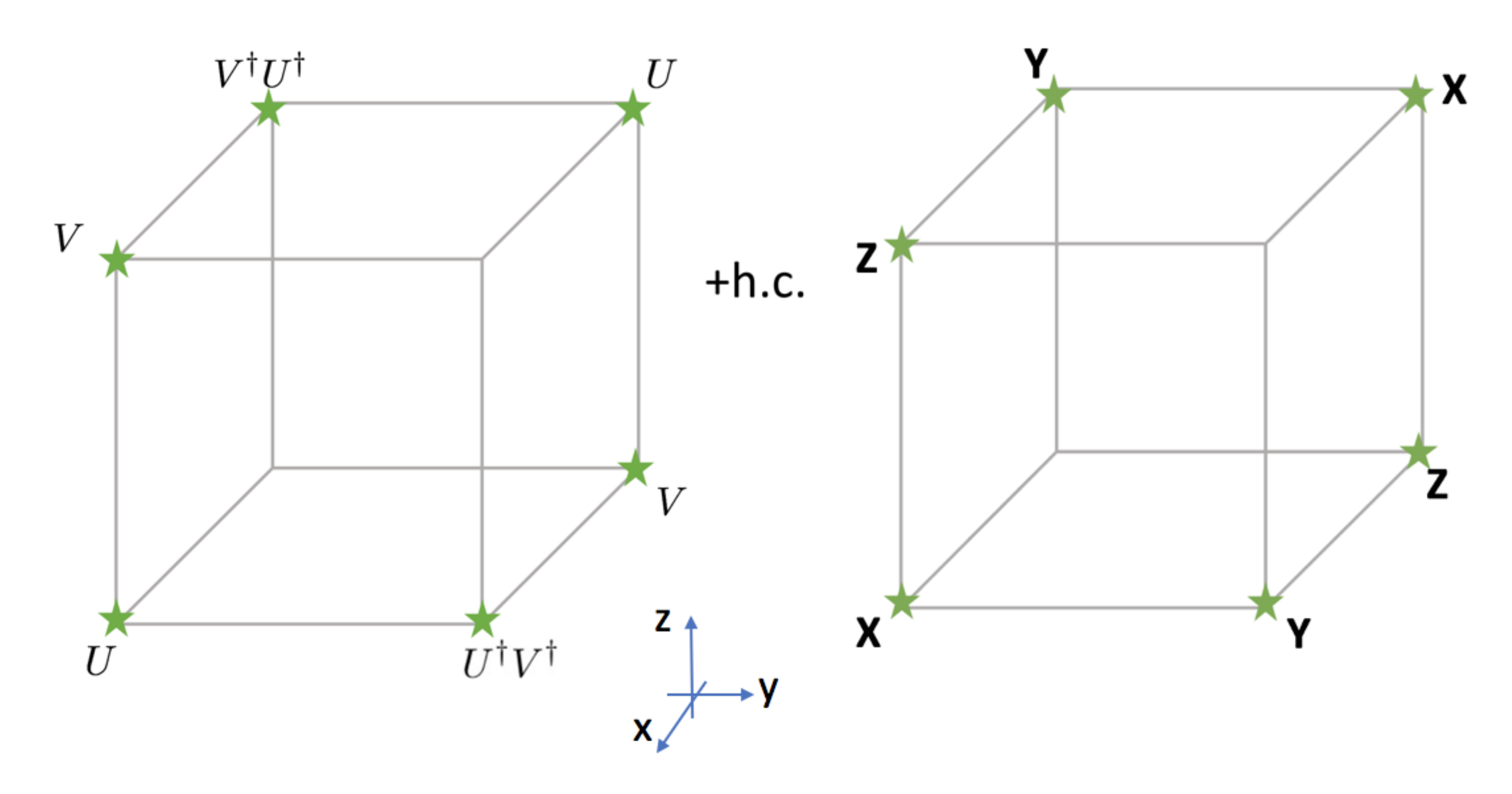}
  \caption{Left: The magnetic field operator in our quantized (but unconstrained) theory.  Adding a mass term for matter fields and imposing the Chern-Simons constraint $B= \rho$ leads to a Hamiltonian that can be expressed as the product of the 6 operators shown, plus its Hermitian conjugate.  Right: For $s=2$, this construction gives the tilted Chamon code. 
}
  \label{cha}
\end{figure}

In terms of the spin matrices identified in Eq.~(\ref{Eq:AtoSpin}), this product can be expressed:
\ba
e^{ i B(\vec{r},t) } &=& U_{\vec{r}-\hat{x}-\hat{y}}^\dagger V_{\vec{r}-\hat{x}-\hat{y}}^\dagger
U_{\vec{r}-\hat{x}} U_{\vec{r}-\hat{y}-\hat{z}} \n
&&\times V_{\vec{r}-\hat{y}} V_{\vec{r}-\hat{x}-\hat{z}} V_{\vec{r}-\hat{z}}^\dagger U_{\vec{r}-\hat{z}}^\dagger
\label{eq:eiB}
\ea
where we have 
used
\be
e^{ -i ( A_1(\vec{r},t) + A_2 (\vec{r},t))} = e^{ - i A_1(\vec{r},t)} e^{ - i A_2 (\vec{r},t)} e^{ - i \pi /s}
\ee
The result is a product of six spin operators at the corners of the cube, as shown in Fig.~\ref{cha}.

The lattice Hamiltonian corresponding to our pure Chern-Simons theory is then: 
\be \label{HLattice}
H = - \frac{1}{2} \sum_{\vec{r}} \left( e^{ i B(\vec{r}, t) }+ e^{ -i B(\vec{r}, t) }\right) \ .
\ee
Clearly, the ground states of this model obey $B(\vec{r}, t)  \equiv 0$, corresponding to the manifold of states of the Chern-Simons theory in the absence of sources.  The excited states can be understood as the result of 
 introducing gapped, non-dynamical matter sources on the sites of our lattice.  After imposing the constraint $ B (\vec{r},t) =\frac{2 \pi }{s} \rho (\vec{r},t) $, the mass gap for these non-dynamical sources leads to a Hamiltonian of the form (\ref{HLattice}).

An interesting example is  $s=2$.  In this case, we have  $\sigma^x=U,\sigma^z=V$, and $U V = - i \sigma^y$, which obeys the required algebra $UV = - VU$.
In this case our Hamiltonian (\ref{HLattice}) becomes
\begin{align}
H
&=-\sum_{\vec{r}} \sigma^x_{\vec{r}-\hat{x}}\sigma^x_{\vec{r}-\hat{y}-\hat{z}}\sigma^y_{\vec{r}-\hat{x}-\hat{y}}\sigma^y_{\vec{r}-\hat{z}}\sigma^z_{\vec{r}-\hat{x}-\hat{z}}\sigma^z_{\vec{r}-\hat{y}}
\label{sta}
\end{align}

This is exactly the Chamon code\cite{Chamon2005-fc} with a tilted geometry\cite{shirley2019universal}. In retrospect, this correspondence is quite natural: the Chamon code has six types of lineon operators, along the three cubic axes and three diagonal directions, each of which creates a distinct lineon-type excitation free to move only along that linear direction. The  mobility of these excitations, together with the Wilson line algebra of the Chamon code, coincide with our Chern-Simons gauge theory at $s=2$. By counting the number of independent stabilizers in Eq.~\ref{sta}, one can also see directly that the ground state degeneracy of the tilted Chamon code is $2^{4L-3}$ on an $L\times L \times L$ lattice with periodic boundary conditions, exactly as predicted for our Chern-Simons theory.

\begin{figure}[h]
  \centering
      \includegraphics[width=0.5\textwidth]{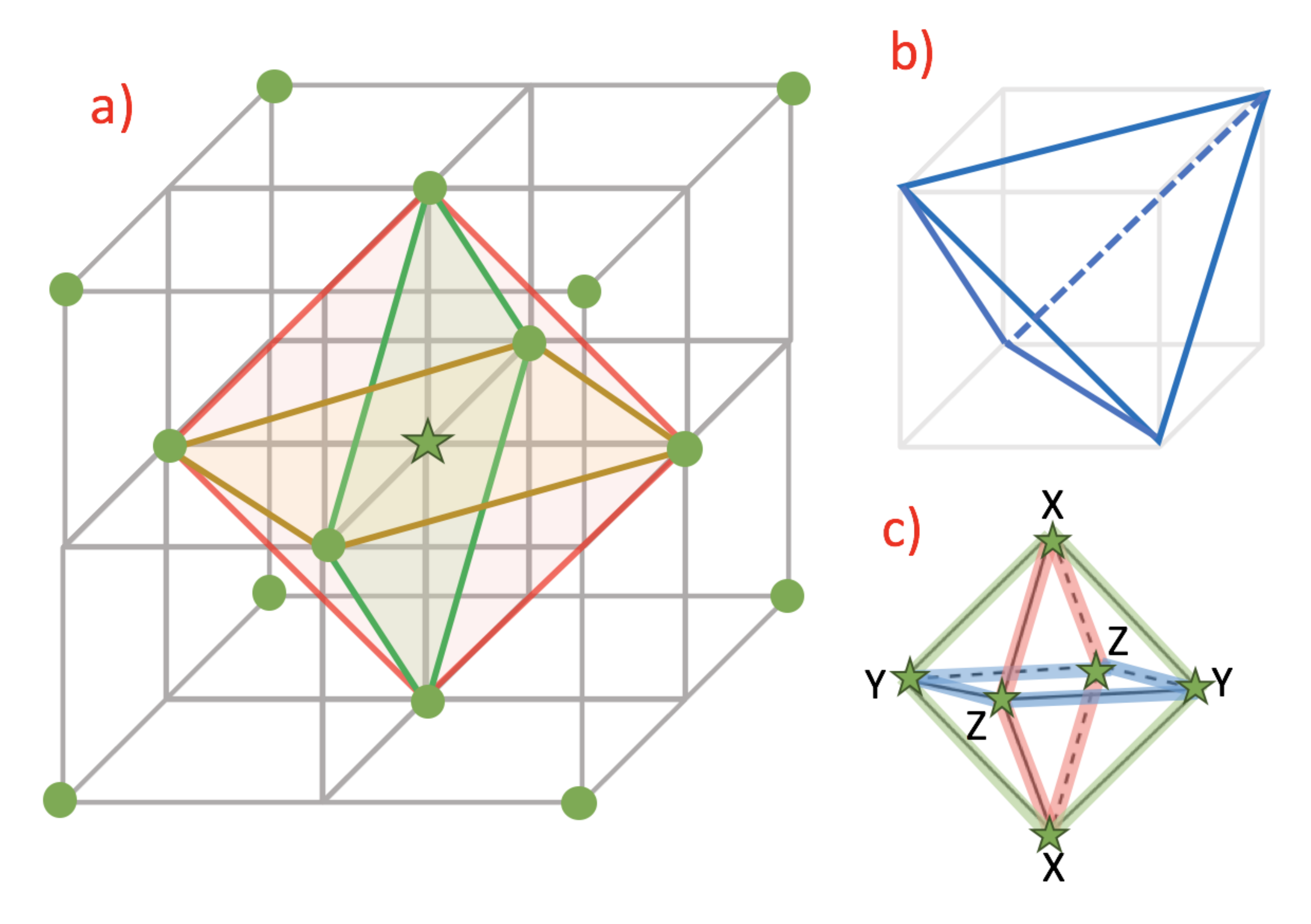}
  \caption{Geometry of the $\alpha=2,\beta=1$ theory corresponding to the Chamon code on the FCC lattice.  a) The gauge fields $A_1,A_2$ lives on the dual lattice (star) at the centers of octahedral plaquettes.  The matter fields live at sites (dot) of the direct lattice. $e^{i A_1}$ is coupled to the four charges at the corners of the red plaquette; $e^{i A_2}$ is coupled to the four charges at the corners of the green plaquette.
b) The cage-net configuration.
  c) The magnetic field operator for $s=2$ (i.e. for the Chamon code).  Sites shown here represent sites on the dual lattice. } 
  \label{one}
\end{figure} 

The original Chamon code on the FCC lattice, which has full cubic symmetry, can be obtained by quantizing the  generalized Fracton Chern-Simons theory  described in Sec.~\ref{c3section}, with $s=2$.  Specifically, by taking differential operators of the form (\ref{Eq:Dgens}),  with  $\alpha=2, \beta=1$, we obtain
\begin{align}
&D_1= (\partial_y+\partial_z)(\partial_y-\partial_z)  = (\partial_y^2 - \partial_z^2)\nonumber\\
& D_2=(\partial_z+\partial_x)(\partial_z-\partial_x) = (\partial_z^2 - \partial_x^2)\ \ .
\end{align}
To get a lattice model in agreement with Chamon's code, we define the discretized differential operators as
\begin{eqnarray}
D_1 = (d_y \hat{d}_y - d_z \hat{d}_z)\\
D_2 =  (d_z \hat{d}_z - d_x \hat{d}_x)
\end{eqnarray}
where $d_i$ and $\hat{d}_i$ are forward and backwards difference operators on the cubic lattice (\ref{Eq:ForwardDiff}).  
As Chamon's code is defined on the FCC lattice, 
we place gauge fields $A_1(\vec{r},t)$ and $A_2(\vec{r},t)$ on the $a$ sublattice of the simple cubic lattice, while the matter fields (and $\alpha(\vec{r},t)$) live on the $b$ sublattice.  
The $a$ sublattice on which $A_1$ and $A_2$ live therefore forms an FCC lattice.
Consequently, the $E$ and $B$ fields of this lattice model live on the $a$ and $b$ sublattice, respectively.
The resulting gauge theory has charge conservation on (111),(11-1),(-111),(1-11) planes. The gauge-invariant cage-nets form symmetric tetraheda, whose edges lie in the $x \pm y,y \pm z,x \pm z$ directions, as shown in Fig.~\ref{one}. The corresponding lattice gauge theory can be obtained by gauging a plaquette ring exchange model with ring exchange terms on the three plaquettes of the Octahedron shown in Fig.~\ref{one}.  These ring-exchange processes conserve charge on (111),(11-1),(-111),(1-11) planes. 
After gauging the resulting U(1) subsystem symmetries following the method described in Section~\ref{dis}, we obtain two gauge fields $A_1,A_2$ at the center of each Octahedron, with gauge transformations,
\begin{align}
&A_1 \rightarrow A_1+ (d_y \hat{d}_y - d_z \hat{d}_z) \alpha \nonumber\\
& A_2 \rightarrow A_2+(d_z \hat{d}_z - d_x \hat{d}_x)\alpha
\end{align}
The resultant gauge theory contains lineon excitations extended along the $\hat{i}\pm \hat{j}$ directions, whose end-points carry dipoles oriented in $\hat{i}\mp \hat{j}$.  With this definition of $D_1, D_2$, 
we obtain a lattice Hamiltonian: 
\begin{align}
&H=-\sum_{\vec{r} }\cos(B(\vec{r}, t))\nonumber\\
&=-\sum_{\vec{r} \in b} \sigma^x_{\vec{r}+\hat{x}}\sigma^x_{\vec{r}-\hat{x}}\sigma^y_{\vec{r}+\hat{y}}\sigma^y_{\vec{r}-\hat{y}}\sigma^z_{\vec{r}+\hat{z}}\sigma^z_{\vec{r}-\hat{z}}
\label{sta2}
\end{align}
which as above, can be viewed as resulting from introducing massive non-dynamical matter sources, and imposing the Chern-Simons constraint $B (\vec{r},t) = \rho ( \vec{r},t)$.
Here $\vec{r} \in b$ sums over $b$ sublattice sites denoted by the dots in Fig.~\ref{one}.
This exactly reproduces the Hamiltonian and 1-dimensional excitations of the Chamon code on FCC lattice formed by the $a$ sublattice of our simple cubic lattice.

One can also define lattice Chern-Simons theories with $s>2$ in this geometry, following the procedure outlined above.  Note, however, that despite the seeming cubic symmetry of this model, for $s>2$ our lattice Chern-Simons theories do not have cubic symmetry.   This is because for $s>2$ the operator $B$ -- and therefore the Chern-Simons action --  is odd under $C_4^i$ rotations. Interestingly, however, in the absence of matter we have $B=0$, and the resulting ground states of the lattice model are invariant under full cubic symmetry\footnote{for $s=2$, the $\pm \pi$ fluxes are equivalent, so in this case the full theory has cubic symmetry.}.

\section{Gapless higher-rank Chern-Simons theories with three gauge fields in 3 dimensions} \label{GaplessCS}

The theories described thus far are tensor gauge theories in the sense that the gauge transformations are quadratic in derivatives; however, they do not correspond to any higher-rank gauge theories discussed in the literature so far.  This is because, if our charge is a scalar, then our Chern-Simons theory is gapped only if we have at most 2 gauge fields.  In three dimensional symmetric tensor gauge theories, the natural number of gauge fields is either 3 (if $A_{ij}$ is an off-diagonal symmetric tensor) or 6 (for a general symmetric tensor).
To make contact with these theories, here we will briefly describe the fate of Chern-Simons theory of the symmetric, off-diagonal tensor gauge theory.  The main interesting feature of this theory is that, unlike the pure Maxwell theory\cite{xu2008resonating}, Maxwell-Chern-Simons theory of off-diagonal symmetric tensor gauge fields in 3 dimensions is deconfined.

We consider a symmetric off-diagonal tensor gauge theory with the gauge fields $A_{xy}, A_{xz}$, and $A_{yz}$.  This off-diagonal tensor structure is not invariant under continuous rotations in 3 dimensions, but rather only under the symmetries of a cubic lattice.  The gauge transformations of this theory are\cite{pretko2017generalized,you2018symmetric,ma2018fracton,bulmash2018higgs} 
\be
A_{ij} \rightarrow A_{ij} + \partial_i \partial_j \alpha \ , \ \ A_0 \rightarrow A_0 + \partial_t \alpha
\ee
Note that our gauge parameter $\alpha$ is a scalar, indicating that this is a scalar charge theory, in the language of Ref. \cite{pretko2017generalized,you2018symmetric,ma2018fracton,bulmash2018higgs}.
The gauge invariant electric and magnetic fields are
\begin{align}
E_{ij}=\partial_t A_{ij} - \partial_i \partial_j A_0  \n
B_x = \partial_y A_{xz} - \partial_z A_{xy} \n
B_y = \partial_z A_{xy} - \partial_x A_{yz} \n
B_z = \partial_x A_{yz} - \partial_y A_{xz} 
\end{align}
The magnetic fields satisfy $\sum_i B_i =0$, such that there are only two independent field components.  The gauge invariant ribbon operators have the form
\be \label{GaplessRibbon}
\oint_{C(x,y) }  \sum A_{i z} {\bf d l}_i
\ee
for $C(x,y)$ any closed curve in the $x,y$ plane, and similarly in the other directions.  Note that unlike the theories discussed in previous sections of this paper, coupling this theory to matter leads to dipolar excitations (planeons) that are mobile in 2-dimensional planes.

In this theory, it is not possible to write a Chern-Simons action imposing the constraint $B = \rho$, since our charge $\rho$ is a scalar, but the magnetic field $B$ is not.  As $\sum_i B_i =0$, the lowest-order constraint that does not violate 3-fold rotational symmetry about the $(1,1,1)$ direction is therefore
\be \label{Eq:CsConst2}
\frac{s}{2 \pi} \partial_i B_i = \rho 
\ee
To enforce this constraint, we choose the Chern-Simons action to be: 
\begin{align} \label{LCS3}
\mac{L}_{\text{CS}} =& \frac{s}{4 \pi } \epsilon^{ijk} (A_{jk}  \partial_t A_{ij} + 2 \partial_j A_0  \partial_i A_{jk } ) \n
&
  - A_0 \rho - A_{ij} J_{ij} 
\end{align}

Because constraint (\ref{Eq:CsConst2}) is not sufficient to fully fix the magnetic field, the pure Chern-Simons theory is unstable, and contains an extensive number of ground states in any geometry.  Instead, we consider a Maxwell-Chern-Simons theory, of the form
\be \label{LMaxCS}
\mac{L} = -\frac{1}{2 g^2} \left( \sum_{ij} E_{ij}^2 + \sum_i B_i^2 \right) + \mac{L}_{CS}
\ee
In the absence of sources, the equations of motion are:
\begin{align} \label{Eqs4}
\sum_{ij} \partial_i \partial_j E_{ij}+ \frac{s}{2 \pi} \sum_i \partial_i B_i  = 0 \n
\partial_t E_{xy} + \frac{s}{2 \pi}(E_{yz} - E_{xz} )  = \partial_z (B_x - B_y) \n
\partial_t E_{yz} + \frac{s}{2 \pi} (E_{xz} - E_{xy} )= \partial_x (B_y - B_z) \n
\partial_t E_{xz} + \frac{s}{2 \pi}(E_{xy} - E_{yz} )= \partial_y (B_z- B_x) \n
\end{align}
In addition,  the analogue of the homogeneous Maxwell equations are:
\begin{align} \label{Eqs3}
\sum_i B_i =0  \n
\partial_z E_{xy} - \partial_y E_{xz} = - \partial_t B_x \n
\partial_y E_{xz} - \partial_x E_{yz} = - \partial_t B_z \n
\partial_x E_{yz} - \partial_z E_{xy} = - \partial_t B_y
\end{align}

One can solve Eq.s (\ref{Eqs3},\ref{Eqs4}) to reveal two modes, with frequencies
\be
\omega^2 _\pm= \vec{k}^2 + \frac{3}{2} s^2 \pm \sqrt{ \Delta^4_{\vec{k}} + s^2 ( k_x + k_y + k_z)^2 + \frac{9}{4} s^4 }
\ee
$\omega_- $ is gapless as $\vec{k} \rightarrow 0$, while $\omega_+$ has a gap proportional to the Chern-Simons coupling $s$.  Thus in the infrared our action (\ref{LMaxCS}) describes a symmetric off-diagonal tensor gauge theory with a single propagating gapless mode.  

To better understand this gapless fixed point, it is convenient to add off-diagonal couplings in the electric fields; the symmetric combination of these violates no lattice symmetries and is thus allowed.  This allows us to consider the action 
\ba \label{Lmin}
\mac{L} =&\frac{s}{4 \pi} \left [  \sqrt{3} \left(  - A_u \partial_t A_v + A_v \partial_t A_u \right )  - 2 A_0 \partial_i B_i   \right ] \n
&+ \frac{1}{2 g^2} \left( E_{\ell}^2 + \sum_i B_i^2 \right )
\ea
where 
\be
E_{\ell} = \sum_{ij} E_{ij}
\ee
We note that this particular choice of Lagrangian has the peculiarity that the massive branch of solutions to Maxwell-Chern-Simons theory are entirely absent; thus we do not need to project out any high-energy modes in order to study the long-wavelength theory.

\subsection{Confinement vs. Chern-Simons terms}

It is known \cite{xu2008resonating} that in the absence of the Chern-Simons term, the Lagrangian (\ref{LMaxCS}) leads to a confining theory for all values of the gauge coupling $g$.  We now show that the Chern-Simons term prevents confinement, by a mechanism similar to that identified by Ref.\cite{fradkin1991chern} in $2+1$ dimensional Chern-Simons theories.  

First, let us review the nature of confinement in the Maxwell theory.  Ref.~\cite{xu2008resonating} showed that in the absence of a Chern-Simons term, if the U(1) gauge field is compact then $2 \pi$ flux defects will proliferate, confining the theory.  These defects correspond to introducing a $2 \pi$ branch plane in the gauge parameter $\alpha$, that emenates from the origin along, for example, the $\hat{x}, \hat{z}$ axes.  We define the branch plane by a singularity in the derivative of $\alpha$, as follows:
\be
 \partial_y \alpha = 2 \pi \delta(y) \theta(z) \theta(x)  
\ee
From this, we see that $\partial_x \partial_y \alpha = 2 \pi \delta (y) \delta(x) \theta(z)$ is well-defined away from $z=0$, and similarly that  $\partial_z \partial_y \alpha = 2 \pi \delta (y) \delta(z) \theta(x)$ is well-defined away from $x=0$.  $\partial_x \partial_z \alpha$ is well-defined (and vanishes) away from the origin.

To see that this branch cut introduces magnetic flux at the origin, consider a region $R$ about the origin bounded a curve $C$ in the $x-y$ plane, and stretching from $-l/2$ to $l/2$ in the $z$ direction.  Assuming that our gauge fields are pure gauge, along such a ribbon, we have
\ba
\int_{R} B_z &=& \oint_C (\int_{-l/2}^{l/2}  A_{iz} dz) d x_i \n
&=& \oint_C  \left( \int_{-l/2}^{l/2}  \partial_z  \partial_i \alpha \right ) d x_i \n
&=&  2 \pi \oint_C \delta(y) \theta(x) d x_i= 2 \pi
\ea
where the last equality holds because the ribbon crosses the $x$ axis at a single point.  For a ribbon that does not enclose the origin, this quantity is 0.  Thus inserting a branch sheet of this type in $\alpha$ can be viewed as a large gauge transformation, which changes the $z$ and $x$ components of the magnetic flux by $\pm 2 \pi$ at the origin.  These are precisely the topological defects that proliferate to drive confinement \cite{xu2008resonating}.  

Next, we show that such defects cannot proliferate in the presence of a Chern-Simons term.  To see this, we first note that the Chern-Simons term is gauge invariant in the bulk only when the homogeneous Maxwell equations (\ref{Eqs3}) are satisfied.  Specifically, under a gauge transformation by $\alpha$ the bulk Chern-Simons action changes according to
\be
\delta \mac{L}_{\text{CS}} = \frac{s}{2 \pi } \alpha \partial_i (\epsilon^{ijk}   \partial_j  E_{ki}   + \partial_t  B^i ) 
\ee

However, large gauge transformations like the one described above stem from processes in which the homogeneous Maxwell equations are violated.  
To see this, we integrate the homogeneous Maxwell equations (\ref{Eqs3}) over some spatial region $R$:
\be
\dot{\Phi}_i(R) = \int_0^l d x_i \int_S \sum_{j,k} \epsilon^{ijk} \partial_j E_{k i}  =  \int_0^l d x_i \oint_{\partial S} E_{i j} (dl)_j
\ee
Here 
\be
\phi_i (R) = \int_R \phi_i (R) d^3 {\bf r}
\ee
and we have taken the region $R$ to have length $l$ in the direction parallel to $i$, and span a surface $S$ in the transverse directions.  We can see that on a closed manifold, or on an infinite manifold with appropriate boundary conditions, if we take $i=x$ and $S$ to be the entire $y-z$ plane, then the right-hand side must vanish.  Thus processes that change the magnetic flux through any planar region (of width $l$) necessarily fail to satisfy the homogeneous Maxwell's equations, and thus are not gauge invariant in the presence of a Chern-Simons term.  Thus, exactly as in the case of usual Chern-Simons theories in $2+1$ dimensions, integrating over the gauge parameter in the partition function suppresses these processes, and thus prevents confinement.

 An example of a continuum spacetime process that inserts a a flux of $ 2 \pi$ in $B_x$ is the monopole- like solution:
 \begin{align} \label{Hom2}
 &(-E_{xz}, E_{xy}, B_x ) = \\
& \frac{1}{2} \left( \frac{y}{(y^2+z^2 +t^2)^{3/2}},\frac{z}{(y^2+z^2 +t^2)^{3/2}},\frac{t}{(y^2+z^2 +t^2)^{3/2}} \right ) \nonumber 
 \end{align}
 It is easy to check that with this solution, $\int d y d z B_x = 2 \pi \theta(t)$, so that the magnetic flux changes by $2 \pi$.  On the other hand, the homogeneous Maxwell equation (\ref{Eqs3}) requires that the divergence of the vector defined in Eq. (\ref{Hom2}) vanish.  For our solution this is the case everywhere except at $y=z=t=0$, where it is singular; one can check that this singularity has the form $\partial_z E_{xy} - \partial_y E_{xz} + \partial_t B = \delta( (y,z,t) - (0,0,0))$, leading to a gauge-dependent contribution to the Chern-Simons action.  

\subsection{quantized gapless theory}

Finally, let us study the properties of our quantized gapless theory.  
The canonical commutation relations from our Lagrangian (\ref{Lmin}) are:
\ba
\left[ A_{xy},  \frac{s}{2 \pi} (A_{xz} - A_{yz} ) + \frac{1}{\sqrt{3} g^2 } E_{\ell} \right] =  i  \n
\left[ A_{xz}, \frac{s}{2 \pi} (A_{yz} - A_{xy}) + \frac{1}{\sqrt{3} g^2 } E_{\ell} \right] =  i  \n
\left[ A_{yz}, \frac{s}{2 \pi} ( A_{xy} - A_{xz} ) + \frac{1}{\sqrt{3} g^2 } E_{\ell}\right] =  i  
\ea
In addition, we have $[ A_{yz} - A_{xz}, E_{\ell} ]=[ A_{xy} - A_{xz}, E_{\ell} ] =0$.  It follows that 
\be
 \frac{1}{\sqrt{3}  g^2 } [A_{ij}, E_{\ell} ] =\frac{ i}{3}  \ \ .
 \ee
   One can use this to determine the remaining commutation relations between the $A_{ij}$:
\ba
\left[ A_{xy},A_{xz}\right] = \frac{2 \pi}{3 s} i   \n
\left[ A_{xz}, A_{yz} \right] = \frac{ 2\pi}{3 s} i \n
\left[  A_{yz},A_{xy}  \right] = \frac{ 2\pi}{3 s} i  
\ea

 In Appendix \ref{lineop}, we show that  line operators of the form (\ref{GaplessRibbon}) commute with the constraint, and are thus all allowed within the low-energy theory. 
However, in general these line operators do not preserve the magnetic field or $E_{\ell}$.  For example,
\ba
\left [ \int d x  A_{xz} ({\bf r}), B_x ({\bf r'}) \right ] = \n
= i s \int dx \partial_y \delta( {\bf r} - {\bf r'}) 
\ea
which is not, in general, $0$.  Thus we conclude that the line operators in general do not commute with the Hamiltonian, and thus do not keep the system in its ground state.   
To see this explicitly, let $\hat{W} = e^{ i \int A_{ij} } $ represent a unitary Wilson ribbon operator.   Since $e^{i A_{ij}}$ is a raising operator for $E_{\ell}$ (for any $i,j$), the energy difference between a state with and without having acted with the Wilson ribbon is:

\ba
\langle \hat{W}^\dag \Psi |  E_{\ell}^2 | \hat{W} \Psi \rangle -\langle \Psi |  E_{\ell}^2 | \Psi \rangle = 
\langle \Psi | \hat{W}^\dag [ E_{\ell}^2, \hat{W} ] | \Psi \rangle \n
 =  \langle \Psi | \hat{ W}^\dag E_{\ell} \hat{W}  + E_{\ell} | \Psi \rangle =  \langle \Psi |\frac{g^2}{\sqrt{3}}  + 2 E_{\ell} | \Psi \rangle  \nonumber 
\ea
Thus  the Wilson ribbon creates a line of electric field along the path of the Wilson ribbon -- much as occurs in ordinary Maxwell theory.  A similar argument shows that a Wilson line also generates magnetic flux.  
Thus, in the presence of low-energy gapless modes, the Wilson line operators do not map between different quantum states of the same energy, in spite of the fact that they relate different classical ground state configurations.  Further, as Wilson lines are not associated with large gauge transformations in the quantum theory, a priori we do not expect the Chern-Simons coefficient to be quantized in this case as the gapless gauge fluctuation could modify the Chern-Simons coupling.  

In our gapless theory, dipoles (which appear at the end of open ribbon operators) create electric and magnetic fields as they move about, and thus have long-ranged Coulomb-like interactions.  However, they also acquire a statistical interaction from the non-trivial commutators of their ribbons.  This statistic is between dipoles of the same orientation (which are restricted to move in the same 2d plane), though there is also a contact interaction (that is not topological in nature) between crossing lines of dipoles with different orientation.  
\section{Outlook} \label{Sec:Outlook}

We have investigated how a field-theoretically motivated approach to constructing TQFT-like actions for higher rank gauge fields in 3 spatial dimensions leads to a number of insights about the possibilities for fractonic tensor gauge theories.  First, we have outlined one general philosophy for writing down such terms, consisting of identifying gauge-invariant (in the bulk) actions that impose a constraint binding charge to the higher-rank gauge flux, and discussed its interplay with symmetry.  We have described both Chern-Simons -like and BF-like versions of this construction, though our main focus has been on the former.  

Second, we have presented a detailed analysis of both the classical and (lattice-regularized) quantum versions of one particular gapped fractonic Chern-Simons theory.  Notably, by analyzing the gauge transformations of Wilson-like oeprators, we have seen how even in the absence of a Gauss' law constraint, the structure of gauge transformations restricts charges' mobility, giving an alternate perspective on the paradigm developed by Ref. \cite{bulmash2018generalized}.  We have also identified several ways in which this theory  is qualitatively distinct from fracton-inspired field theories in the existing literature, including the presence time-reversal-symmetry-breaking galpess boundary modes in the classical theory, and self-statistics for charged particles upon quantization.  Finally, we have established a strong correspondence between our theory and fracton order, both in terms of its physical properties (such as restricted quasiparticle mobility and the scaling of the ground state degeneracy with system size), and by establishing a correspondence to a lattice Hamiltonian that can be viewed as a generalized Chamon code.  To the best of our knowledge, this $Z_s$ generalization of the Chamon code has not previously appeared in the literature.

Third, we have briefly described a scenario that has not previously been considered in the context of higher-rank gauge theories, of a gapless higher-rank Maxwell Chern-Simons theory, with both spontaneously broken time reversal symmetry and long-ranged Coulomb-like interactions.  This scenario is interesting in part because it demonstrates how a higher-rank Chern-Simons term can prevent confinement.  

Our work raises several interesting questions.  First, the observation that our fractonic Chern-Simons theory is gauge invariant only up to a boundary term is surprising in light of the fact that our quantized lattice gauge theory can be mapped exactly onto a commuting projector Hamiltonian.  In particular, this suggests that in spite of the chiral nature of the resulting pattern of dipolar current flow on the boundary, our fractonic Chern-Simons theory may not have {\it ungappable} chiral boundary states, as are present in 2+1 D quantum Hall systems.  We defer a better understanding of how our lattice model preserves gauge invariance at the boundary for future work.

Second, here we have used a lattice regularization in order to quantize our theory.  This avoids several issues that arise in the continuum with a compact U(1) theory.  Further, a lattice regularization naturally captures the geometrical aspects of fracton theories, such as the dependence of the ground state degeneracy on system size.  Nevertheless, it would be interesting to study other possible regularizations, and understand whether -- or in what sense -- a truly continuum version of these quantum field theories exists.

Third, it is clear that the construction described here admits several generalizations.   First, the higher-rank BF theories described in Section \ref{Sec:GeneralCS} can be constructed for theories whose gauge transformations are not purely second order in derivatives, but more general polynomials in the momenta.  This framework can potentially be used to explore field theories with fractonic current conservation laws that result from general subsystem symmetries\cite{gromov2018towards}, including type-II Fracton theories\cite{bulmash2018generalized,Haah2011-ny}.
Second, our construction can be extended to vector-charge theories, or (in the case of BF-like actions) to hybrid theories with both vector and scalar charges.  This allows one to contemplate actions that impose a much broader class of constraints.  
Such generalizations are clearly necessary to capture many known fracton orders, and replicate existing field theories such as  Ref. \cite{slagle2017fracton}'s description of the X-cube model.

\section{Acknowledgements}
We are grateful to Kevin Slagle, Cenke Xu, Xie Chen, Andrey Gromov and Mike Hermele for helpful discussions. YY is supported by PCTS Fellowship at Princeton University. FJB is grateful for the financial support of nsf-dmr 1352271 and the Sloan Foundation FG-2015-65927. YY and FJB performed part of this work at the Aspen Center for Physics, which is supported by National Science Foundation grant PHY-1607611.

\appendix

\section{Allowed directions of line operators} \label{LineApp}

To see that the line operators must run along certain directions, we consider an arbitrary linear combination of the two gauge fields
\be
A_l = a A_u + b A_v 
\ee
Under gauge transformations,
\be
A_l \rightarrow A_l  +  (a+b) \partial_x \partial_y \alpha -  (a-b) \partial_x \partial_z \alpha - 2 b   \partial_y \partial_z \alpha
\ee
In order for a line operator of the form $\int A_0 d x_l$ to be gauge invariant, it must be the case that $A_l \rightarrow A_l +( \sum_k a_k  \partial_k )\partial_l \alpha$ -- in other words, the gauge transformation must factor.  However, since the gauge transformation contains no terms quadratic in derivatives along any of the cubic axes, it must have the form
\be
(a+b) \partial_x \partial_y \alpha -  (a-b) \partial_x \partial_z \alpha - 2 b   \partial_y \partial_z \alpha = (c \partial_x + d \partial_y) \partial_z \alpha
\ee
or similarly with the $x,y,$ and $z$ labels permuted on the right-hand side.  It is easy to check that the only solutions to this equation have $d= - c$, for which we obtain the linear combinations of gauge fields $A_u, A_{u^*}$, and $A_{u^{**}}$ given above.  Thus the 6 line operators identified in the text exhaust all gauge-invariant line operators in our model.

\section{Irreducible Representations of $C_3$}\label{irrepApp}

The rotation group $C_3$ in 3 dimensions has four irreducible representations: two scalar representations $\Gamma_{A_1} \ ,  \Gamma_{A_2}$, and two 2-dimensional representations which we will call $\Gamma_{a}  \ ,  \Gamma_{b}$.  These descend from the vector $(L_{\ell} = 1$) and tensor ($L_\ell = 2$) representations of continuous rotations about the $(1,1,1)$ axis, respectively; in this case since angular momentum $L_\ell = 3$ represents a state invariant under $C_3$ rotations, these are analogous to two vector representations.  

For our purposes, it is convenient to describe these irreducible representations using differential operators of the form $\partial_i \partial_j$.  
Specifically, we represent the matrix element in the basis of $\partial_l^2,(\partial_{\hat{u}_{\perp}}^2 +\partial_u^2),\partial_l \partial_u,\partial_l \partial_{\hat{u}_{\perp}},(\partial_u^2 -\partial_{\hat{u}_{\perp}}^2),\partial_u \partial_{\hat{u}_{\perp}}$ which form irreducible representations of the quadratic derivative operator.

In this notation, we have
\begin{align}
&\Gamma_{A_1}=| \partial_l^2 \rangle \langle \partial_l^2  |,~\Gamma_{A_2}=| \partial_u^2 +\partial_{\hat{u}_{\perp}}^2 \rangle \langle \partial_u^2 +\partial_{\hat{u}_{\perp}}^2  |   \nonumber\\
&\Gamma_{a}=\frac{-1}{2}( | \partial_l \partial_u \rangle \langle \partial_l \partial_u  |-| \partial_l \partial_{\hat{u}_{\perp}} \rangle \langle \partial_l \partial_{\hat{u}_{\perp}}|)\nonumber\\
&+ \frac{\sqrt{3}}{2} | \partial_l \partial_{\hat{u}_{\perp}} \rangle \langle \partial_l \partial_u  |- \frac{\sqrt{3}}{2} | \partial_l \partial_u \rangle \langle \partial_l \partial_{\hat{u}_{\perp}}|  \nonumber\\
&\Gamma_{b}=\frac{-1}{2} (| \partial_u^2 -\partial_{\hat{u}_{\perp}}^2 \rangle \langle  \partial_u^2 -\partial_{\hat{u}_{\perp}}^2 |+| 2 \partial_u \partial_{\hat{u}_{\perp}} \rangle \langle  2 \partial_u \partial_{\hat{u}_{\perp}} | )\nonumber\\
& - \frac{\sqrt{3}}{2} | 2 \partial_u \partial_{\hat{u}_{\perp}} \rangle \langle  \partial_u^2 -\partial_{\hat{u}_{\perp}}^2 |+ \frac{\sqrt{3}}{2} | \partial_u^2 -\partial_{\hat{u}_{\perp}}^2 \rangle \langle  2 \partial_u \partial_{\hat{u}_{\perp}}   |
\nonumber\\
\end{align}
From these, we deduce the form (\ref{Eq:Dirreps}) for the operators $D_i^a, D_i^b$ transforming in the 2D irreps $\Gamma_a $ and $\Gamma_b$.

\section{The case $\beta = 0$} \label{StackedApp}

As discussed in the main text, taking $D_i$ to transform purely in the $\Gamma^a$ irreducible representation of $C_3$ yields a theory that is reminiscent of a stack of decoupled 2D  layers.  Here we give a few more details on the nature of the gauge-invariant operators and mobility of sources in this theory.  

The $\Gamma^a$ irreducible representation leads to $D_i$ operators of the form
\be  D_1 = d_{\ell} d_{u}, D_2 = d_{\ell} d_{u_{\perp}}
\ee
In this case the line operator
\be
\oint (A_1 d u + A_2 d u_\perp )
\ee
is gauge invariant for any closed curve  in the $(u, u_\perp)$ plane.  The charge in -- and consequently dipole moment perpendicular to -- each  $(u, u_\perp)$  plane is also conserved.  

In addition, the line operators 
\be
\oint A_1 d \ell \ , \ \ \oint A_2 d \ell
\ee
are also gauge invariant.    Open line operators of this type can be made gauge invariant by binding dipoles along the $u$ and $u_{\perp}$ directions, respectively.  However, these are 1-dimensional line operators, in the sense that they are not free to bend.  To see this, consider 
\be
\delta( \int_{x}^y A_1 d \ell ) = \partial_1 \alpha_{x}^y \ , \ \ \delta( \int_{x}^y A_2 d \ell ) = \partial_2 \alpha_{x}^y
\ee
By contrast, we have
\be
\delta( \int_{x}^y A_1 d u ) =\delta( \int_{x}^y A_2 d u_{\perp} ) = \partial_{\ell} \alpha_{x}^y
\ee
A similar result holds if we replace $u, u_{\perp}, \ell$ with any three distinct directions.  Correspondingly, in 3 dimensions we also find that $\oint B d \ell =0$, so the charge along and dipole moment perpendicular to each $\ell$-line is also conserved.   

If $l_1, l_2$, and $l_3$ are not linearly independent, we can however have trivalent junctions between one line in the $l_1, l_2$ plane and two lines in the $l_3$ direction.  But this is redundant, since in this case $l_3$ lies in the plane spanned by $l_1$ and $l_2$.

Thus this theory resembles a stack of decoupled vecctor gauge theories, but with extra conservation laws pertaining to dipole moments in the $(u, u_\perp)$ plane.  We expect objects with a dipole moment along $\ell$ that are free to move in the $(u, u_\perp)$ planes, and objects with dipole moment along $u$ or $u_{\perp}$ can move only in the $\ell$ direction.  

We finish by noting that the Chern-Simons term described here does not fully gap this theory.  Specifically, in this case we may add a Maxwell term that is of the same order in derivatives as the Chern-Simons term, since the operator $b = \partial_{u_\perp} A_1 - \partial u A_2$ is gauge invariant.  Adding a term of the form $\sum_i E_i^2 + b^2$, where $E_i$ is the electric field defined in Eq. (\ref{Eq:Efield}), leads to a gapless theory due to the presence of a collective mode with $k_{\ell} =0$ that is not affected by the Chern-Simons constraint.

\section{Line operators in the gapless Chern-Simons theory}\label{lineop}

 The constraint obtained by taking the variation of our Lagrangian with respect to $A_0$ is:
\be
 \frac{s}{2 \pi } \partial_i B_i + \frac{1}{\sqrt{3} g^2} \sum_{ij} \partial_i \partial_j E_{\ell} = \rho
\ee

In order to calculate the commutators of our line operators with the constraint, it is useful to switch to commutators between fields in momentum space. We have
\ba
\left[ A_{xy}( {\bf q}), A_{xz}( {\bf q'}) \right ]&=& i k \delta( {\bf q}+ {\bf q'}) \n
\left[ A_{xy}( {\bf q}), A_{yz}( {\bf q'}) \right ] &=& -i k \delta( {\bf q}+ {\bf q'}) \n
\ea
where here 
\be
k = - i \left[ A_{xy},A_{xz}\right] = 2 \pi/ (3 s) \ \ .
\ee
Then 
\ba
\left[ A_{xy}( {\bf q}), B_{x}( {\bf q'}) \right ]&=& i k q'_y\delta( {\bf q}+ {\bf q'}) \n
\left[ A_{xy}( {\bf q}), B_{y}( {\bf q'}) \right ]&=& i k q'_x \delta( {\bf q}+ {\bf q'}) \n
\left[ A_{xy}( {\bf q}), B_{z}( {\bf q'}) \right ]&=& -i k ( q'_y+ q'_x) \delta( {\bf q}+ {\bf q'}) \n
\ea
Similarly,
\ba
\left[ A_{xz}( {\bf q}), B_{x}( {\bf q'}) \right ]&=& i k q'_z\delta( {\bf q}+ {\bf q'}) \n
\left[ A_{xz}( {\bf q}), B_{y}( {\bf q'}) \right ]&=&- i k (q'_z +q'_x) \delta( {\bf q}+ {\bf q'}) \n
\left[ A_{xz}( {\bf q}), B_{z}( {\bf q'}) \right ]&=& i k  q'_x \delta( {\bf q}+ {\bf q'}) \n
\ea
and
\ba
\left[ A_{yz}( {\bf q}), B_{x}( {\bf q'}) \right ]&=&- i k ( q'_y+ q'_z) \delta( {\bf q}+ {\bf q'}) \n
\left[ A_{yz}( {\bf q}), B_{y}( {\bf q'}) \right ]&=& i k q'_z \delta( {\bf q}+ {\bf q'}) \n
\left[ A_{yz}( {\bf q}), B_{z}( {\bf q'}) \right ]&=& i kq'_y \delta( {\bf q}+ {\bf q'}) \n
\ea

From these, we obtain that 
\be
 \frac{s}{2 \pi } [A_{xy}({\bf q'}), q_i B_i ( {\bf q})  ] = \frac{i}{3} ( 2 q_x q_y - q_xq_z - q_y q_z ) \delta( {\bf q} + {\bf q'})
\ee
where we have used the fact that $k s/ 2 \pi = 1/3$.  
Similarly,
\be
 \frac{1}{\sqrt{3} g^2} [A_{xy}({\bf q'}), \sum_{ij} q_i q_j  E_{\ell} ( {\bf q})  ] = \frac{i}{3}   (  q_x q_y +  q_x q_z + q_y q_z  ) \delta( {\bf q} + {\bf q'})
\ee
Combining these, we find that
\be
[ A_{xy}({\bf q'}),   \frac{s}{2 \pi } \partial_i B_i + \frac{1}{\sqrt{3} g^2} \sum_{ij} \partial_i \partial_j E_{\ell}  ] =
\frac{ i}{3} \left(  3  q_x q_y   \right  )  \delta( {\bf q} + {\bf q'})
\ee
Thus $A_{xy}$ per se does not commute with the constraint.  However, the line operators $\int A_{xy} dx, \int A_{xy} dy$ do.  
Letting $C$ denote our constraint, we have
\ba
\left [ \int d x  A_{xy} ({\bf r}), C ({\bf r'}) \right] = \n
 \int dx \int d^3 {\bf q} \int d^3 {\bf q'} [ A_{xy}({\bf q'}), C ({\bf q}) ] e^{i ( {\bf q'} \cdot {\bf r} + {\bf q} \cdot {\bf r'}) } \n
=i    \int dx \int d^3 {\bf q}  e^{i  {\bf q} \cdot ({\bf r'} - {\bf r}) } q_x q_y \n
= i  \int dx  \partial_{y'}  \partial_{x'}  \delta( {\bf r} - {\bf r'}) 
\ea
If we allow ourselves to interchange the derivatives and the integration, this clearly vanishes, since integral is $1$ no matter what $x'$ is.  Though this sounds questionable, I believe that the derivative of the $\delta$ function is defined by integrating by parts under the integral, which would give the same result.  (In contrast, the derivative of a $\delta$ function that is not under an integral cannot be defined in this way).  
Thus, we conclude that the line operators do commute with the constraint.  In general, this result will hold for any line integral of $A_{xy}$ along a curve in the $xy$- plane.

\bibliography{biblio.bib}

\begin{thebibliography}{72}%
\makeatletter
\providecommand \@ifxundefined [1]{%
 \@ifx{#1\undefined}
}%
\providecommand \@ifnum [1]{%
 \ifnum #1\expandafter \@firstoftwo
 \else \expandafter \@secondoftwo
 \fi
}%
\providecommand \@ifx [1]{%
 \ifx #1\expandafter \@firstoftwo
 \else \expandafter \@secondoftwo
 \fi
}%
\providecommand \natexlab [1]{#1}%
\providecommand \enquote  [1]{``#1''}%
\providecommand \bibnamefont  [1]{#1}%
\providecommand \bibfnamefont [1]{#1}%
\providecommand \citenamefont [1]{#1}%
\providecommand \href@noop [0]{\@secondoftwo}%
\providecommand \href [0]{\begingroup \@sanitize@url \@href}%
\providecommand \@href[1]{\@@startlink{#1}\@@href}%
\providecommand \@@href[1]{\endgroup#1\@@endlink}%
\providecommand \@sanitize@url [0]{\catcode `\\12\catcode `\$12\catcode
  `\&12\catcode `\#12\catcode `\^12\catcode `\_12\catcode `\%12\relax}%
\providecommand \@@startlink[1]{}%
\providecommand \@@endlink[0]{}%
\providecommand \url  [0]{\begingroup\@sanitize@url \@url }%
\providecommand \@url [1]{\endgroup\@href {#1}{\urlprefix }}%
\providecommand \urlprefix  [0]{URL }%
\providecommand \Eprint [0]{\href }%
\providecommand \doibase [0]{http://dx.doi.org/}%
\providecommand \selectlanguage [0]{\@gobble}%
\providecommand \bibinfo  [0]{\@secondoftwo}%
\providecommand \bibfield  [0]{\@secondoftwo}%
\providecommand \translation [1]{[#1]}%
\providecommand \BibitemOpen [0]{}%
\providecommand \bibitemStop [0]{}%
\providecommand \bibitemNoStop [0]{.\EOS\space}%
\providecommand \EOS [0]{\spacefactor3000\relax}%
\providecommand \BibitemShut  [1]{\csname bibitem#1\endcsname}%
\let\auto@bib@innerbib\@empty
\bibitem [{\citenamefont {Wen}(1990)}]{wen1990topological}%
  \BibitemOpen
  \bibfield  {author} {\bibinfo {author} {\bibfnamefont {X.-G.}\ \bibnamefont
  {Wen}},\ }\href@noop {} {\bibfield  {journal} {\bibinfo  {journal} {Int. J.
  Mod. Phys. B}\ }\textbf {\bibinfo {volume} {4}},\ \bibinfo {pages} {239}
  (\bibinfo {year} {1990})}\BibitemShut {NoStop}%
\bibitem [{\citenamefont {Wen}(2003)}]{wen2003quantum}%
  \BibitemOpen
  \bibfield  {author} {\bibinfo {author} {\bibfnamefont {X.-G.}\ \bibnamefont
  {Wen}},\ }\href@noop {} {\bibfield  {journal} {\bibinfo  {journal} {Physical
  review letters}\ }\textbf {\bibinfo {volume} {90}},\ \bibinfo {pages}
  {016803} (\bibinfo {year} {2003})}\BibitemShut {NoStop}%
\bibitem [{\citenamefont {Willett}\ \emph {et~al.}(1987)\citenamefont
  {Willett}, \citenamefont {Eisenstein}, \citenamefont {St{\"o}rmer},
  \citenamefont {Tsui}, \citenamefont {Gossard},\ and\ \citenamefont
  {English}}]{willett1987observation}%
  \BibitemOpen
  \bibfield  {author} {\bibinfo {author} {\bibfnamefont {R.}~\bibnamefont
  {Willett}}, \bibinfo {author} {\bibfnamefont {J.}~\bibnamefont {Eisenstein}},
  \bibinfo {author} {\bibfnamefont {H.}~\bibnamefont {St{\"o}rmer}}, \bibinfo
  {author} {\bibfnamefont {D.}~\bibnamefont {Tsui}}, \bibinfo {author}
  {\bibfnamefont {A.}~\bibnamefont {Gossard}}, \ and\ \bibinfo {author}
  {\bibfnamefont {J.}~\bibnamefont {English}},\ }\href@noop {} {\bibfield
  {journal} {\bibinfo  {journal} {Phys. Rev. Lett.}\ }\textbf {\bibinfo
  {volume} {59}},\ \bibinfo {pages} {1776} (\bibinfo {year}
  {1987})}\BibitemShut {NoStop}%
\bibitem [{\citenamefont {Dijkgraaf}\ and\ \citenamefont
  {Witten}(1990)}]{dijkgraaf1990topological}%
  \BibitemOpen
  \bibfield  {author} {\bibinfo {author} {\bibfnamefont {R.}~\bibnamefont
  {Dijkgraaf}}\ and\ \bibinfo {author} {\bibfnamefont {E.}~\bibnamefont
  {Witten}},\ }\href@noop {} {\bibfield  {journal} {\bibinfo  {journal} {Comm.
  Math. Phys.}\ }\textbf {\bibinfo {volume} {129}},\ \bibinfo {pages} {393}
  (\bibinfo {year} {1990})}\BibitemShut {NoStop}%
\bibitem [{\citenamefont {Wen}\ and\ \citenamefont
  {Zee}(1992)}]{wen1992classification}%
  \BibitemOpen
  \bibfield  {author} {\bibinfo {author} {\bibfnamefont {X.-G.}\ \bibnamefont
  {Wen}}\ and\ \bibinfo {author} {\bibfnamefont {A.}~\bibnamefont {Zee}},\
  }\href@noop {} {\bibfield  {journal} {\bibinfo  {journal} {Phys. Rev. B}\
  }\textbf {\bibinfo {volume} {46}},\ \bibinfo {pages} {2290} (\bibinfo {year}
  {1992})}\BibitemShut {NoStop}%
\bibitem [{\citenamefont {Bernevig}\ \emph {et~al.}(2006)\citenamefont
  {Bernevig}, \citenamefont {Hughes},\ and\ \citenamefont
  {Zhang}}]{bernevig2006quantum}%
  \BibitemOpen
  \bibfield  {author} {\bibinfo {author} {\bibfnamefont {B.~A.}\ \bibnamefont
  {Bernevig}}, \bibinfo {author} {\bibfnamefont {T.~L.}\ \bibnamefont
  {Hughes}}, \ and\ \bibinfo {author} {\bibfnamefont {S.-C.}\ \bibnamefont
  {Zhang}},\ }\href@noop {} {\bibfield  {journal} {\bibinfo  {journal}
  {Science}\ }\textbf {\bibinfo {volume} {314}},\ \bibinfo {pages} {1757}
  (\bibinfo {year} {2006})}\BibitemShut {NoStop}%
\bibitem [{\citenamefont {Fu}\ \emph {et~al.}(2007)\citenamefont {Fu},
  \citenamefont {Kane},\ and\ \citenamefont {Mele}}]{Fu2007-xo}%
  \BibitemOpen
  \bibfield  {author} {\bibinfo {author} {\bibfnamefont {L.}~\bibnamefont
  {Fu}}, \bibinfo {author} {\bibfnamefont {C.~L.}\ \bibnamefont {Kane}}, \ and\
  \bibinfo {author} {\bibfnamefont {E.~J.}\ \bibnamefont {Mele}},\ }\href@noop
  {} {\bibfield  {journal} {\bibinfo  {journal} {Phys. Rev. Lett.}\ }\textbf
  {\bibinfo {volume} {98}},\ \bibinfo {pages} {106803} (\bibinfo {year}
  {2007})}\BibitemShut {NoStop}%
\bibitem [{\citenamefont {Chen}\ \emph {et~al.}(2011)\citenamefont {Chen},
  \citenamefont {Liu},\ and\ \citenamefont {Wen}}]{chen2011two}%
  \BibitemOpen
  \bibfield  {author} {\bibinfo {author} {\bibfnamefont {X.}~\bibnamefont
  {Chen}}, \bibinfo {author} {\bibfnamefont {Z.-X.}\ \bibnamefont {Liu}}, \
  and\ \bibinfo {author} {\bibfnamefont {X.-G.}\ \bibnamefont {Wen}},\
  }\href@noop {} {\bibfield  {journal} {\bibinfo  {journal} {Physical Review
  B}\ }\textbf {\bibinfo {volume} {84}},\ \bibinfo {pages} {235141} (\bibinfo
  {year} {2011})}\BibitemShut {NoStop}%
\bibitem [{\citenamefont {Chen}\ \emph {et~al.}(2012)\citenamefont {Chen},
  \citenamefont {Gu}, \citenamefont {Liu},\ and\ \citenamefont
  {Wen}}]{chen2012symmetry}%
  \BibitemOpen
  \bibfield  {author} {\bibinfo {author} {\bibfnamefont {X.}~\bibnamefont
  {Chen}}, \bibinfo {author} {\bibfnamefont {Z.-C.}\ \bibnamefont {Gu}},
  \bibinfo {author} {\bibfnamefont {Z.-X.}\ \bibnamefont {Liu}}, \ and\
  \bibinfo {author} {\bibfnamefont {X.-G.}\ \bibnamefont {Wen}},\ }\href@noop
  {} {\bibfield  {journal} {\bibinfo  {journal} {Science}\ }\textbf {\bibinfo
  {volume} {338}},\ \bibinfo {pages} {1604} (\bibinfo {year}
  {2012})}\BibitemShut {NoStop}%
\bibitem [{\citenamefont {Kitaev}(2003)}]{kitaev2003fault}%
  \BibitemOpen
  \bibfield  {author} {\bibinfo {author} {\bibfnamefont {A.~Y.}\ \bibnamefont
  {Kitaev}},\ }\href@noop {} {\bibfield  {journal} {\bibinfo  {journal} {Ann.
  Phys.}\ }\textbf {\bibinfo {volume} {303}},\ \bibinfo {pages} {2} (\bibinfo
  {year} {2003})}\BibitemShut {NoStop}%
\bibitem [{\citenamefont {Haah}(2011)}]{Haah2011-ny}%
  \BibitemOpen
  \bibfield  {author} {\bibinfo {author} {\bibfnamefont {J.}~\bibnamefont
  {Haah}},\ }\href@noop {} {\bibfield  {journal} {\bibinfo  {journal} {Phys.
  Rev. A}\ }\textbf {\bibinfo {volume} {83}},\ \bibinfo {pages} {042330}
  (\bibinfo {year} {2011})}\BibitemShut {NoStop}%
\bibitem [{\citenamefont {Hal\'asz}\ \emph {et~al.}(2017)\citenamefont
  {Hal\'asz}, \citenamefont {Hsieh},\ and\ \citenamefont
  {Balents}}]{Halasz2017-ov}%
  \BibitemOpen
  \bibfield  {author} {\bibinfo {author} {\bibfnamefont {G.~B.}\ \bibnamefont
  {Hal\'asz}}, \bibinfo {author} {\bibfnamefont {T.~H.}\ \bibnamefont {Hsieh}},
  \ and\ \bibinfo {author} {\bibfnamefont {L.}~\bibnamefont {Balents}},\
  }\href@noop {} {\bibfield  {journal} {\bibinfo  {journal} {Phys. Rev. Lett.}\
  }\textbf {\bibinfo {volume} {119}},\ \bibinfo {pages} {257202} (\bibinfo
  {year} {2017})}\BibitemShut {NoStop}%
\bibitem [{\citenamefont {Vijay}\ \emph {et~al.}(2016)\citenamefont {Vijay},
  \citenamefont {Haah},\ and\ \citenamefont {Fu}}]{Vijay2016-dr}%
  \BibitemOpen
  \bibfield  {author} {\bibinfo {author} {\bibfnamefont {S.}~\bibnamefont
  {Vijay}}, \bibinfo {author} {\bibfnamefont {J.}~\bibnamefont {Haah}}, \ and\
  \bibinfo {author} {\bibfnamefont {L.}~\bibnamefont {Fu}},\ }\href@noop {}
  {\bibfield  {journal} {\bibinfo  {journal} {Phys. Rev. B}\ }\textbf {\bibinfo
  {volume} {94}},\ \bibinfo {pages} {235157} (\bibinfo {year}
  {2016})}\BibitemShut {NoStop}%
\bibitem [{\citenamefont {Vijay}\ \emph {et~al.}(2015)\citenamefont {Vijay},
  \citenamefont {Haah},\ and\ \citenamefont {Fu}}]{Vijay2015-jj}%
  \BibitemOpen
  \bibfield  {author} {\bibinfo {author} {\bibfnamefont {S.}~\bibnamefont
  {Vijay}}, \bibinfo {author} {\bibfnamefont {J.}~\bibnamefont {Haah}}, \ and\
  \bibinfo {author} {\bibfnamefont {L.}~\bibnamefont {Fu}},\ }\href@noop {}
  {\bibfield  {journal} {\bibinfo  {journal} {Phys. Rev. B}\ }\textbf {\bibinfo
  {volume} {92}},\ \bibinfo {pages} {235136} (\bibinfo {year}
  {2015})}\BibitemShut {NoStop}%
\bibitem [{\citenamefont {Chamon}(2005)}]{Chamon2005-fc}%
  \BibitemOpen
  \bibfield  {author} {\bibinfo {author} {\bibfnamefont {C.}~\bibnamefont
  {Chamon}},\ }\href@noop {} {\bibfield  {journal} {\bibinfo  {journal} {Phys.
  Rev. Lett.}\ }\textbf {\bibinfo {volume} {94}},\ \bibinfo {pages} {040402}
  (\bibinfo {year} {2005})}\BibitemShut {NoStop}%
\bibitem [{\citenamefont {Hsieh}\ and\ \citenamefont
  {Hal{\'a}sz}(2017{\natexlab{a}})}]{hsieh2017fractons}%
  \BibitemOpen
  \bibfield  {author} {\bibinfo {author} {\bibfnamefont {T.~H.}\ \bibnamefont
  {Hsieh}}\ and\ \bibinfo {author} {\bibfnamefont {G.~B.}\ \bibnamefont
  {Hal{\'a}sz}},\ }\href@noop {} {\bibfield  {journal} {\bibinfo  {journal}
  {Phys. Rev. B}\ }\textbf {\bibinfo {volume} {96}},\ \bibinfo {pages} {165105}
  (\bibinfo {year} {2017}{\natexlab{a}})}\BibitemShut {NoStop}%
\bibitem [{\citenamefont {Slagle}\ and\ \citenamefont
  {Kim}(2017{\natexlab{a}})}]{Slagle2017-ne}%
  \BibitemOpen
  \bibfield  {author} {\bibinfo {author} {\bibfnamefont {K.}~\bibnamefont
  {Slagle}}\ and\ \bibinfo {author} {\bibfnamefont {Y.~B.}\ \bibnamefont
  {Kim}},\ }\href@noop {} {\  (\bibinfo {year} {2017}{\natexlab{a}})},\ \Eprint
  {http://arxiv.org/abs/1704.03870} {arXiv:1704.03870 [cond-mat.str-el]}
  \BibitemShut {NoStop}%
\bibitem [{\citenamefont {Hsieh}\ and\ \citenamefont
  {Hal{\'a}sz}(2017{\natexlab{b}})}]{Hsieh2017-sc}%
  \BibitemOpen
  \bibfield  {author} {\bibinfo {author} {\bibfnamefont {T.~H.}\ \bibnamefont
  {Hsieh}}\ and\ \bibinfo {author} {\bibfnamefont {G.~B.}\ \bibnamefont
  {Hal{\'a}sz}},\ }\href@noop {} {\  (\bibinfo {year} {2017}{\natexlab{b}})},\
  \Eprint {http://arxiv.org/abs/1703.02973} {arXiv:1703.02973
  [cond-mat.str-el]} \BibitemShut {NoStop}%
\bibitem [{\citenamefont {Shirley}\ \emph
  {et~al.}(2018{\natexlab{a}})\citenamefont {Shirley}, \citenamefont {Slagle},\
  and\ \citenamefont {Chen}}]{shirley2018fractional}%
  \BibitemOpen
  \bibfield  {author} {\bibinfo {author} {\bibfnamefont {W.}~\bibnamefont
  {Shirley}}, \bibinfo {author} {\bibfnamefont {K.}~\bibnamefont {Slagle}}, \
  and\ \bibinfo {author} {\bibfnamefont {X.}~\bibnamefont {Chen}},\ }\href@noop
  {} {\bibfield  {journal} {\bibinfo  {journal} {arXiv:1806.08625}\ } (\bibinfo
  {year} {2018}{\natexlab{a}})}\BibitemShut {NoStop}%
\bibitem [{\citenamefont {Yoshida}(2013{\natexlab{a}})}]{yoshida2013exotic}%
  \BibitemOpen
  \bibfield  {author} {\bibinfo {author} {\bibfnamefont {B.}~\bibnamefont
  {Yoshida}},\ }\href@noop {} {\bibfield  {journal} {\bibinfo  {journal} {Phys.
  Rev. B}\ }\textbf {\bibinfo {volume} {88}},\ \bibinfo {pages} {125122}
  (\bibinfo {year} {2013}{\natexlab{a}})}\BibitemShut {NoStop}%
\bibitem [{\citenamefont {Yoshida}(2013{\natexlab{b}})}]{Yoshida2013-of}%
  \BibitemOpen
  \bibfield  {author} {\bibinfo {author} {\bibfnamefont {B.}~\bibnamefont
  {Yoshida}},\ }\href@noop {} {\bibfield  {journal} {\bibinfo  {journal} {Phys.
  Rev. B Condens. Matter}\ }\textbf {\bibinfo {volume} {88}},\ \bibinfo {pages}
  {125122} (\bibinfo {year} {2013}{\natexlab{b}})}\BibitemShut {NoStop}%
\bibitem [{\citenamefont {Ma}\ \emph {et~al.}(2017{\natexlab{a}})\citenamefont
  {Ma}, \citenamefont {Lake}, \citenamefont {Chen},\ and\ \citenamefont
  {Hermele}}]{Ma2017-qq}%
  \BibitemOpen
  \bibfield  {author} {\bibinfo {author} {\bibfnamefont {H.}~\bibnamefont
  {Ma}}, \bibinfo {author} {\bibfnamefont {E.}~\bibnamefont {Lake}}, \bibinfo
  {author} {\bibfnamefont {X.}~\bibnamefont {Chen}}, \ and\ \bibinfo {author}
  {\bibfnamefont {M.}~\bibnamefont {Hermele}},\ }\href@noop {} {\  (\bibinfo
  {year} {2017}{\natexlab{a}})},\ \Eprint {http://arxiv.org/abs/1701.00747}
  {arXiv:1701.00747 [cond-mat.str-el]} \BibitemShut {NoStop}%
\bibitem [{\citenamefont {Vijay}(2017)}]{Vijay2017-ey}%
  \BibitemOpen
  \bibfield  {author} {\bibinfo {author} {\bibfnamefont {S.}~\bibnamefont
  {Vijay}},\ }\href@noop {} {\bibfield  {journal} {\bibinfo  {journal}
  {arXiv:1701.00762}\ } (\bibinfo {year} {2017})}\BibitemShut {NoStop}%
\bibitem [{\citenamefont {Slagle}\ and\ \citenamefont
  {Kim}(2017{\natexlab{b}})}]{Slagle2017-gk}%
  \BibitemOpen
  \bibfield  {author} {\bibinfo {author} {\bibfnamefont {K.}~\bibnamefont
  {Slagle}}\ and\ \bibinfo {author} {\bibfnamefont {Y.~B.}\ \bibnamefont
  {Kim}},\ }\href@noop {} {\bibfield  {journal} {\bibinfo  {journal} {Phys.
  Rev. B Condens. Matter}\ }\textbf {\bibinfo {volume} {96}},\ \bibinfo {pages}
  {195139} (\bibinfo {year} {2017}{\natexlab{b}})}\BibitemShut {NoStop}%
\bibitem [{\citenamefont {Ma}\ \emph {et~al.}(2017{\natexlab{b}})\citenamefont
  {Ma}, \citenamefont {Schmitz}, \citenamefont {Parameswaran}, \citenamefont
  {Hermele},\ and\ \citenamefont {Nandkishore}}]{Ma2017-cb}%
  \BibitemOpen
  \bibfield  {author} {\bibinfo {author} {\bibfnamefont {H.}~\bibnamefont
  {Ma}}, \bibinfo {author} {\bibfnamefont {A.~T.}\ \bibnamefont {Schmitz}},
  \bibinfo {author} {\bibfnamefont {S.~A.}\ \bibnamefont {Parameswaran}},
  \bibinfo {author} {\bibfnamefont {M.}~\bibnamefont {Hermele}}, \ and\
  \bibinfo {author} {\bibfnamefont {R.~M.}\ \bibnamefont {Nandkishore}},\
  }\href@noop {} {\  (\bibinfo {year} {2017}{\natexlab{b}})},\ \Eprint
  {http://arxiv.org/abs/1710.01744} {arXiv:1710.01744 [cond-mat.str-el]}
  \BibitemShut {NoStop}%
\bibitem [{\citenamefont {Slagle}\ and\ \citenamefont
  {Kim}(2017{\natexlab{c}})}]{Slagle2017-la}%
  \BibitemOpen
  \bibfield  {author} {\bibinfo {author} {\bibfnamefont {K.}~\bibnamefont
  {Slagle}}\ and\ \bibinfo {author} {\bibfnamefont {Y.~B.}\ \bibnamefont
  {Kim}},\ }\href@noop {} {\  (\bibinfo {year} {2017}{\natexlab{c}})},\ \Eprint
  {http://arxiv.org/abs/1712.04511} {arXiv:1712.04511 [cond-mat.str-el]}
  \BibitemShut {NoStop}%
\bibitem [{\citenamefont {Shirley}\ \emph {et~al.}(2017)\citenamefont
  {Shirley}, \citenamefont {Slagle}, \citenamefont {Wang},\ and\ \citenamefont
  {Chen}}]{shirley2017fracton}%
  \BibitemOpen
  \bibfield  {author} {\bibinfo {author} {\bibfnamefont {W.}~\bibnamefont
  {Shirley}}, \bibinfo {author} {\bibfnamefont {K.}~\bibnamefont {Slagle}},
  \bibinfo {author} {\bibfnamefont {Z.}~\bibnamefont {Wang}}, \ and\ \bibinfo
  {author} {\bibfnamefont {X.}~\bibnamefont {Chen}},\ }\href@noop {} {\bibfield
   {journal} {\bibinfo  {journal} {arXiv preprint arXiv:1712.05892}\ }
  (\bibinfo {year} {2017})}\BibitemShut {NoStop}%
\bibitem [{\citenamefont {Pretko}\ and\ \citenamefont
  {Radzihovsky}(2018)}]{pretko2017fracton}%
  \BibitemOpen
  \bibfield  {author} {\bibinfo {author} {\bibfnamefont {M.}~\bibnamefont
  {Pretko}}\ and\ \bibinfo {author} {\bibfnamefont {L.}~\bibnamefont
  {Radzihovsky}},\ }\href@noop {} {\bibfield  {journal} {\bibinfo  {journal}
  {Phys. Rev. Lett.}\ }\textbf {\bibinfo {volume} {120}},\ \bibinfo {pages}
  {195301} (\bibinfo {year} {2018})}\BibitemShut {NoStop}%
\bibitem [{\citenamefont {Ma}\ \emph {et~al.}(2018)\citenamefont {Ma},
  \citenamefont {Hermele},\ and\ \citenamefont {Chen}}]{ma2018fracton}%
  \BibitemOpen
  \bibfield  {author} {\bibinfo {author} {\bibfnamefont {H.}~\bibnamefont
  {Ma}}, \bibinfo {author} {\bibfnamefont {M.}~\bibnamefont {Hermele}}, \ and\
  \bibinfo {author} {\bibfnamefont {X.}~\bibnamefont {Chen}},\ }\href@noop {}
  {\bibfield  {journal} {\bibinfo  {journal} {Phys. Rev. B}\ }\textbf {\bibinfo
  {volume} {98}},\ \bibinfo {pages} {035111} (\bibinfo {year}
  {2018})}\BibitemShut {NoStop}%
\bibitem [{\citenamefont {Prem}\ \emph
  {et~al.}(2017{\natexlab{a}})\citenamefont {Prem}, \citenamefont {Pretko},\
  and\ \citenamefont {Nandkishore}}]{prem2017emergent}%
  \BibitemOpen
  \bibfield  {author} {\bibinfo {author} {\bibfnamefont {A.}~\bibnamefont
  {Prem}}, \bibinfo {author} {\bibfnamefont {M.}~\bibnamefont {Pretko}}, \ and\
  \bibinfo {author} {\bibfnamefont {R.}~\bibnamefont {Nandkishore}},\
  }\href@noop {} {\bibfield  {journal} {\bibinfo  {journal} {arXiv preprint
  arXiv:1709.09673}\ } (\bibinfo {year} {2017}{\natexlab{a}})}\BibitemShut
  {NoStop}%
\bibitem [{\citenamefont
  {Pretko}(2017{\natexlab{a}})}]{pretko2017subdimensional}%
  \BibitemOpen
  \bibfield  {author} {\bibinfo {author} {\bibfnamefont {M.}~\bibnamefont
  {Pretko}},\ }\href@noop {} {\bibfield  {journal} {\bibinfo  {journal}
  {Physical Review B}\ }\textbf {\bibinfo {volume} {95}},\ \bibinfo {pages}
  {115139} (\bibinfo {year} {2017}{\natexlab{a}})}\BibitemShut {NoStop}%
\bibitem [{\citenamefont {Bulmash}\ and\ \citenamefont
  {Barkeshli}(2018{\natexlab{a}})}]{bulmash2018higgs}%
  \BibitemOpen
  \bibfield  {author} {\bibinfo {author} {\bibfnamefont {D.}~\bibnamefont
  {Bulmash}}\ and\ \bibinfo {author} {\bibfnamefont {M.}~\bibnamefont
  {Barkeshli}},\ }\href@noop {} {\bibfield  {journal} {\bibinfo  {journal}
  {Phys. Rev. B}\ }\textbf {\bibinfo {volume} {97}},\ \bibinfo {pages} {235112}
  (\bibinfo {year} {2018}{\natexlab{a}})}\BibitemShut {NoStop}%
\bibitem [{\citenamefont {Prem}\ \emph
  {et~al.}(2017{\natexlab{b}})\citenamefont {Prem}, \citenamefont {Haah},\ and\
  \citenamefont {Nandkishore}}]{Prem2017-ql}%
  \BibitemOpen
  \bibfield  {author} {\bibinfo {author} {\bibfnamefont {A.}~\bibnamefont
  {Prem}}, \bibinfo {author} {\bibfnamefont {J.}~\bibnamefont {Haah}}, \ and\
  \bibinfo {author} {\bibfnamefont {R.}~\bibnamefont {Nandkishore}},\
  }\href@noop {} {\bibfield  {journal} {\bibinfo  {journal} {Phys. Rev. B}\
  }\textbf {\bibinfo {volume} {95}},\ \bibinfo {pages} {155133} (\bibinfo
  {year} {2017}{\natexlab{b}})}\BibitemShut {NoStop}%
\bibitem [{\citenamefont {Bulmash}\ and\ \citenamefont
  {Barkeshli}(2018{\natexlab{b}})}]{bulmash2018generalized}%
  \BibitemOpen
  \bibfield  {author} {\bibinfo {author} {\bibfnamefont {D.}~\bibnamefont
  {Bulmash}}\ and\ \bibinfo {author} {\bibfnamefont {M.}~\bibnamefont
  {Barkeshli}},\ }\href@noop {} {\bibfield  {journal} {\bibinfo  {journal}
  {arXiv preprint arXiv:1806.01855}\ } (\bibinfo {year}
  {2018}{\natexlab{b}})}\BibitemShut {NoStop}%
\bibitem [{\citenamefont {You}\ \emph {et~al.}(2018{\natexlab{a}})\citenamefont
  {You}, \citenamefont {Devakul}, \citenamefont {Burnell},\ and\ \citenamefont
  {Sondhi}}]{you2018subsystem}%
  \BibitemOpen
  \bibfield  {author} {\bibinfo {author} {\bibfnamefont {Y.}~\bibnamefont
  {You}}, \bibinfo {author} {\bibfnamefont {T.}~\bibnamefont {Devakul}},
  \bibinfo {author} {\bibfnamefont {F.}~\bibnamefont {Burnell}}, \ and\
  \bibinfo {author} {\bibfnamefont {S.}~\bibnamefont {Sondhi}},\ }\href@noop {}
  {\bibfield  {journal} {\bibinfo  {journal} {Physical Review B}\ }\textbf
  {\bibinfo {volume} {98}},\ \bibinfo {pages} {035112} (\bibinfo {year}
  {2018}{\natexlab{a}})}\BibitemShut {NoStop}%
\bibitem [{\citenamefont {Devakul}\ \emph {et~al.}(2018)\citenamefont
  {Devakul}, \citenamefont {You}, \citenamefont {Burnell},\ and\ \citenamefont
  {Sondhi}}]{devakul2018fractal}%
  \BibitemOpen
  \bibfield  {author} {\bibinfo {author} {\bibfnamefont {T.}~\bibnamefont
  {Devakul}}, \bibinfo {author} {\bibfnamefont {Y.}~\bibnamefont {You}},
  \bibinfo {author} {\bibfnamefont {F.}~\bibnamefont {Burnell}}, \ and\
  \bibinfo {author} {\bibfnamefont {S.}~\bibnamefont {Sondhi}},\ }\href@noop {}
  {\bibfield  {journal} {\bibinfo  {journal} {arXiv preprint arXiv:1805.04097}\
  } (\bibinfo {year} {2018})}\BibitemShut {NoStop}%
\bibitem [{\citenamefont {You}\ \emph {et~al.}(2018{\natexlab{b}})\citenamefont
  {You}, \citenamefont {Devakul}, \citenamefont {Burnell},\ and\ \citenamefont
  {Sondhi}}]{you2018symmetric}%
  \BibitemOpen
  \bibfield  {author} {\bibinfo {author} {\bibfnamefont {Y.}~\bibnamefont
  {You}}, \bibinfo {author} {\bibfnamefont {T.}~\bibnamefont {Devakul}},
  \bibinfo {author} {\bibfnamefont {F.}~\bibnamefont {Burnell}}, \ and\
  \bibinfo {author} {\bibfnamefont {S.}~\bibnamefont {Sondhi}},\ }\href@noop {}
  {\bibfield  {journal} {\bibinfo  {journal} {arXiv preprint arXiv:1805.09800}\
  } (\bibinfo {year} {2018}{\natexlab{b}})}\BibitemShut {NoStop}%
\bibitem [{\citenamefont {Shirley}\ \emph
  {et~al.}(2018{\natexlab{b}})\citenamefont {Shirley}, \citenamefont {Slagle},\
  and\ \citenamefont {Chen}}]{shirley2018foliated}%
  \BibitemOpen
  \bibfield  {author} {\bibinfo {author} {\bibfnamefont {W.}~\bibnamefont
  {Shirley}}, \bibinfo {author} {\bibfnamefont {K.}~\bibnamefont {Slagle}}, \
  and\ \bibinfo {author} {\bibfnamefont {X.}~\bibnamefont {Chen}},\ }\href@noop
  {} {\bibfield  {journal} {\bibinfo  {journal} {arXiv:1806.08679}\ } (\bibinfo
  {year} {2018}{\natexlab{b}})}\BibitemShut {NoStop}%
\bibitem [{\citenamefont {Song}\ \emph {et~al.}(2018)\citenamefont {Song},
  \citenamefont {Prem}, \citenamefont {Huang},\ and\ \citenamefont
  {Martin-Delgado}}]{song2018twisted}%
  \BibitemOpen
  \bibfield  {author} {\bibinfo {author} {\bibfnamefont {H.}~\bibnamefont
  {Song}}, \bibinfo {author} {\bibfnamefont {A.}~\bibnamefont {Prem}}, \bibinfo
  {author} {\bibfnamefont {S.-J.}\ \bibnamefont {Huang}}, \ and\ \bibinfo
  {author} {\bibfnamefont {M.~A.}\ \bibnamefont {Martin-Delgado}},\ }\href@noop
  {} {\bibfield  {journal} {\bibinfo  {journal} {arXiv preprint
  arXiv:1805.06899}\ } (\bibinfo {year} {2018})}\BibitemShut {NoStop}%
\bibitem [{\citenamefont {Bulmash}\ and\ \citenamefont
  {Iadecola}(2018)}]{bulmash2018braiding}%
  \BibitemOpen
  \bibfield  {author} {\bibinfo {author} {\bibfnamefont {D.}~\bibnamefont
  {Bulmash}}\ and\ \bibinfo {author} {\bibfnamefont {T.}~\bibnamefont
  {Iadecola}},\ }\href@noop {} {\bibfield  {journal} {\bibinfo  {journal}
  {arXiv preprint arXiv:1810.00012}\ } (\bibinfo {year} {2018})}\BibitemShut
  {NoStop}%
\bibitem [{\citenamefont {Prem}\ \emph {et~al.}(2018)\citenamefont {Prem},
  \citenamefont {Vijay}, \citenamefont {Chou}, \citenamefont {Pretko},\ and\
  \citenamefont {Nandkishore}}]{prem2018pinch}%
  \BibitemOpen
  \bibfield  {author} {\bibinfo {author} {\bibfnamefont {A.}~\bibnamefont
  {Prem}}, \bibinfo {author} {\bibfnamefont {S.}~\bibnamefont {Vijay}},
  \bibinfo {author} {\bibfnamefont {Y.-Z.}\ \bibnamefont {Chou}}, \bibinfo
  {author} {\bibfnamefont {M.}~\bibnamefont {Pretko}}, \ and\ \bibinfo {author}
  {\bibfnamefont {R.~M.}\ \bibnamefont {Nandkishore}},\ }\href@noop {}
  {\bibfield  {journal} {\bibinfo  {journal} {arXiv preprint arXiv:1806.04148}\
  } (\bibinfo {year} {2018})}\BibitemShut {NoStop}%
\bibitem [{\citenamefont {Cho}\ \emph {et~al.}(2015)\citenamefont {Cho},
  \citenamefont {Parrikar}, \citenamefont {You}, \citenamefont {Leigh},\ and\
  \citenamefont {Hughes}}]{cho2015condensation}%
  \BibitemOpen
  \bibfield  {author} {\bibinfo {author} {\bibfnamefont {G.~Y.}\ \bibnamefont
  {Cho}}, \bibinfo {author} {\bibfnamefont {O.}~\bibnamefont {Parrikar}},
  \bibinfo {author} {\bibfnamefont {Y.}~\bibnamefont {You}}, \bibinfo {author}
  {\bibfnamefont {R.~G.}\ \bibnamefont {Leigh}}, \ and\ \bibinfo {author}
  {\bibfnamefont {T.~L.}\ \bibnamefont {Hughes}},\ }\href@noop {} {\bibfield
  {journal} {\bibinfo  {journal} {Physical Review B}\ }\textbf {\bibinfo
  {volume} {91}},\ \bibinfo {pages} {035122} (\bibinfo {year}
  {2015})}\BibitemShut {NoStop}%
\bibitem [{\citenamefont {Pretko}\ and\ \citenamefont
  {Radzihovsky}(2017)}]{Pretko2017-ej}%
  \BibitemOpen
  \bibfield  {author} {\bibinfo {author} {\bibfnamefont {M.}~\bibnamefont
  {Pretko}}\ and\ \bibinfo {author} {\bibfnamefont {L.}~\bibnamefont
  {Radzihovsky}},\ }\href@noop {} {\  (\bibinfo {year} {2017})},\ \Eprint
  {http://arxiv.org/abs/1711.11044} {arXiv:1711.11044 [cond-mat.str-el]}
  \BibitemShut {NoStop}%
\bibitem [{\citenamefont {Slagle}\ \emph
  {et~al.}(2018{\natexlab{a}})\citenamefont {Slagle}, \citenamefont {Prem},\
  and\ \citenamefont {Pretko}}]{slagle2018symmetric}%
  \BibitemOpen
  \bibfield  {author} {\bibinfo {author} {\bibfnamefont {K.}~\bibnamefont
  {Slagle}}, \bibinfo {author} {\bibfnamefont {A.}~\bibnamefont {Prem}}, \ and\
  \bibinfo {author} {\bibfnamefont {M.}~\bibnamefont {Pretko}},\ }\href@noop {}
  {\bibfield  {journal} {\bibinfo  {journal} {arXiv preprint arXiv:1807.00827}\
  } (\bibinfo {year} {2018}{\natexlab{a}})}\BibitemShut {NoStop}%
\bibitem [{\citenamefont {Gromov}(2017)}]{gromov2017fractional}%
  \BibitemOpen
  \bibfield  {author} {\bibinfo {author} {\bibfnamefont {A.}~\bibnamefont
  {Gromov}},\ }\href@noop {} {\bibfield  {journal} {\bibinfo  {journal} {arXiv
  preprint arXiv:1712.06600}\ } (\bibinfo {year} {2017})}\BibitemShut {NoStop}%
\bibitem [{\citenamefont {{Pretko}}(2018)}]{2018arXiv180711479P}%
  \BibitemOpen
  \bibfield  {author} {\bibinfo {author} {\bibfnamefont {M.}~\bibnamefont
  {{Pretko}}},\ }\href@noop {} {\bibfield  {journal} {\bibinfo  {journal}
  {ArXiv e-prints}\ } (\bibinfo {year} {2018})},\ \Eprint
  {http://arxiv.org/abs/1807.11479} {arXiv:1807.11479 [cond-mat.str-el]}
  \BibitemShut {NoStop}%
\bibitem [{\citenamefont {Pai}\ and\ \citenamefont
  {Pretko}(2018)}]{pai2018fractonic}%
  \BibitemOpen
  \bibfield  {author} {\bibinfo {author} {\bibfnamefont {S.}~\bibnamefont
  {Pai}}\ and\ \bibinfo {author} {\bibfnamefont {M.}~\bibnamefont {Pretko}},\
  }\href@noop {} {\bibfield  {journal} {\bibinfo  {journal} {arXiv preprint
  arXiv:1804.01536}\ } (\bibinfo {year} {2018})}\BibitemShut {NoStop}%
\bibitem [{\citenamefont {Ma}\ and\ \citenamefont
  {Pretko}(2018)}]{ma2018higher}%
  \BibitemOpen
  \bibfield  {author} {\bibinfo {author} {\bibfnamefont {H.}~\bibnamefont
  {Ma}}\ and\ \bibinfo {author} {\bibfnamefont {M.}~\bibnamefont {Pretko}},\
  }\href@noop {} {\bibfield  {journal} {\bibinfo  {journal} {arXiv preprint
  arXiv:1803.04980}\ } (\bibinfo {year} {2018})}\BibitemShut {NoStop}%
\bibitem [{\citenamefont {Pretko}(2017{\natexlab{b}})}]{pretko2017generalized}%
  \BibitemOpen
  \bibfield  {author} {\bibinfo {author} {\bibfnamefont {M.}~\bibnamefont
  {Pretko}},\ }\href@noop {} {\bibfield  {journal} {\bibinfo  {journal}
  {Physical Review B}\ }\textbf {\bibinfo {volume} {96}},\ \bibinfo {pages}
  {035119} (\bibinfo {year} {2017}{\natexlab{b}})}\BibitemShut {NoStop}%
\bibitem [{\citenamefont {Pretko}(2017{\natexlab{c}})}]{pretko2017finite}%
  \BibitemOpen
  \bibfield  {author} {\bibinfo {author} {\bibfnamefont {M.}~\bibnamefont
  {Pretko}},\ }\href@noop {} {\bibfield  {journal} {\bibinfo  {journal}
  {Physical Review B}\ }\textbf {\bibinfo {volume} {96}},\ \bibinfo {pages}
  {115102} (\bibinfo {year} {2017}{\natexlab{c}})}\BibitemShut {NoStop}%
\bibitem [{\citenamefont {Yan}\ \emph {et~al.}(2019)\citenamefont {Yan},
  \citenamefont {Benton}, \citenamefont {Jaubert},\ and\ \citenamefont
  {Shannon}}]{yan2019rank}%
  \BibitemOpen
  \bibfield  {author} {\bibinfo {author} {\bibfnamefont {H.}~\bibnamefont
  {Yan}}, \bibinfo {author} {\bibfnamefont {O.}~\bibnamefont {Benton}},
  \bibinfo {author} {\bibfnamefont {L.~D.}\ \bibnamefont {Jaubert}}, \ and\
  \bibinfo {author} {\bibfnamefont {N.}~\bibnamefont {Shannon}},\ }\href@noop
  {} {\bibfield  {journal} {\bibinfo  {journal} {arXiv preprint
  arXiv:1902.10934}\ } (\bibinfo {year} {2019})}\BibitemShut {NoStop}%
\bibitem [{\citenamefont {You}\ and\ \citenamefont {von
  Oppen}(2018)}]{you2018majorana}%
  \BibitemOpen
  \bibfield  {author} {\bibinfo {author} {\bibfnamefont {Y.}~\bibnamefont
  {You}}\ and\ \bibinfo {author} {\bibfnamefont {F.}~\bibnamefont {von
  Oppen}},\ }\href@noop {} {\bibfield  {journal} {\bibinfo  {journal} {arXiv
  preprint arXiv:1812.06091}\ } (\bibinfo {year} {2018})}\BibitemShut {NoStop}%
\bibitem [{\citenamefont {Gromov}(2018)}]{gromov2018towards}%
  \BibitemOpen
  \bibfield  {author} {\bibinfo {author} {\bibfnamefont {A.}~\bibnamefont
  {Gromov}},\ }\href@noop {} {\bibfield  {journal} {\bibinfo  {journal} {arXiv
  preprint arXiv:1812.05104}\ } (\bibinfo {year} {2018})}\BibitemShut {NoStop}%
\bibitem [{\citenamefont {Hansson}\ \emph {et~al.}(2004)\citenamefont
  {Hansson}, \citenamefont {Oganesyan},\ and\ \citenamefont
  {Sondhi}}]{hansson2004superconductors}%
  \BibitemOpen
  \bibfield  {author} {\bibinfo {author} {\bibfnamefont {T.}~\bibnamefont
  {Hansson}}, \bibinfo {author} {\bibfnamefont {V.}~\bibnamefont {Oganesyan}},
  \ and\ \bibinfo {author} {\bibfnamefont {S.}~\bibnamefont {Sondhi}},\
  }\href@noop {} {\bibfield  {journal} {\bibinfo  {journal} {Annals of
  Physics}\ }\textbf {\bibinfo {volume} {313}},\ \bibinfo {pages} {497}
  (\bibinfo {year} {2004})}\BibitemShut {NoStop}%
\bibitem [{Note1()}]{Note1}%
  \BibitemOpen
  \bibinfo {note} {More generally, the term `higher rank gauge theory' refers
  to any symmetric gauge structure whose gauge transformation contains higher
  order differential forms.}\BibitemShut {Stop}%
\bibitem [{\citenamefont {Pretko}(2018)}]{pretko2018fracton}%
  \BibitemOpen
  \bibfield  {author} {\bibinfo {author} {\bibfnamefont {M.}~\bibnamefont
  {Pretko}},\ }\href@noop {} {\bibfield  {journal} {\bibinfo  {journal}
  {Physical Review B}\ }\textbf {\bibinfo {volume} {98}},\ \bibinfo {pages}
  {115134} (\bibinfo {year} {2018})}\BibitemShut {NoStop}%
\bibitem [{\citenamefont {Slagle}\ \emph
  {et~al.}(2018{\natexlab{b}})\citenamefont {Slagle}, \citenamefont {Aasen},\
  and\ \citenamefont {Williamson}}]{slagle2018foliated}%
  \BibitemOpen
  \bibfield  {author} {\bibinfo {author} {\bibfnamefont {K.}~\bibnamefont
  {Slagle}}, \bibinfo {author} {\bibfnamefont {D.}~\bibnamefont {Aasen}}, \
  and\ \bibinfo {author} {\bibfnamefont {D.}~\bibnamefont {Williamson}},\
  }\href@noop {} {\bibfield  {journal} {\bibinfo  {journal} {arXiv preprint
  arXiv:1812.01613}\ } (\bibinfo {year} {2018}{\natexlab{b}})}\BibitemShut
  {NoStop}%
\bibitem [{\citenamefont {Pai}\ and\ \citenamefont
  {Hermele}(2019)}]{pai2019fracton}%
  \BibitemOpen
  \bibfield  {author} {\bibinfo {author} {\bibfnamefont {S.}~\bibnamefont
  {Pai}}\ and\ \bibinfo {author} {\bibfnamefont {M.}~\bibnamefont {Hermele}},\
  }\href@noop {} {\bibfield  {journal} {\bibinfo  {journal} {arXiv preprint
  arXiv:1903.11625}\ } (\bibinfo {year} {2019})}\BibitemShut {NoStop}%
\bibitem [{\citenamefont {Xu}\ and\ \citenamefont
  {Wu}(2008)}]{xu2008resonating}%
  \BibitemOpen
  \bibfield  {author} {\bibinfo {author} {\bibfnamefont {C.}~\bibnamefont
  {Xu}}\ and\ \bibinfo {author} {\bibfnamefont {C.}~\bibnamefont {Wu}},\
  }\href@noop {} {\bibfield  {journal} {\bibinfo  {journal} {Physical Review
  B}\ }\textbf {\bibinfo {volume} {77}},\ \bibinfo {pages} {134449} (\bibinfo
  {year} {2008})}\BibitemShut {NoStop}%
\bibitem [{\citenamefont {Huang}\ \emph {et~al.}(2018)\citenamefont {Huang},
  \citenamefont {Prem}, \citenamefont {Song},\ and\ \citenamefont
  {Hermele}}]{huang2018cage}%
  \BibitemOpen
  \bibfield  {author} {\bibinfo {author} {\bibfnamefont {S.-J.}\ \bibnamefont
  {Huang}}, \bibinfo {author} {\bibfnamefont {A.}~\bibnamefont {Prem}},
  \bibinfo {author} {\bibfnamefont {H.}~\bibnamefont {Song}}, \ and\ \bibinfo
  {author} {\bibfnamefont {M.}~\bibnamefont {Hermele}},\ }\href@noop {}
  {\bibfield  {journal} {\bibinfo  {journal} {Bulletin of the American Physical
  Society}\ } (\bibinfo {year} {2018})}\BibitemShut {NoStop}%
\bibitem [{\citenamefont {Fradkin}\ and\ \citenamefont
  {Schaposnik}(1991)}]{fradkin1991chern}%
  \BibitemOpen
  \bibfield  {author} {\bibinfo {author} {\bibfnamefont {E.}~\bibnamefont
  {Fradkin}}\ and\ \bibinfo {author} {\bibfnamefont {F.~A.}\ \bibnamefont
  {Schaposnik}},\ }\href@noop {} {\bibfield  {journal} {\bibinfo  {journal}
  {Physical review letters}\ }\textbf {\bibinfo {volume} {66}},\ \bibinfo
  {pages} {276} (\bibinfo {year} {1991})}\BibitemShut {NoStop}%
\bibitem [{Note2()}]{Note2}%
  \BibitemOpen
  \bibinfo {note} {This only applies to the case where $D_1,D_2$ do not share
  any common factor. Otherwise, even the magnetic flux fluctuation is fixed,
  there might exist some local operator with lower order exhibiting a
  dispersive gapless mode.}\BibitemShut {Stop}%
\bibitem [{Note3()}]{Note3}%
  \BibitemOpen
  \bibinfo {note} {For general differential polynomial $D_i^{(e)}, D_i^{(o)}$,
  the coefficient in each differential term is dimensionful so such operator is
  only well-defined on the lattice}\BibitemShut {NoStop}%
\bibitem [{\citenamefont {Slagle}\ and\ \citenamefont
  {Kim}(2017{\natexlab{d}})}]{slagle2017fracton}%
  \BibitemOpen
  \bibfield  {author} {\bibinfo {author} {\bibfnamefont {K.}~\bibnamefont
  {Slagle}}\ and\ \bibinfo {author} {\bibfnamefont {Y.~B.}\ \bibnamefont
  {Kim}},\ }\href@noop {} {\bibfield  {journal} {\bibinfo  {journal} {Phys.
  Rev. B}\ }\textbf {\bibinfo {volume} {96}},\ \bibinfo {pages} {165106}
  (\bibinfo {year} {2017}{\natexlab{d}})}\BibitemShut {NoStop}%
\bibitem [{\citenamefont {{You}}\ \emph {et~al.}(2018)\citenamefont {{You}},
  \citenamefont {{Devakul}}, \citenamefont {{Burnell}},\ and\ \citenamefont
  {{Sondhi}}}]{2018arXiv180302369Y}%
  \BibitemOpen
  \bibfield  {author} {\bibinfo {author} {\bibfnamefont {Y.}~\bibnamefont
  {{You}}}, \bibinfo {author} {\bibfnamefont {T.}~\bibnamefont {{Devakul}}},
  \bibinfo {author} {\bibfnamefont {F.~J.}\ \bibnamefont {{Burnell}}}, \ and\
  \bibinfo {author} {\bibfnamefont {S.~L.}\ \bibnamefont {{Sondhi}}},\
  }\href@noop {} {\bibfield  {journal} {\bibinfo  {journal} {ArXiv e-prints}\ }
  (\bibinfo {year} {2018})},\ \Eprint {http://arxiv.org/abs/1803.02369}
  {arXiv:1803.02369 [cond-mat.str-el]} \BibitemShut {NoStop}%
\bibitem [{\citenamefont {{Song}}\ \emph {et~al.}(2018)\citenamefont {{Song}},
  \citenamefont {{Prem}}, \citenamefont {{Huang}},\ and\ \citenamefont
  {{Martin-Delgado}}}]{2018arXiv180506899S}%
  \BibitemOpen
  \bibfield  {author} {\bibinfo {author} {\bibfnamefont {H.}~\bibnamefont
  {{Song}}}, \bibinfo {author} {\bibfnamefont {A.}~\bibnamefont {{Prem}}},
  \bibinfo {author} {\bibfnamefont {S.-J.}\ \bibnamefont {{Huang}}}, \ and\
  \bibinfo {author} {\bibfnamefont {M.~A.}\ \bibnamefont {{Martin-Delgado}}},\
  }\href@noop {} {\bibfield  {journal} {\bibinfo  {journal} {ArXiv e-prints}\ }
  (\bibinfo {year} {2018})},\ \Eprint {http://arxiv.org/abs/1805.06899}
  {arXiv:1805.06899 [cond-mat.str-el]} \BibitemShut {NoStop}%
\bibitem [{\citenamefont {Eliezer}\ and\ \citenamefont
  {Semenoff}(1992)}]{eliezer199266}%
  \BibitemOpen
  \bibfield  {author} {\bibinfo {author} {\bibfnamefont {D.}~\bibnamefont
  {Eliezer}}\ and\ \bibinfo {author} {\bibfnamefont {G.}~\bibnamefont
  {Semenoff}},\ }\href {\doibase https://doi.org/10.1016/0003-4916(92)90339-N}
  {\bibfield  {journal} {\bibinfo  {journal} {Annals of Physics}\ }\textbf
  {\bibinfo {volume} {217}},\ \bibinfo {pages} {66 } (\bibinfo {year}
  {1992})}\BibitemShut {NoStop}%
\bibitem [{\citenamefont {Sun}\ \emph {et~al.}(2015)\citenamefont {Sun},
  \citenamefont {Kumar},\ and\ \citenamefont {Fradkin}}]{PhysRevB.92.115148}%
  \BibitemOpen
  \bibfield  {author} {\bibinfo {author} {\bibfnamefont {K.}~\bibnamefont
  {Sun}}, \bibinfo {author} {\bibfnamefont {K.}~\bibnamefont {Kumar}}, \ and\
  \bibinfo {author} {\bibfnamefont {E.}~\bibnamefont {Fradkin}},\ }\href
  {\doibase 10.1103/PhysRevB.92.115148} {\bibfield  {journal} {\bibinfo
  {journal} {Phys. Rev. B}\ }\textbf {\bibinfo {volume} {92}},\ \bibinfo
  {pages} {115148} (\bibinfo {year} {2015})}\BibitemShut {NoStop}%
\bibitem [{\citenamefont {Witten}(1991)}]{witten1991quantization}%
  \BibitemOpen
  \bibfield  {author} {\bibinfo {author} {\bibfnamefont {E.}~\bibnamefont
  {Witten}},\ }\href@noop {} {\bibfield  {journal} {\bibinfo  {journal}
  {Communications in Mathematical Physics}\ }\textbf {\bibinfo {volume}
  {137}},\ \bibinfo {pages} {29} (\bibinfo {year} {1991})}\BibitemShut
  {NoStop}%
\bibitem [{\citenamefont {{Wang}}\ \emph {et~al.}()\citenamefont {{Wang}},
  \citenamefont {{Shirley}},\ and\ \citenamefont {{Chen}}}]{chen2019}%
  \BibitemOpen
  \bibfield  {author} {\bibinfo {author} {\bibfnamefont {T.}~\bibnamefont
  {{Wang}}}, \bibinfo {author} {\bibfnamefont {W.}~\bibnamefont {{Shirley}}}, \
  and\ \bibinfo {author} {\bibfnamefont {X.}~\bibnamefont {{Chen}}},\
  }\href@noop {} {\bibinfo  {journal} {arXiv preprint arXiv:1904.01111}\
  }\BibitemShut {NoStop}%
\bibitem [{\citenamefont {Shirley}\ \emph {et~al.}(2019)\citenamefont
  {Shirley}, \citenamefont {Slagle},\ and\ \citenamefont
  {Chen}}]{shirley2019universal}%
  \BibitemOpen
\bibfield  {journal} {  }\bibfield  {author} {\bibinfo {author} {\bibfnamefont
  {W.}~\bibnamefont {Shirley}}, \bibinfo {author} {\bibfnamefont
  {K.}~\bibnamefont {Slagle}}, \ and\ \bibinfo {author} {\bibfnamefont
  {X.}~\bibnamefont {Chen}},\ }\href@noop {} {\bibfield  {journal} {\bibinfo
  {journal} {SciPost Physics}\ }\textbf {\bibinfo {volume} {6}},\ \bibinfo
  {pages} {015} (\bibinfo {year} {2019})}\BibitemShut {NoStop}%
\bibitem [{Note4()}]{Note4}%
  \BibitemOpen
  \bibinfo {note} {For $s=2$, the $\pm \pi $ fluxes are equivalent, so in this
  case the full theory has cubic symmetry.}\BibitemShut {Stop}%
\end{thebibliography}%
\end{document}